\documentclass[aps,prb,twocolumn,amsmath,amssymb,superscriptaddress,nofootinbib]{revtex4-2}
\usepackage{titlesec}
\usepackage{graphicx}
\usepackage{subfigure}
\usepackage{mathrsfs}
\usepackage{amsthm}
\usepackage{amsmath}
\usepackage{float}
\usepackage{color}
\usepackage{braket}
\usepackage[
            pdfstartview=FitH,
            CJKbookmarks=true,
            bookmarksnumbered=true,
            bookmarksopen=true,
            colorlinks, 
            pdfborder=001,
            linkcolor=blue,
            anchorcolor=blue,
            citecolor=blue
            ]{hyperref}

\begin{document}

\title{Fermi Energy Sensitive Universal Conductance Fluctuations in Anisotropic Materials}

\author{Qiang Yang}
\affiliation{School of Physics and Technology, Nanjing Normal University, Nanjing 210023, People’s Republic of China}

\author{Yayun Hu}
\email{yyhu@zhejianglab.edu.cn}
\affiliation{Zhejiang Lab, Hangzhou 311100, People’s Republic of China}

\author{Zhe Hou}
\affiliation{School of Physics and Technology, Nanjing Normal University, Nanjing 210023, People’s Republic of China}

\author{Peiqing Tong}
\email{pqtong@njnu.edu.cn}
\affiliation{School of Physics and Technology, Nanjing Normal University, Nanjing 210023, People’s Republic of China}
\affiliation{Ministry of Education Key Laboratory of Numerical Simulation of Large Scale Complex Systems, Nanjing Normal University, Nanjing 210023, People’s Republic of China}

\date{\today}

\begin{abstract}
Universal conductance fluctuations (UCF) are a hallmark of quantum interference in mesoscopic devices. According to the Altshuler-Lee-Stone theory, the amplitude of UCF remains independent of system parameters such as Fermi energy and disorder strength. However, recent experiments have demonstrated a significant variation in UCF with respect to Fermi energy in the anisotropic Dirac semimetal $\mathrm{Cd_3As_2}$, suggesting a dependence on band anisotropy. In this work, we reconcile the discrepancy between theoretical predictions and experimental observations through a detailed study of UCF versus Fermi energy using a tight-binding model with tunable anisotropy parameters. Near the band edge, the Hamiltonian is simplified to an anisotropic free electron gas model, recovering the generalized Altshuler-Lee-Stone theory. However, as the Fermi energy shifts toward the band center, where rotational symmetry breaks into $C_4$ (four-fold rotational) symmetry, the UCF amplitude deviates from the standard theory. Our findings reveal that UCF becomes increasingly sensitive to Fermi energy as the anisotropy grows stronger. Furthermore, using realistic parameters for $\mathrm{Cd_3As_2}$, our calculations demonstrate an increase in UCF away from the Dirac point, in qualitative agreement with experimental results. The enhancement of UCF occurs in two perpendicular transport directions that we have calculated, albeit with quantitative differences in magnitude, which can be tested in future experiments. Given the prevalence of anisotropic materials and technical advances in engineering anisotropy through strain or twist, our results offer a valuable reference for characterizing intrinsic electronic properties via UCF.
\end{abstract}

\maketitle

\section{Introduction}
Universal conductance fluctuation (UCF) manifests itself as aperiodic and reproducible fluctuations superimposed on the ohmic resistance of disordered mesoscopic devices when scanning the gate voltage or external magnetic field~\cite{altshuler1985fluctuations,lee1985universal,stone1985magnetoresistance}. The aperiodic fluctuations in diffusive transport stem from the fundamental quantum interference among all the phase-coherent Feynman paths of the traversing electrons. Thus, UCF serves as fingerprints for metallic materials due to their sensitivity to complex impurity configurations and sample-specific characteristics~\cite{ALLEN2006,akkermans2007mesoscopic,Saurav2023benchmarking}. Despite that, the Altshuler-Lee-Stone theory has established a connection between the universal statistical properties of conductance fluctuations and the valuable information of materials, such as the dimensions and symmetry classes~\cite{lee1987universal,altshuler1986repulsion,beenakker1997random}. For example, in a standard two-terminal setup, diagrammatic calculations using a free electron-gas model have found that the intrinsic amplitude of UCF at zero-temperature is determined by the number of uncorrelated bands $k$, the level degeneracy $s$, the symmetry parameter $\beta$, as well as the shape of the material, which reads~\cite{lee1987universal}:
\begin{equation}\label{eq:deltaGdepend}
	\Delta G = c_d\sqrt{\frac{ks^2}{\beta}}
\end{equation}
in units of $\frac{e^2}{h}$. Here $\beta = 1, 2, 4$ for orthogonal, unitary and symplectic systems, respectively, and the effect of dimension is incorporated in the prefactor $c_d$. Equation ~\ref{eq:deltaGdepend} highlights the UCF in characterizing the materials' symmetry class and their dimension properties in the diffusive regime. Its value is universal and independent of the materials' details, such as the magnitude of conductance, system size, Fermi energy, disorder strength, and the strength of magnetic fields applied to samples.

According to Eq.~\ref{eq:deltaGdepend}, the integers $k,s,\beta$ and the prefactor $c_d$ contribute to the UCF amplitude independently. The formers' contribution is evidenced by the experimentally observed decay of the UCF amplitude upon the application of an external magnetic field, which suppresses the Cooperon mode and turns $\beta=1$ or $4$ into $\beta=2$~\cite{debray1989reduction,moon1997observation,pal2012direct,wang2016universal,zhang2018}. The latter contribution from $c_d$ has a much more complex parameter dependence. Before touching on the main focus of our work on the Fermi energy dependence of $c_d$, we briefly review several known aspects. In an isotropic free electron gas considered by the Altshuler-Lee-Stone theory, the coefficient $c_d$ is a function of the system size $L_{\alpha}(\alpha=x,y,z)$ for metallic samples. It becomes quite sensitive to $L_{x/y}$ when the transversal directions $L_{x/y}$ are longer than the longitudinal transport direction $L_z$, and diverges in the limit of $L_{x/y}\gg L_z$~\cite{Hu2017,ruhlander2001symmetry}. Moreover, the value of $c_d$ also varies significantly for systems beyond the isotropic free-electron gas model. Further developments of the theory have addressed several aspects of interest in the conductance fluctuations, including the proximity to a superconductor~\cite{takane1992conductance,ryu2007conductance,antonenko2020mesoscopic}, the positions of leads and geometry of the probing devices~\cite{zyuzin1990mesoscopic,chandrasekhar1991weak}, boundary conditions~\cite{ruhlander2001symmetry}, the non-integer dimensions~\cite{travvenec2004universal,amin2022multifractal}, the near-ballistic regime~\cite{asano1996conductance}, spin-orbital coupling~\cite{adam2006mesoscopic,bardarson2007mesoscopic,scheid2009anisotropic,kaneko2010symmetry} and band anisotropy~\cite{Hu2017}, etc. Despite the factors that modify $c_d$, it is generally considered as an irrelevant constant of order unity for a given material in the literature for several reasons. First, for a given experimental device or material, it is not easy to calculate the specific value of $c_d$ using realistic parameters. Instead, applying the Altshuler-Lee-Stone theory to estimate its magnitude in experiments is convenient because of its simplicity and broad applicability. Second, the relative variance of $c_d$ is usually not sensitive to parameters such as Fermi energy or magnetic field, partially because these fields only vary within a small range. Third, for most experimentally relevant quasi-1D nanowire devices, $c_d$ is pinned at 0.365, which is robust against other perturbations. Above all, the role of $c_d$ is not as much emphasized in most of the literature. 

Thus, it becomes essential to recalibrate $c_d$ when it is sufficiently sensitive to external parameters like the Fermi energy and induces changes comparable to those caused by the symmetry indices. One such case is the study of UCF in topological materials, which has witnessed considerable theoretical interest~\cite{ren2006universal,takagaki2012conductance,zhang2014universal,choe2015universal,vasconcelos2016universal,kechedzhi2008quantum,kharitonov2008universal,qiao2008universal,rycerz2007anomalously,rossi2012universal,Hu2017,hsu2018conductance,aslani2019conductance,li2009dorokhov,Louis2020,shafiei2024tailoring} as well as experimental progress~\cite{chen2010magnetoresistance,kandala2013surface,li2014indications,trivedi2016weak,wang2016universal,Islam2018,andersen2023universal,pal2012direct,matsuo2013experimental,Xiao23}. In topological materials, the manipulation of integers $k, s, \beta$ by sweeping across a significant range of Fermi energy and magnetic field allows the amplitude of UCF to be adjusted between various regimes. A celebrated example is the 2D Dirac semimetal graphene, where transitions between UCF plateaus have been observed when the Fermi energy is gated away from the Dirac point, due to distinct valley and spin degeneracies~\cite{pal2012direct,kharitonov2008universal,kechedzhi2008quantum,staley2008suppression}. 
Subsequent work has also predicted Fermi-energy-dependent UCF in several topological materials~\cite{Hu2017a,Hu2017,hsu2018conductance,aslani2019conductance}. In particular, Ref.~\cite{Hu2017a} has predicted the increase of $c_d$ when the Fermi energy moves away from the Dirac point and attributed it to the effects of band anisotropy of the 3D Dirac semimetal $\mathrm{Cd_3As_2}$. Such an increase of UCF has been observed in a recent experiment in Ref.~\cite{Xiao23}. Although one can not exclude possible contributions to this phenomenon from other mechanism~\cite{hsu2018conductance,aslani2019conductance,wang2019non,chen2015disorder,liu2016conductance,alagha2010universal,li2014indications}, it is believed that the main reason stems from the band anisotropy.

The experimental observation~\cite{Xiao23} raises a natural and unaddressed question about the sensitivity of $c_d$ to Fermi energy in anisotropic materials. This question is relevant to future studies on UCF because anisotropic materials are abundant in nature. Advanced technologies are well developed to engineer anisotropy through strain or twist ~\cite{cao2018unconventional,cao2018correlated,edelberg2020tunable,bai2020excitons,kennes2021moire,shafiei2024tailoring,Hou24}. For example, the UCF of the twisted bilayer graphene quantum dot ~\cite{Hou24} shows a negligible dependence on the twisting angles, suggesting that materials with stronger anisotropy than graphene are necessary for a UCF dependent on the twist angle. This is also a nontrivial problem because it contradicts popular beliefs that the UCF is Fermi energy independent in general, according to the Altshuler-Lee-Stone theory. 

This paper focuses on the Fermi energy dependence of $c_d$ in anisotropic materials. For this purpose, we study the nature of UCF in a single-band tight-binding model in the presence of band anisotropy. This model captures the topic in two aspects. First, it simplifies the analysis of symmetry indices by fixing $k=s=\beta=1$ regardless of the Fermi energy. Second, it connects the ellipsoidal Fermi surface to a more complex Fermi surface that can only be dealt with numerically by merely tuning the Fermi energy from the band edge to the band center.

Starting from a continuous free electron-gas model, we analytically determine that $c_d$ in Eq.~\ref{eq:deltaGdepend} should be replaced by $\tilde{c}_d(t_x,t_y,t_z)$. This new term is a function of the diffusion time $t_{\alpha}$ $(\alpha=x,y,x)$ along the dimension $L_{\alpha}$ in the presence of anisotropic dispersion. Numerical evidence and physical arguments are provided for the analytic results. Our analysis reveals an equivalence between the band anisotropy and the spatial anisotropy in $L_{\alpha}$, and identifies regions of the anisotropy parameter where $\tilde{c}_d$ is sensitive or insensitive to the band anisotropy. Then, we show numerically that, in the parameter regions of sensitivity, the coefficient $\tilde{c}_d$, and thus the UCF amplitude is sensitive to the Fermi energy, which doubles the UCF for realistic anisotropy parameters accessible to ordinary materials. These conclusions hold for 2D and 3D materials. Finally, we use realistic material parameters for $\mathrm{Cd_3As_2}$, and demonstrate the Fermi energy dependence of UCF, in consistent with the experiments. We also find that the increase of the UCF versus Fermi energy is a robust feature in $\mathrm{Cd_3As_2}$, not necessarily requiring transport along the $L_z$ direction, which was assumed in the previous theory~\cite{Xiao23,Hu2017a}. These results should be amenable to future experimental verifications. Combined with the well-developed method to tune the Fermi energy through gate voltages, our findings will allow for delicate control of the UCF in future device design using materials with anisotropic band structures.

The organization of this paper is as follows. Section ~\ref{modelmethod} gives the standard tight-binding model and the methods to calculate the conductance. Section ~\ref{anisotropicTheoryData} generalizes the UCF theory to an anisotropic free electron-gas model and verifies the theory using numerical calculations. Section ~\ref{FermiEnergyDependence} showcases the significant Fermi energy dependence of UCF as found in both 2D and 3D tight-binding models with band anisotropy. Section ~\ref{ExpAndMaterial} examines the Fermi energy dependence of UCF in $\mathrm{Cd_3As_2}$ using realistic parameters. Section~\ref{conclusion} concludes the paper with discussions.

\section{Model and method}\label{modelmethod}
To study the effect of anisotropy on UCF, we first use a free electron-gas model with tunable anisotropic dispersions, described by the Hamiltonian: 
\begin{equation}\label{h0_continous}
	H_0 (\mathbf{k})= M_0 + M_xk_x^2 + M_yk_y^2 + M_zk_z^2,
\end{equation}
where $M_{0}$ and $M_{\alpha}$ ($\alpha=x,y,z$) are tunable parameters, and $\mathbf{k}=(k_x, k_y, k_z)$ is the wavevector. 

Next, we discretize the continuum model into a lattice system as follows
\begin{equation}\label{h0_discrete}
	H_0 = V\sum_\mathbf{r}a_\mathbf{r}^\dagger a_\mathbf{r} + \Big(\sum_{\mathbf{r},\alpha=x,y,z}T_\alpha a_\mathbf{r}^\dagger a_{\mathbf{r}+\delta\boldsymbol{\alpha}} + H.c.\Big),
\end{equation} 
with
\begin{align}
	V &= M_0 + \frac{2M_x}{a^2} + \frac{2M_y}{b^2} + \frac{2M_z}{c^2},  \\
	T_x &= \frac{M_x}{a^2}, T_y = \frac{M_y}{b^2}, T_z = \frac{M_z}{c^2}, \label{hopping}
\end{align}
where $a_\mathbf{r}(a_\mathbf{r}^\dagger)$ represents the electron annihilation(creation) operators on site $\mathbf{r}$. $\delta\boldsymbol{\alpha}$ is the primitive lattice vector along the $\alpha$ direction and $|\delta\boldsymbol{\alpha}|=a$, $b$ and $c$ for $\alpha=x$, $y$ and $z$, respectively. $T_{\alpha}$ represents the hopping term along the $\alpha$ direction, and $V$ is the on-site energy. 

To study electronic transport, we consider a two-terminal device consisting of a central region and two semi-infinite leads. The leads can be modeled by $H_0$, whereas the central region is modeled by: $H^{cen} = H_0 + U(\mathbf{r})$, where $U(\mathbf{r})=\sum_{\mathbf{r}}\varepsilon(\mathbf{r})a_{\mathbf{r}}^{\dagger}a_{\mathbf{r}}$ are the uniform random potentials on site, with $\varepsilon(\mathbf{r})$ distributed uniformly within $[\frac{-W}{2}, \frac{W}{2}]$, and $W$ the strength of the disorder. Throughout this paper, we fix $M_z=0.5$ and $a=b=c=1$, with the unit of energy and length set to unity unless otherwise noted. For simplicity, we neglect the spin degree of freedom. The band edges of Eq.~\ref{h0_discrete} are zero energy at the bottom and $4(M_x+M_y+M_z)$ at the top, respectively. A discussion of its Fermi surface and density of states is presented in Appendix ~\ref{fermisurfacefree}. 

Numerically, we compute the zero-temperature conductance by the Landauer-B$\ddot{\mathrm{u}}$ttiker formula that can be derived from the non-equilibrium Green's function method as~\cite{Meir92Landauer, Dat, Landauer, Fisher, Butti, zhang2019electronic, Mac, Met}, 
\begin{equation}
    G = \frac{e^2}{h}{\rm Tr}[\Gamma_LG^r\Gamma_RG^a] \label{nonequilibrium}
\end{equation}
where $G^{r/a}(E_F) =[(E_F \pm i\eta) {\textbf I} - {\textbf H}^{cen} - \Sigma_L - \Sigma_R]^{-1}$ is the retarded(advanced) Green function. ${\textbf H}^{cen}$ is the Hamiltonian matrix of the central region, $\textbf I$ is the identity matrix, $\eta$ is an infinitesimal positive number, $\Gamma_{L/R} = i(\Sigma_{L/R}^r - \Sigma_{L/R}^a)$ is the line width function, and $\Sigma_{L/R}$ is the self-energy of the left (right) lead. In our work, ${\textbf H}^{cen}$ is taken to be either Eq.~\ref{h0_discrete} or Eq.~\ref{cd3as2_h_discretize} below with disorder. The conductance fluctuation $\Delta G$ is the standard deviation of conductance $G$, defined as $\Delta G=\sqrt{\langle (G - \langle G \rangle)^2\rangle }$, calculated over an ensemble of more than 1200 disorder configurations.

\section{Dependence of UCF on Band Anisotropy}\label{anisotropicTheoryData}

The diagrammatic calculations of UCF for isotropic systems with $M_x=M_y=M_z$ in Eq.~\ref{h0_continous} have been derived in previous works~\cite{altshuler1985fluctuations,lee1985universal,lee1987universal}. In the following, we generalize these results to anisotropic systems with unequal $M_{\alpha}$. The same issue has been addressed in the Appendix of ~\cite{Hu2017}; here, we present a more comprehensive analysis with insightful physical pictures and careful numerical verifications. In Section ~\ref{anisotropicBand}, we argue that such generalizations only replace the eigenvalues of the propagators in the original results in Ref.~\cite{lee1987universal} and are equivalent to introducing an anisotropy in the spatial dimension, i.e., unequal $L_{\alpha}$ in the system size. In Section ~\ref{anisotropicBandNumerical}, we provide numerical calculations of UCF to test against and confirm the generalized theory directly.

\subsection{Theoretical results of UCF for anisotropic band dispersion}\label{anisotropicBand}
\begin{figure}
	\centering
	\includegraphics[width=1.0\linewidth]{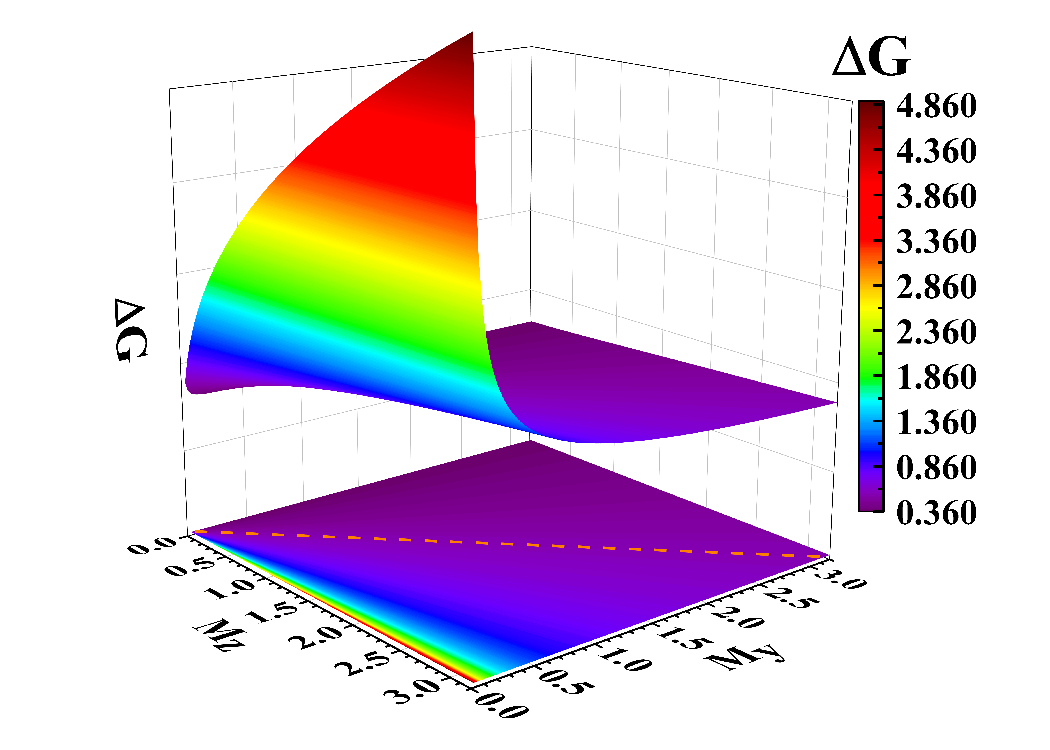}
	\caption{(Color online) Variation of Universal Conductance Fluctuations $\Delta G$ (in units of $\frac{e^2}{h}$) versus anisotropy in momentum space, as determined using Eq.~\ref{dgformula1} in Appendix ~\ref{CalculateDeltaG}. This analysis maintains isotropy in 3D real space ($L_x = L_y = L_z$). Momentum space anisotropy is modulated by $M_y$ and $M_z$ in a step size of 0.02, with $M_x = M_y$. The transport direction is along the $z$-axis. The UCF is sensitive (insensitive) to anisotropy for $M_z/M_y\geq1$ ($M_z/M_y<1$). The dashed line separates the region of sensitivity and insensitivity. }\label{FigDeltaGvsMzy}
\end{figure}

\begin{figure*}
	\centering
	\includegraphics[width=1.0\linewidth]{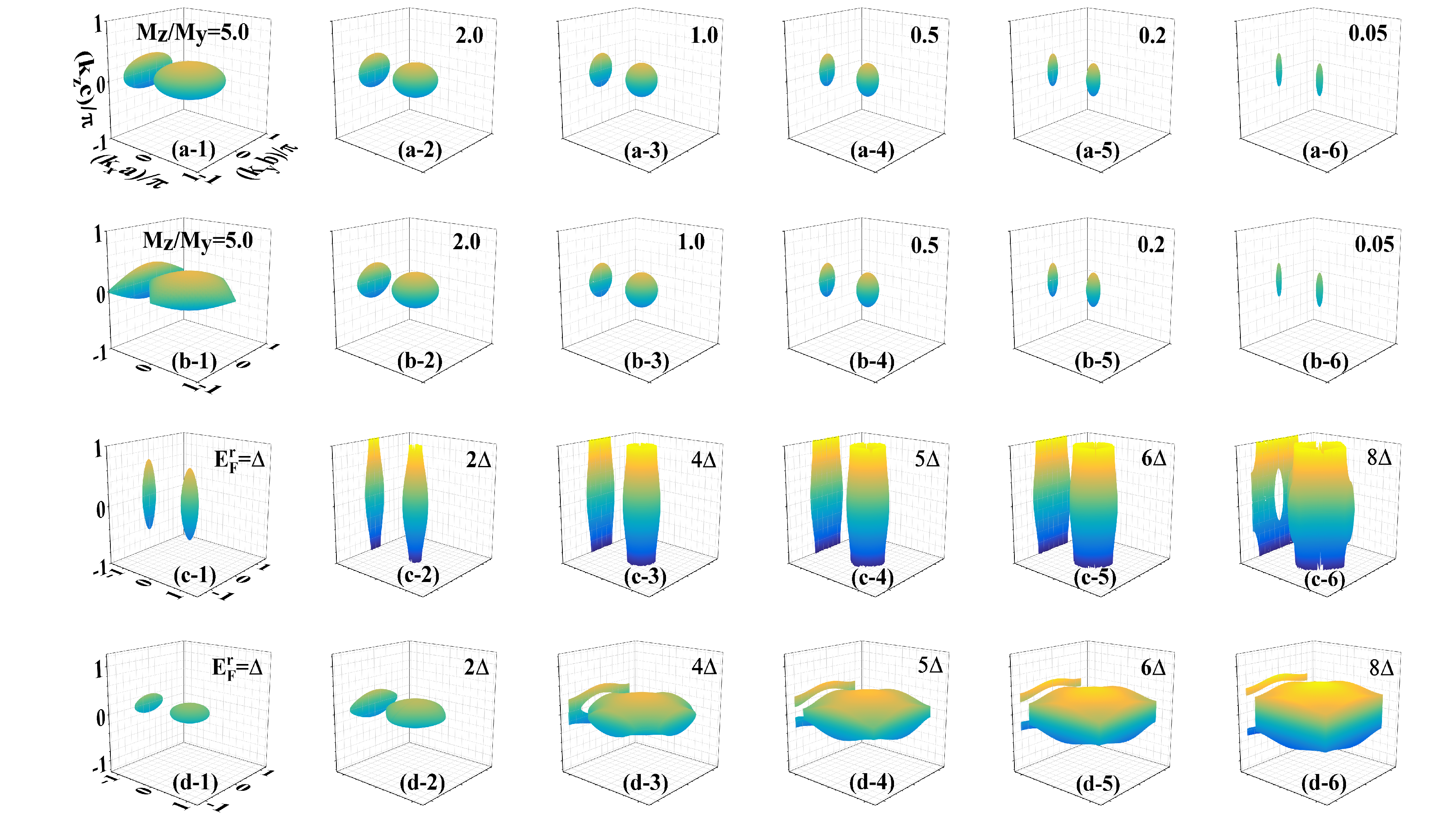}
\caption{ (Color online) The shapes of the Fermi surface in the momentum space $(k_x, k_y, k_z)$, calculated for the (a) continuous Hamiltonian in Eq.~\ref{h0_continous} and (b-d) discretized Hamiltonian in Eq.~\ref{h0_discrete}, as well as their projections onto the $k_y-k_z$ plane. $M_z=0.5$ is fixed, and $M_x=M_y$. (a-1)-(a-6)  The Fermi energy $E_F=0.4$ is fixed for different anisotropic parameters $\frac{M_z}{M_y}=5.0, 2.0, 1.0, 0.5, 0.2, 0.05$. (b-1)-(b-6) The same parameters as in (a) but for the discretized Hamiltonian. (c-1)-(c-6) The anisotropy parameter $\frac{M_z}{M_y}=0.2$ is fixed for different reduced Fermi energy $E^r_F=\Delta, 2\Delta, 4\Delta, 5\Delta, 6\Delta$ and $8\Delta$, respectively. $\Delta=(M_x+M_y+M_z)/4$ such that the bandwidth is $16\Delta$ and $8\Delta$ is at the band center. (d-1)-(d-6) Same as in (c) but for the anisotropy parameter $\frac{M_z}{M_y}=5.0$. In (a-d), the labeling ticks of the axis are the same as those in (a-1).}\label{fermisurfaces}
\end{figure*}

In this section, we obtain the UCF for an anisotropic band following Ref.~\cite{lee1987universal}, whose starting point is the impurity-averaged perturbation theory and does not assume the isotropy of the system as a precondition. Without going into the details of the derivations, we list several key modifications due to anisotropic dispersions and use the existing result in Ref.~\cite{lee1987universal} to simplify the deduction. 

We first note that the diagrammatic techniques remain valid if we substitute the original isotropic Hamiltonian in the building blocks of the derivation, i.e., the disorder averaged the retarded/advanced Green function~\cite{abrikosov2012methods} $\langle G^{r/a}(E, \mathbf{k})\rangle=\langle(E-H_0(\mathbf{k}) \pm \frac{i}{2\tau} )^{-1}\rangle$, by our anisotropic Hamiltonian in Eq.~\ref{h0_continous}. For anisotropic systems with quadratic energy dispersion and random disorder, the elastic scattering time $\tau$ remains constant~\cite{Rudenko2024,ALLEN2006,Yuan15}. As a result, the diffusion propagator $P(\mathbf{r}, \mathbf{r}^\prime)$ in Eq.(A4) of Ref.~\cite{lee1987universal} now satisfies an anisotropic diffusion equation as
\begin{equation}
    (-D_x\nabla_x^2-D_y\nabla_y^2-D_z\nabla_z^2)P(\mathbf{r}, \mathbf{r}^\prime)=(1/\tau)\delta(\mathbf{r}-\mathbf{r}^\prime),
\end{equation}
where $D_\alpha=(v^{\alpha}_{F})^2/d$ and $v^{\alpha}_{F}$ the maximal Fermi velocity along the $\alpha$ direction. Here, we have taken the correlation energy $\Delta E=0$ as is defined for UCF and omitted the inelastic scattering term, which vanishes in the zero temperature limit considered in this paper.  
The replacement of the diffusion constant by an anisotropic diffusion vector $\mathbf{D}=(D_x, D_y, D_z)$ immediately leads to  
the following modifications to the eigenvalues in Eq.(A10) of Ref.~\cite{lee1987universal}:
\begin{equation}\label{modifiedeigenvalue}
\lambda_n = \tau D_z(\frac{\pi}{L_z})^2[n_z^2 + n_x^2\frac{D_x}{D_z}\frac{L_z^2}{L_x^2} + n_y^2\frac{D_y}{D_z}\frac{L_z^2}{L_y^2}].
\end{equation}
where $n_\alpha$ are integer labels of the eigenvalues.

Modifying the energy scale in Eq.(A12) of Ref.~\cite{lee1987universal} as $E_c=D_z(\pi/L_z)^2$ we can redefine the reduced eigenvalue 
\begin{align}\label{modifiedeigenvalue_new}
	\tilde{\lambda}_n = \lambda_n/(E_c\tau) &= n_z^2 + n_x^2\frac{D_x}{D_z}\frac{L_z^2}{L_x^2} + n_y^2\frac{D_y}{D_z}\frac{L_z^2}{L_y^2} \\ \notag
     &= n_z^2 + n_x^2\frac{t_z}{t_x} + n_y^2\frac{t_z}{t_y},
\end{align}
where we have used the relation $t_{\alpha}=L_{\alpha}^2/(2D_{\alpha})$ in the last line. As shown in Appendix ~\ref{CalculateDeltaG}, the reduced eigenvalues $\tilde{\lambda}_n$ in Eq.~\ref{modifiedeigenvalue_new} are sufficient to determine $c_d$ in Eq.~\ref{eq:deltaGdepend}, which in turn gives the UCF amplitude $\Delta G$.  

The comments on each line in Eq.~\ref{modifiedeigenvalue_new} are presented as follows. For isotropic systems with $D_\alpha=D$, the prefactor $c_d(L_x, L_y, L_z)$ is a function of the ratios of the transverse system sizes $L_x$ and $L_y$ over the longitudinal system size $L_z$ in the transport direction. This is a well-known fact of the spatial dimension dependence of UCF. For anisotropic systems, however, the anisotropic diffusion vector effectively modifies the $L_\alpha$ dependence, and $c_d$ is replaced by $\tilde{c}_d(t_x, t_y, t_z)$ that explicitly depends on the ratios of diffusion time across the three dimensions of the sample.

The diffusion constant is inversely proportional to the effective band mass~\cite{ashcroft1976solid}, given by $\frac{1}{2M_\alpha}$ according to Eq.~\ref{h0_continous}. We have $D_\alpha\propto M_{\alpha}$ (note that in our definition, the coefficient $M_\alpha$ is inversely proportional to the effective band mass). We can rewrite $\tilde{\lambda}_n$ in Eq.~\ref{modifiedeigenvalue_new} as
\begin{equation}\label{modifiedeigenvalue_M}
	\tilde{\lambda}_n = n_z^2 + n_x^2\frac{M_x}{M_z}\frac{L_z^2}{L_x^2} + n_y^2\frac{M_y}{M_z}\frac{L_z^2}{L_y^2}.
\end{equation}
Following the above arguments, the theoretical results of UCF amplitude $\Delta G$ are obtained by replacing $\tilde{\lambda}_n$ in Eq.~\ref{modifiedeigenvalue_M} by those in Eq.(2.9) in the standard formula of Ref.~\cite{lee1987universal}.

The theoretical results of $\Delta G=\tilde{c}_d(t_x, t_y, t_z)$ versus combinations of $(M_y, M_z)$ for $M_y, M_z \in [0, 3]$ have been calculated [see Appendix ~\ref{CalculateDeltaG} for details] and plotted in Fig.~\ref{FigDeltaGvsMzy}. It can be seen that when $M_z/M_y<1$, the UCF amplitude approaches $0.365$, the quasi-1D value for $c_d$. When $M_z/M_y>1$, the UCF amplitude increases rapidly and shoots up by more than ten times around the corner $(0, 3)$ in the $M_y-M_z$ plane. This indicates that the UCF amplitude can be extremely sensitive to band anisotropy and will cause significantly measurable effects in experiments.

The above scenario is similar to the sensitivity of $c_d$ to the system size $L_{\alpha}(\alpha=x,y,z)$. For transport along z direction, the typical behavior of $c_d$ is as follows. For samples of quasi-1D shape, $c_d$ is pinned around $0.365$ and is insensitive to the transverse sizes $L_{x,y}$ as long as $L_z/L_{x,y}\gg 1$. Its value slightly increases to 0.431 in quasi-2D ($L_{z}/L_{x}=L_{y}/L_{x}\gg 1$) and 0.548 in quasi-3D ($L_{z}=L_{y}=L_{x}$) samples~\cite{lee1987universal,Hu2017}. But $c_d$ becomes extremely sensitive to the transverse sizes once the sample is wide enough ($L_{z}/L_{x,y}< 1$) and diverges for ($L_{z}/L_{x,y}\ll 1$)~\cite{Hu2017,ruhlander2001symmetry}.

To visualize the band anisotropy in momentum space for several typical cases in Fig.~\ref{FigDeltaGvsMzy}, we plot the Fermi surfaces of the free electron gas in Eq.~\ref{h0_continous} in Fig.~\ref{fermisurfaces}(a-1)-(a-6). The conductance fluctuations continuously decrease from the enhanced value in (a-1) to a minimal quasi-1D value in (a-6), as the shape of the Fermi surface also approaches a quasi-1D ellipsoid in momentum space. For $\frac{M_z}{M_y}>1$ in Fig.~\ref{fermisurfaces}(a-1)-(a-2), the Fermi surface is compressed along the transport direction. The diffusion along the $L_z$ direction takes much shorter $t_z$ than $t_{x/y}$ along the transverse direction due to the larger diffusion coefficient $D_z>D_{x/y}$. Presumably, one can picture that in the limit of $t_z/t_{x/y}\rightarrow 0$, the electrons fly through the sample without fully self-averaging along the transverse paths, thus leading to conductances with more significant variance. In contrast, when $\frac{M_z}{M_y}<1$ in Fig.~\ref{fermisurfaces}(a-4)-(a-6), the Fermi surface is stretched along the transport direction, forming a quasi-1D ellipsoid. The diffusion along the $L_z$ direction takes a much longer time $t_z$ than $t_{x/y}$ because the electron diffuses slower along the transport direction, i.e., $D_z<D_{x/y}$. In the limit of $t_z/t_{x/y}\rightarrow \infty$, the electrons have fully traversed the $L_{x/y}$ several times before leaving the central region. Thus, the transmission after ergodic scattering shows only minimal conductance fluctuations. 

We conclude this section by noting the equivalent roles of the spatial size $L_\alpha$ and the anisotropic parameters $M_\alpha \propto D_\alpha$ in Eq.~\ref{modifiedeigenvalue_M}. Another way to understand the equivalence is to absorb the anisotropy parameters in the Hamiltonian in Eq.~\ref{h0_continous} into the Hamiltonian
in Eq.~\ref{h0_discrete} by rescaling the lattice constant in the hopping term in Eq.~\ref{hopping} as $\hat{a}_\alpha\equiv a_\alpha/\sqrt{M_\alpha}$. This way, one can interpret that the system has isotropic dispersion but with anisotropic hopping terms due to the choice of lattice constants. Meanwhile, the system size under this interpretation is rescaled to $\hat{L}_\alpha \equiv L_\alpha/\sqrt{M_\alpha}$, that is, the same number of lattice sites multiplied by the lattice constant $\hat{a}_\alpha$. The fact that band anisotropy and spatial anisotropy are transferable justifies their equivalence. In analogy with the dependence of $c_d$ on spatial anisotropy, we define the regions of sensitivity ($M_z/M_y\geq1$) and insensitivity($M_z/M_y<1$) for $\tilde{c}_d$ in Fig.~\ref{FigDeltaGvsMzy}, as separated by the dashed line. Within the region of (in)sensitivity, $\tilde{c}_d$ is (in)sensitive to the changes in anisotropy parameter $M_z/M_y$. In summary, in the presence of both anisotropies in $L_\alpha$ and $M_\alpha$, it is convenient to use the relative diffusion time $t_{z}/t_{x/y}$ that converts all these anisotropies to estimate the UCF amplitude in the diffusive free-electron gas. 

\subsection{Numerical results of UCF for the anisotropic free electron gas on a 2D lattice}\label{anisotropicBandNumerical}

In this subsection, we compare the numerical results of UCF from ensemble-averaged calculations to test the validity of our theoretical results. Throughout this paper, we fix the infinite left and right leads to have isotropic dispersion to avoid the influence of lead anisotropy, with the transport direction along $z$ unless otherwise stated. The UCF theory applies in any dimensions~\cite{travvenec2004universal,amin2022multifractal}; however, to lower the calculation cost, we only perform detailed calculations for Hamiltonian $H^{cen}$ on a 2D lattice in this subsection. (A representative result showing comparisons between theory and numerics in 3D lattices has also been given in Appendix \ref{anisotropicBandNumerical3D}.)

To focus on the effect of anisotropy in momentum space, we first consider a square central region with $L_y=L_z=L, L_x=1$ to remove the spatial anisotropy. The Fermi energy in lead is $E_{F}^{lead}=0.6$; in the central region, it is $E_{F}=0.4$. We emphasize that choosing Fermi energy near the band bottom is crucial for simulating the free electron gas model considered in the Altshuler-Lee-Stone theory. In addition, $M_0=0.0$, $M_x=0.0$, and $M_z=0.5$ are fixed in both the lead and central regions. The zero-temperature conductance along the $z$-direction is numerically calculated by recursive Green's function method~\cite{zhang2019electronic,Mac,Met} and ensemble-averaged to get the conductance fluctuations. 

\begin{figure*}
	\centering
	\includegraphics[width=1.0\linewidth]{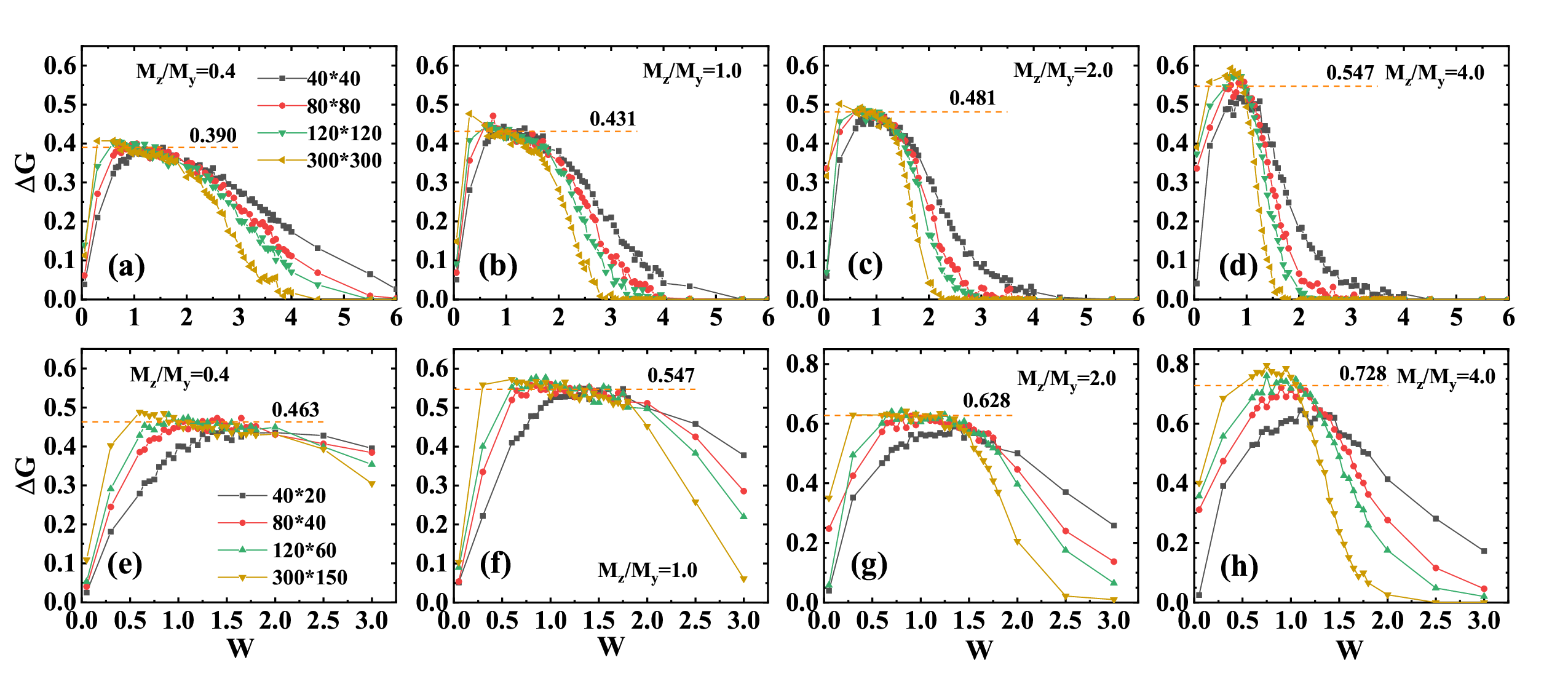}
	\caption{(Color online) Numerical results of the conductance fluctuations $\Delta G$ versus the disorder strength W for different anisotropy parameters $\frac{M_z}{M_y}$ in the 2D (a-d) square system with $L_y/L_z=1, L_x=1$ and (e-h) rectangular system with $L_y/L_z=2, L_x=1$. (a) $\frac{M_z}{M_y}=0.4$. (b) $\frac{M_z}{M_y}=1.0$. (c) $\frac{M_z}{M_y}=2.0$. (d) $\frac{M_z}{M_y}=4.0$. The Horizontal dashed lines in (a-d) correspond to the theoretical values of 0.390, 0.431, 0.481, and 0.547, respectively.  (e) $\frac{M_z}{M_y}=0.4$. (f) $\frac{M_z}{M_y}=1.0$. (g) $\frac{M_z}{M_y}=2.0$. (h) $\frac{M_z}{M_y}=4.0$. The Horizontal dashed lines in (e-h) correspond to the theoretical values of 0.463, 0.547, 0.628, and 0.728, respectively. The open boundary conditions are applied in the numerical calculations. The Fermi energy is $E_F=0.4$ in the central region and $E_F=0.6$ in the lead. The transport direction is along z, and the ensembles for the average are 1200.}
    \label{2d_squ}
\end{figure*}

The conductance fluctuation $\Delta G$ as a function of disorder strength $W$ is shown in Fig.~\ref{2d_squ}. With the increase of W, transport in the central region experiences transitions from a ballistic regime to a diffusive regime. Then the conductance and its fluctuation eventually vanish after a metal-insulator transition for large enough $W$. In the diffusive regime, the conductance fluctuation plateau versus $W$ is a signature of the disorder-independent UCF and is used as a prescription to determine the amplitude of UCF. For example, in Fig.~\ref{2d_squ}(a), we take the average value on the plateau of the $\Delta G$ vs $W$ trace to be the UCF amplitude. This value is close to the theoretical result of $0.390$ (marked by the horizontal dashed line) calculated using the formula in Appendix ~\ref{CalculateDeltaG}, where we have used $\Delta G=\tilde{c}_d\sqrt{\frac{ks^2}{\beta}}$ and the symmetry indices $k=1, \beta=1, s=1$ for the orthogonal ensemble under consideration. We have tested several sizes of $L_y=L_z=40, 80, 120, 300$ to ensure that the influence of the finite-size effect is negligible. The UCF amplitude converges quickly to the theoretical value when the system size reaches $80\times80$. Similar agreements have been observed for various degrees of band anisotropy, which is modulated by the anisotropy parameters $\frac{M_z}{M_y}=1.0, 2.0, 4.0$ in Fig.~\ref{2d_squ} (b-d). The overall increase of UCF with the increasing of $\frac{M_z}{M_y}$ is also consistent with our analysis in the subsection~\ref{anisotropicBand}.

Next, we introduce anisotropy in the real space of the 2D system to study the combined effects of spatial and momentum anisotropy. Specifically, we set the system size to rectangular with $L_y=2L_z$ and performed the calculations with the other parameters unchanged. We consider system sizes of $40\times20, 80\times40, 120\times60$, and $300\times150$ to ensure the convergence of UCF in numerical computations. In the rectangular central region, the theoretical values of UCF for $\frac{M_z}{M_y}=0.4, 1.0, 2.0$ and $4.0$ are $\Delta G=0.463, 0.547, 0.628$ and $0.728$, respectively. These UCF amplitudes are slightly larger than those for the square central region since $L_y/L_z=2$ is equivalent to multiplying $\frac{M_z}{M_y}$ by a factor of $(L_y/L_z)^2=4$ according to Eq.~\ref{modifiedeigenvalue_M}. These results are shown in Fig.~\ref{2d_squ}(e-h), to be compared with Fig.~\ref{2d_squ}(a-d) for the same anisotropy parameters, respectively. Again, we see good consistency between the theoretical values and the UCF plateaus obtained from the numerical results. In particular, the combination of $L_y/L_z=2$ and $\frac{M_z}{M_y}=1$ will give the same UCF amplitude as that for $L_y/L_z=1$ and $\frac{M_z}{M_y}=4$, consistent with Eq.~\ref{modifiedeigenvalue_M}.  

We conclude this section with comments on the finite-size effect in the numerical calculation of UCF. Fig.~\ref{2d_squ} suggests that the finite-size effect is more evident for systems with a larger $\frac{M_z}{M_y}$. By comparing Fig.~\ref{2d_squ}(a-d) with Fig.~\ref{2d_squ}(e-h), it is noticed that for this rectangular central region, it takes a larger system size to obtain converged UCF amplitude, especially for the larger values of $\frac{M_z}{M_y}>1$ [see Fig.~\ref{2d_squ}(c-d)]. In addition, Fig.~\ref{2d_squ} (d) and (h) both suggest that the converged UCF in these strongly anisotropic cases is slightly larger than the theoretical result. This is indeed not a finite-size effect. As will be discussed in detail in the next section, this is caused by the deviation of the Fermi surface from being perfectly ellipsoidal due to the artificial anisotropy inherent in the discretized model.

\section{Dependence of UCF on Fermi energy}\label{FermiEnergyDependence}
The free electron gas model exhibits parabolic dispersion and forms an ellipsoidal Fermi surface, independent of the Fermi energy. However, given the diversity of Fermi surface shapes in realistic materials and the ability to tune the Fermi energy experimentally, it is natural to ask how the UCF amplitude changes versus the Fermi energy when the Fermi surface deviates from an ideal ellipsoid. The tight-binding model is a convenient platform to address this question because it incorporates perturbative distortions to the ideal electron gas model.

The tight-binding model accurately approximates parabolic dispersion only near the band edges. When the Fermi energy $E_F$ is away from the band edge, the Fermi surface is no longer a perfect ellipsoid, and the degree of band anisotropy cannot be faithfully described by the parameters $M_\alpha$. In comparison with the ideal ellipsoidal Fermi surface of Eq.~\ref{h0_continous} in Fig.~\ref{fermisurfaces}(a-1)-(a-6), we show in Fig.~\ref{fermisurfaces}(b)(c) and (d) the Fermi surface of Eq.~\ref{h0_discrete} [see Appendix ~\ref{fermisurfacefree} for details of calculation]. As expected, Fig.~\ref{fermisurfaces}(c)(d) demonstrates that the Fermi surface does show apparent deviations from an ellipsoid, especially for Fermi energy away from the band edge. In contrast, the Altshuler-Lee-Stone theory implicitly assumes a perfect ellipsoid Fermi surface and thus can not apply in those cases.

The tight-binding model, therefore, provides a platform to test the deviation from the Altshuler-Lee-Stone theory when the Fermi surface deviates from the ideal ellipsoid. The single-band model in Eq.~\ref{h0_discrete} is also advantageous because it has a definite orthogonal symmetry class with $\beta=1$ and no spin degeneracy ($s=1$, $k=1$), allowing us to attribute the change in UCF to the prefactor $c_d$ according to Eq.~\ref{eq:deltaGdepend}. In this section, we scan the Fermi energy over the entire energy spectrum and find that the UCF amplitude shows a Fermi-energy dependence in tight binding models. We present the numerical results for 2D and 3D models below. 

\subsection{Results for 2D models}\label{2D_EF_dep}
\begin{figure*}
\centering
\includegraphics[width=1.0\linewidth]{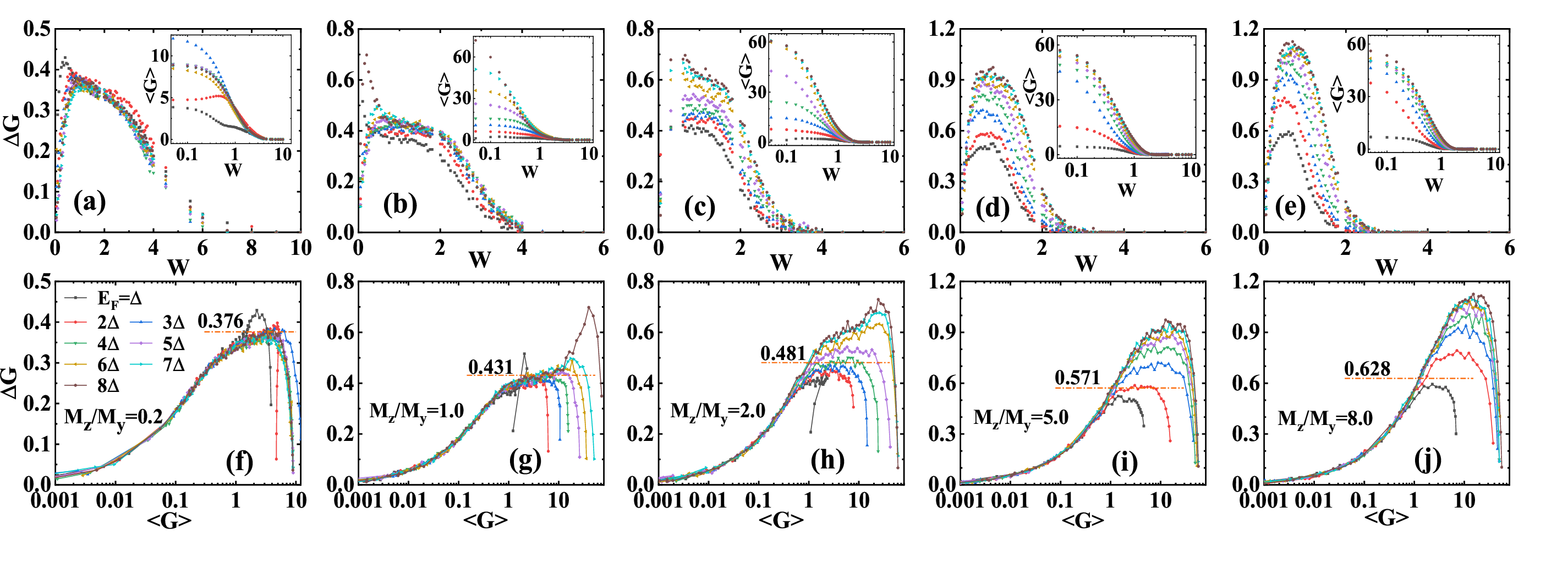}
\caption{(Color online) Evolution of (a-e) the conductance fluctuations $\Delta G$ and the average conductance $\langle G\rangle$ (both in units of $\frac{e^2}{h}$) versus the disorder strength $W$, and (f-j) the corresponding $\Delta G$ versus $\langle G\rangle$. Anisotropy parameters $\frac{M_z}{M_y}=0.2, 1.0, 2.0, 5.0$, and $8.0$, respectively, have been selected for each column of subplots, with $M_z=0.5$ is fixed. The theoretical values of UCF are $0.376$, $0.431$, $0.481$, $0.571$ and $0.628$, respectively. For each $\frac{M_z}{M_y}$, Fermi energy in the central region starts from the band bottom of $E_F=\Delta$, and increases in steps of $\Delta$ to the band center at $E_F=8\Delta$, with $\Delta = (M_y+M_z)/4$. All the subplots share the same legend of $E_F$ as those in (e). The central region size is $L_y=L_z=80$ and $L_x=1$. The lead has an isotropic band $\frac{M_z}{M_y}=1.0$ with a fixed Fermi energy $E_F^{lead}=2.0$ at its band center. }\label{dG_EF}
\end{figure*}

We first present results for 2D square and rectangular central regions, studied in Section ~\ref{anisotropicBandNumerical} with a fixed $E_F=0.4$ near the band edge. Next, we will place the Fermi energy across the entire band. In Fig.~\ref{dG_EF}(a-e), we plot $\Delta G$ versus disorder strength W for $\frac{M_z}{M_y}=0.2, 1.0, 2.0, 5.0$, and $8.0$. The central region is a square with a fixed size of $L_y=L_z=80$, ensuring no spatial anisotropy. For the lead, we use an isotropic band with $\frac{M_z}{M_y}=1.0$ to minimize the lead anisotropy effects. Unlike in Fig.~\ref{2d_squ}, the Fermi energy in the lead is fixed at the band center with $E_F^{lead}=2.0$. This value is picked to maximize the density of states in the lead [see Fig.~\ref{DOS_appendix}]. As will be seen, Fermi energy in the lead does not influence the overall trend of UCF dependence on central region $E_F$, but it does modify the specific UCF amplitude. For Fermi energy in the central region, we scan the entire band in fixed steps of $\Delta$, which is set to be the full bandwidth divided by 16. The calculation is performed from the band bottom of $E_F=\Delta$ to the band top of $E_F=15\Delta$. Only results for the lower half of the band are shown in Fig.~\ref{dG_EF} because the upper half should give, in principle, the same results, as the band top and bottom are equivalent after putting an overall minus sign to Eq.~\ref{h0_discrete}. Experimentally, Fermi energy in the lead and the central region are usually close; it can be tuned by gate voltage or by doping in the material. Our setup here is equivalent to keeping the same Fermi energy in the leads and the central region but applying different gate voltages, i.e., by assigning different $M_0$ in the leads/central regions. 

\begin{figure}
\centering
\includegraphics[width=1.0\linewidth]{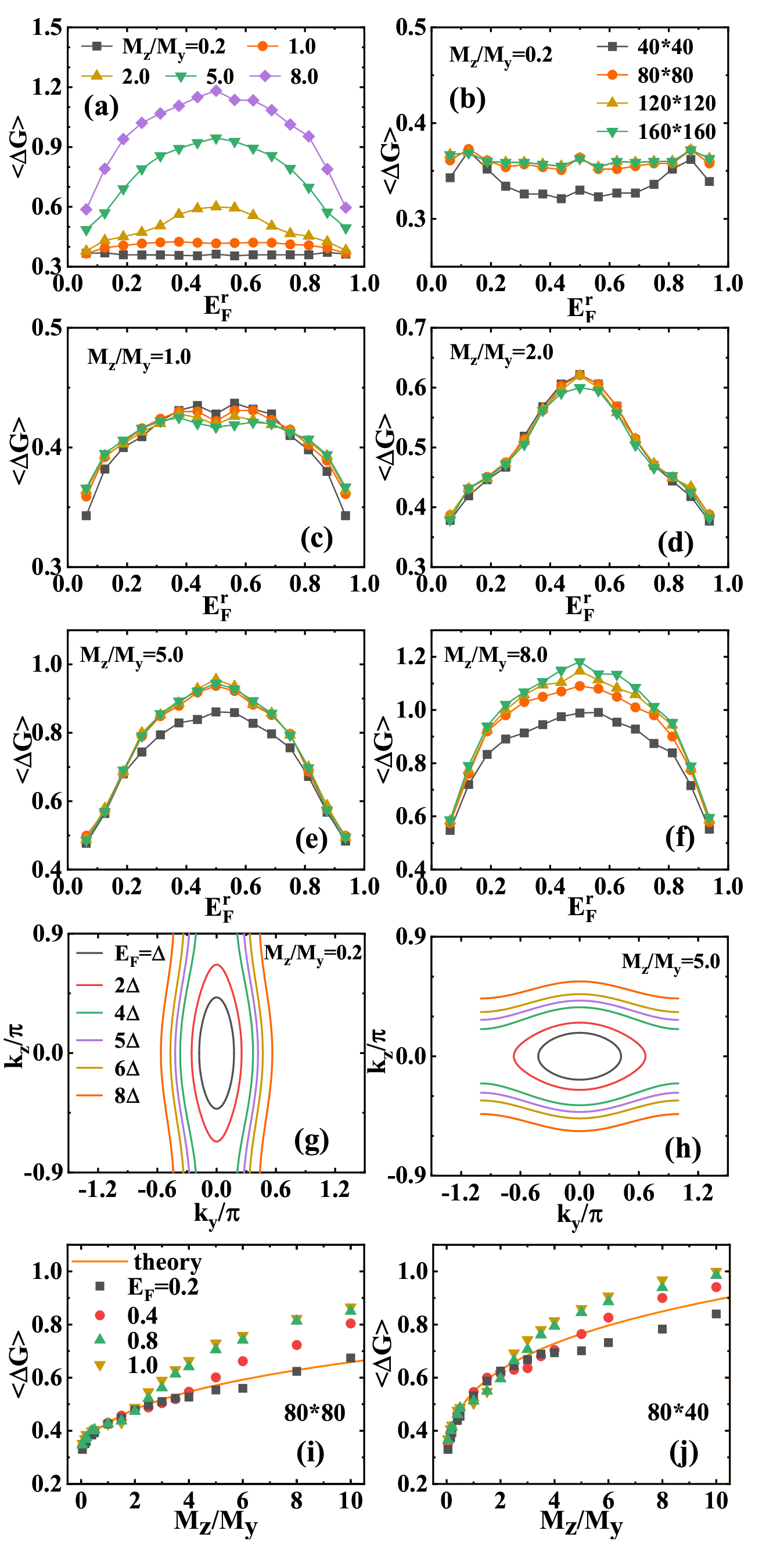}
\caption{(Color online) (a) Evolution of the averaged plateau values of conductance fluctuations $\langle\Delta G\rangle$ versus the reduced Fermi energy $E^r_F$ for different anisotropy parameters $\frac{M_z}{M_y}=0.2, 1.0, 2.0, 5.0$ and $8.0$, respectively. For each $\frac{M_z}{M_y}$, the reduced Fermi energy is $E^r_F\equiv E_F/[4(M_y+M_z)]$, that is $E_F$ divided by the bandwidth. The system size is $L_y=L_z=160$. (b-f) For each $\frac{M_z}{M_y}$, $\langle\Delta G\rangle$ versus the reduced Fermi energy $E^r_F$ for different system sizes of $L_y=L_z=40, 80, 120$ and $160$, respectively. Other parameters are the same as in Fig.~\ref{dG_EF}. (g-h) Evolutions of the Fermi surface of 2D systems for anisotropy parameters $\frac{M_z}{M_y}=0.2$ and $5.0$, respectively. The legend of Fermi energy $E_F$ in (h) is the same as in (g). (i-j) Evolutions of the $\langle\Delta G\rangle$ versus the anisotropy parameters $\frac{M_z}{M_y}$ for the (i) square with $L_y=L_z = 80$ and (j) rectangle with $L_y=2L_z = 80$ central regions, with different Fermi energy $E_F=0.2, 0.4, 0.8,$ and $1.0$. The orange line marks the theoretical value. The Fermi energy in the lead is fixed at $E_F^{lead}=0.6$. Other parameters include $M_0=0.0, M_x=0.0, M_z=0.5$, and the transport is along the $z$ direction.}\label{ucf_EF}
\end{figure}

Figure.~\ref{dG_EF}(a) shows the result for $\frac{M_z}{M_y}=0.2$. This is indeed a special and yet typical case. From Fig.~\ref{FigDeltaGvsMzy}, the ratio $\frac{M_z}{M_y}=0.2$ lies deep in the region of $\Delta G \approx 0.365$, corresponding to the quasi-1D value of UCF. From the $k_y-k_z$ projections of the 3D Fermi surface in Fig.~\ref{fermisurfaces}(c-1)-(c-6), one can see that its Fermi surface is largely in a quasi-1D shape. We collect these projections for different Fermi energy and plot them in Fig.~\ref{ucf_EF}(g) to compare them more clearly. From Eq.~\ref{modifiedeigenvalue_M}, it is equivalent to an isotropic band dispersion but with spatial anisotropy $L_z=\sqrt{\frac{M_y}{M_z}}L_y=\sqrt{5}L_y$, forming a quasi-1D shape that remains robust to perturbative distortions from changes in Fermi energy. As expected, when scanning the Fermi energy, the UCF plateau is around $\Delta G \approx 0.365$. The conductance fluctuations decays rapidly as the system enters the localization regime and vanishes when the disorder strength $W$ approaches the bandwidth of $4(M_y+M_z)=12$. In contrast, the inset of Fig.~\ref{dG_EF}(a) shows that the average conductance changes significantly with Fermi energy due to the varying density of states. This is a defining characteristic of UCF, which is independent of the magnitude of the conductance itself. In Fig.~\ref{dG_EF}(f), we plot $\Delta G$ versus the average conductance $\langle G\rangle$. For small conductances from the localized regime, the lines largely overlap~\cite{giordano1988conductance}. For conductances larger than unity, the UCF plateau emerges when the system enters the diffusive regime. The conductance fluctuations take non-universal values for large conductances in the ballistic regime.

In Fig.~\ref{dG_EF}(b), the anisotropy parameter is $\frac{M_z}{M_y}=1.0$. It is found that the UCF plateau increases slightly when the Fermi energy moves up in the central region, and reaches a maximum around the band center. This contrasts with the case of a square sample and an ellipsoidal Fermi surface in the Altshuler-Lee-Stone theory, where the theoretical value of UCF is $\Delta G=0.431$ independent of Fermi energy. We claim that according to Eq.~\ref{eq:deltaGdepend}, the only reason for the dependence on $E_F$ is that the deviation from the ellipsoidal Fermi surface changes the prefactor $c_d$, and the symmetry indices $\beta=1$ remain unchanged. But to obtain the theoretical value of $c_d$ in the tight-binding model is more complex than a direct replacement of the diffusion constant by the diffusion vector as is done in Section ~\ref{anisotropicBand}. Nonetheless, we can still study its behavior numerically. In the inset, the average conductance shows a similar behavior, mainly because the density of states increases with Fermi energy. It is well-known that the increase in UCF on the plateau is independent of the increase in conductances. However, the larger conductance before the plateau is related to the enhancement of conductance fluctuation in the ballistic regime around $W=0$. After the plateau, the conductance fluctuation vanishes when $W$ exceeds the bandwidth of $4(M_y+M_z)=4$. In Fig.~\ref{dG_EF}(g), it is confirmed that the UCF plateaus all appear in the diffusive regime with $\langle G\rangle >1$.

For higher degrees of anisotropy in Fig.~\ref{dG_EF}(c-e), the conductance fluctuations rise to the plateau with the onset of the disorder and vanish for disorder strengths close to bandwidth. Fig.~\ref{dG_EF}(c-e) shows a more significant increase in UCF, which almost doubles, with the increase of the Fermi energy from the edge to the center of the band. As shown in Fig.~\ref{fermisurfaces}(d-1)-(d-6) for $\frac{M_z}{M_y}=5.0$, such anisotropy compresses the Fermi surface along the $z$ direction of transport. The projections on the $k_y-k_z$ plane show that the Fermi surface width along the $z$ direction is further compressed relative to the y direction as the Fermi energy increases from the edge of the band to the center. For comparison, these projections are collected and plotted in Fig.~\ref{ucf_EF}(h). Fig.~\ref{dG_EF}(h-j) shows that these UCFs occur in the diffusive regime. In Fig.~\ref{dG_EF}(c-e), the UCF increases with increasing $\frac{M_z}{M_y}$. Moreover, the $E_F$ dependence of UCF is more evident for systems with a larger UCF amplitude induced by more substantial band anisotropy. 

To compare directly, we extract the average $\Delta G$ across the plateau and plot the data in Fig.~\ref{ucf_EF}(a). The ranges of $W$ to obtain the averaged $\Delta G$ in Fig.~\ref{dG_EF} are $W\in[0.6, 1.8]$, $[0.55, 1.8]$, $[0.6, 1.6]$, $[0.6, 0.95]$, and $[0.45, 0.8]$ for $M_z/M_y=0.2, 1.0, 2.0, 5.0$, and $8.0$. The range of $W$ for averaging is chosen similarly for other system sizes and is not shown here. Ideally, the UCF should be symmetric around the band center, and the off-matches are due to numerical reasons. Near the band edges of $E_F=\Delta$ and $E_F=15\Delta$, the tight-binding model dispersion mimics the continuum model in Eq.~\ref{h0_continous}, which is the basis of the theory in Section ~\ref{anisotropicBand}. The UCF is close to the theoretical predictions of $\Delta G=0.376, 0.431, 0.481, 0.571$ and $0.628$ for each choice of $\frac{M_z}{M_y}$, respectively. When scanning the Fermi energy, the magnitude of the UCF amplitude sorts in the same order as $\frac{M_z}{M_y}$, that is, from $\frac{M_z}{M_y}=0.2$ for the bottom line upward to $\frac{M_z}{M_y}=8.0$ in the top line. The UCF stays at the quasi-1D value for $\frac{M_z}{M_y}=0.2$ with a significantly stretched Fermi surface along $k_z$ while showing a clear increase around the center of the band for all other cases of $\frac{M_z}{M_y}\geq 1$. 

To ensure that this is not a finite-size effect discussed at the end of Section ~\ref{anisotropicBandNumerical}, we tested the convergence of UCF for system sizes of $40, 80, 120$, and $160$, and the results are shown in Fig.~\ref{ucf_EF}(b-f). In Fig.~\ref{ucf_EF}(b), the UCF remains nearly constant for system sizes larger than $80$. In Fig.~\ref{ucf_EF}(c-f), the UCF converges faster at the edge of the band and slower at the center of the band. Its amplitude approaches a convergence when the system size is $160$. This confirms that the results in Fig.~\ref{ucf_EF}(a) are primarily converged. Moreover, the dependence of the UCF on $E_F$, such as its increase near the band center, is already evident for system sizes larger than $40\times40$.

A closer investigation of the evolution of UCF versus the band anisotropy parameter $\frac{M_z}{M_y}$ is done by fixing the Fermi energy in both the leads at $E_F^{lead}=0.6$ and in the central region at $E_F=0.2, 0.4, 0.8$ and $1.0$, respectively. To reduce computational cost, we consider a system of size $L_y=L_z=80$, which may not yield fully converged UCF results due to the finite-size effect. However, for a fixed system size, Fig.~\ref{ucf_EF}(i) shows that for all $E_F$ considered, the UCF increases monotonically with $M_z/M_y$. This trend aligns with the theoretical prediction, as the orange line shows. In particular, close to the band bottom at $E_F=0.2$, the UCF is close to theoretical values across a wide range of $\frac{M_z}{M_y}$ for the system size of $80\times80$ under consideration. However, for $E_F=0.4$, numerics deviate from theory beyond $\frac{M_z}{M_y}>4$. From Fig.~\ref{2d_squ}, this deviation can be expected to be more significant for a system size larger than $80$ when the converged UCF is obtained. An earlier deviation from the theory is seen for $E_F=0.8$ and $1.0$ at $\frac{M_z}{M_y}>2$. Fig.~\ref{ucf_EF}(j) shows the results for a rectangular lattice of $L_y=80, L_z=40$, where both spatial anisotropy and band anisotropy are present. The numerical and theoretical data show that the additional spatial anisotropy leads to enhanced conductance fluctuation. However, the conductance fluctuations do not converge for the system size $80\times40$. Specifically, conductance fluctuations at $E_F=0.2$ drop below the theoretical prediction for $\frac{M_z}{M_y}>4$. In the thermodynamic limit, it is expected that the UCFs shift above the numerical values in Fig.~\ref{ucf_EF}(j), but the overall trend will be similar. 

\subsection{Results for 3D models}\label{3D_EF_dep}
\begin{figure*}
\centering
\includegraphics[width=1.0\linewidth]{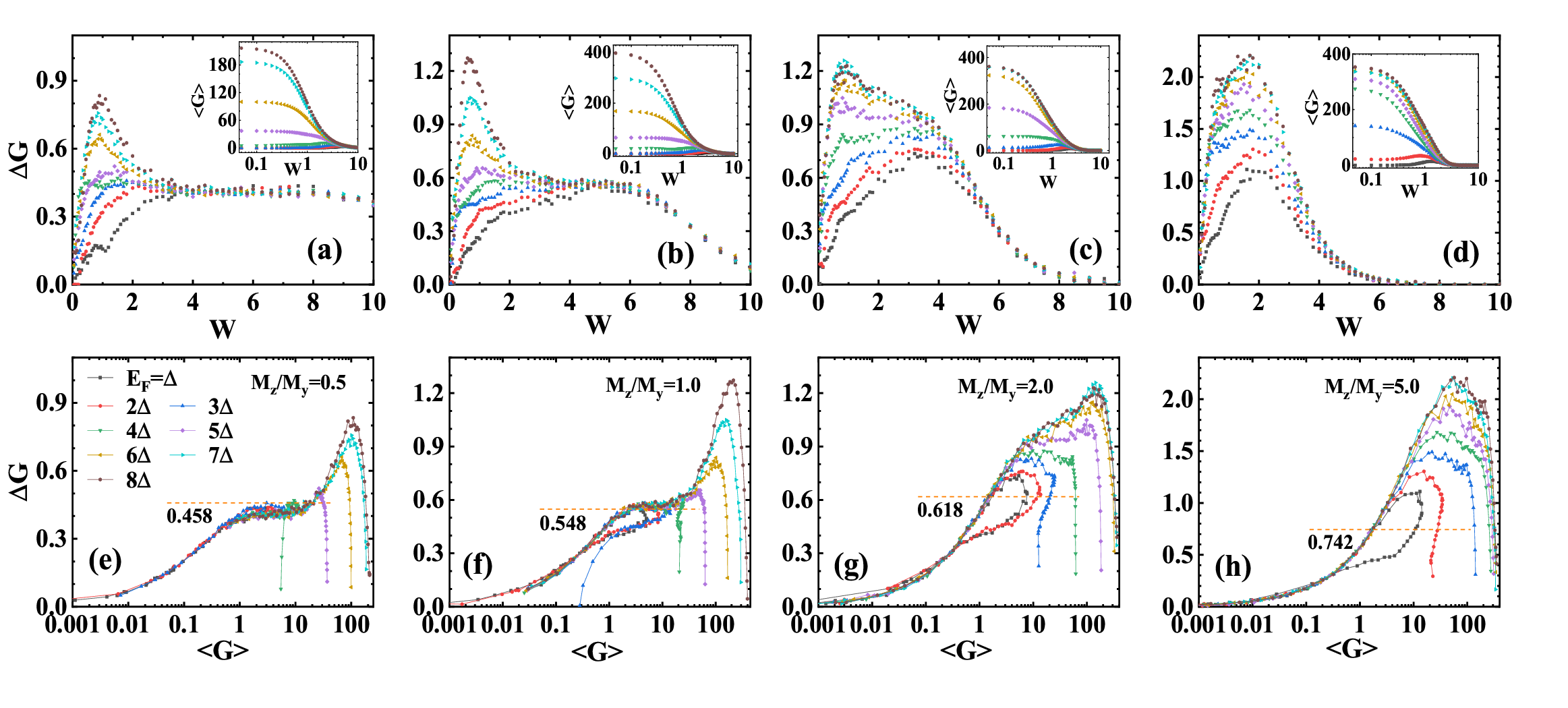}
\caption{(Color online) Evolution of (a-d) the conductance fluctuations $\Delta G$ and the average conductance $\langle G\rangle$ versus disorder strength $W$, and (e-h) the corresponding $\Delta G$ versus $\langle G\rangle$. For each column of subplots, anisotropy parameters $\frac{M_z}{M_y}=0.5, 1.0, 2.0$, and $5.0$, respectively, with $M_z=0.5$, $M_x=M_y$. The theoretical values of UCF are $0.458, 0.548, 0.618$ and $0.742$, respectively. For each $\frac{M_z}{M_y}$, Fermi energy in the central region starts from the band bottom of $E_F=\Delta$ and increases in steps of $\Delta$ to the band center at $E_F=8\Delta$, with $\Delta = (M_x+M_y+M_z)/4$. All the subplots share the same legend of $E_F$ as those in (e). The central region size is fixed at $L_x=L_y=L_z=25$. The lead is set to $\frac{M_z}{M_y}=1.0$ with a fixed Fermi energy $E_F^{lead}=3.0$ at its band center.} \label{deltaG3D}
\end{figure*}
\begin{figure}
\centering
\includegraphics[width=1.0\linewidth]{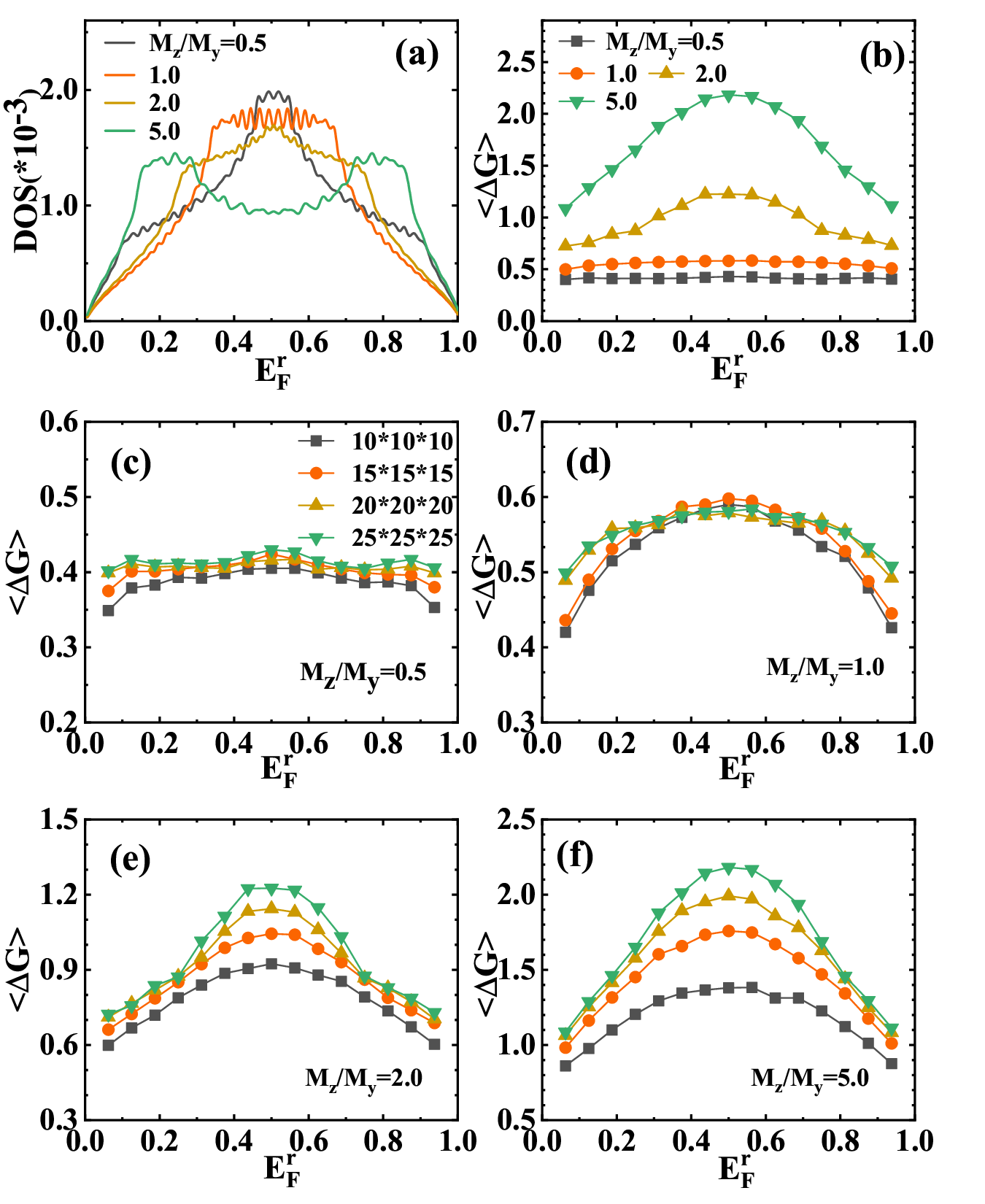}
\caption{(Color online) (a) Density of states for different anisotropy parameters $\frac{M_z}{M_y}=0.5, 1.0, 2.0$ and $5.0$, respectively. (b) Evolution of the averaged plateau values of conductance fluctuations $\langle\Delta G\rangle$ versus the reduced Fermi energy $E^r_F$ for each $\frac{M_z}{M_y}$, where the data to determine $\langle\Delta G\rangle$ is from Fig.~\ref{deltaG3D}. The reduced Fermi energy is $E^r_F\equiv E_F/[4(M_x+M_y+M_z)]$, that is $E_F$ divided by the bandwidth, and the system size is $L_x=L_y=L_z=25$. (c-f) For each $\frac{M_z}{M_y}$, $\langle\Delta G\rangle$ versus the reduced Fermi energy $E^r_F$ for different system sizes of $L_x=L_y=L_z=10, 15, 20$ and $25$, respectively. Other parameters are the same as in Fig.~\ref{deltaG3D}. }\label{ucf_EF_3d}
\end{figure}
In this subsection, we study the Fermi-energy dependence of UCF in a 3D cubic central region with $L_x=L_y=L_z$. The density of states is calculated using a Gaussian broadening, as in Ref.~\cite{Hu2021b}, and plotted in Fig.~\ref{ucf_EF_3d}(a). For the lead, we use an isotropic band with $M_\alpha=0.5$ $(\alpha=x, y, z)$ and fix the Fermi energy $E_F^{lead}=3.0$ at the band center with a large density of states. Similarly to 2D models in Section ~\ref{2D_EF_dep}, we scan the Fermi energy in the central region from the band bottom at $E_F=\Delta$ to the band top at $E_F=15\Delta$, with $\Delta=(M_x+M_y+M_z)/4$ for $\frac{M_y}{M_z}=0.5, 1.0, 2.0$ and $5.0$.

The results for $L_x=L_y=L_z=25$ are presented in Fig.~\ref{deltaG3D}. Similar to the 2D case in Fig.~\ref{dG_EF}(a), though we have a 3D cubic central region, the UCF is pinned close to the quasi-1D value of $\Delta G \approx 0.365$ on the plateau between $W\approx3$ and $W\approx8$, regardless of the significant variance of the density of states or conductance. The conductance fluctuations decay after $W$ exceeds the bandwidth $4(M_x+M_y+M_z)=10$. The onset of the UCF plateau when $\langle G\rangle>1$ is also evident in Fig.~\ref{deltaG3D}(e). The Fermi surface is significantly stretched along the $k_z$ direction and maintains a quasi-1D shape regardless of the Fermi energy. In addition to the UCF plateau, the conductance fluctuation peaks around $W=1$ when Fermi energy is close to the band center. Such peaks also appear and behave similarly in the 2D case. They are more pronounced in the 3D case. This peak lies within the ballistic regime~\cite{asano1996conductance,rycerz2007anomalously} and is distinct from the UCF plateau. However, the peak persists for $M_y/M_z=1.0$ and $2.0$ in Fig.~\ref{deltaG3D}(b-c), before merging into the UCF plateau in Fig.~\ref{deltaG3D}(d) for $M_y/M_z=5.0$. Another plateau for $W$ around the bandwidth $4(M_x+M_y+M_z)=6$ in Fig.~\ref{deltaG3D}(b) is attributed to the metal-insulator transition in the literature and is not the UCF plateau~\cite{qiao2010universal,Hu2017}. The crossover among the various peaks and the UCF plateau makes it tricky to determine the conductance plateau for UCF, especially in Fig.~\ref{deltaG3D}(b-c). Despite these complexities, the conductance fluctuation increases and almost doubles as $E_F$ approaches the band center for all $M_y/M_z>1.0$. 

In Fig.~\ref{ucf_EF_3d}(b), the UCF amplitude averaged over the conductance fluctuation in the plateau from Fig.~\ref{deltaG3D}(a-d) is plotted as a function of the reduced Fermi energy $E^r_F$. For example, we list the ranges of $W$ for obtaining the averaged $\Delta G$ in Fig.~\ref{deltaG3D}. That is, $W\in[2.8, 6.0], [2.4, 5.0], [1.4, 2.0]$ for $M_z/M_y=0.5, 1.0$ and $5.0$. For $M_z/M_y=2.0$, averages are taken within $W\in[3.0, 3.8]$ and $[0.6, 1.2]$ for $E_F=\Delta\sim4\Delta$ and $5\Delta\sim8\Delta$, respectively. Like the 2D case, UCF shows a clear $E_F$ dependence, with the largest amplitude near the band center at $E^r_F=0.5$. To test convergence, the evolution of UCF versus the system size of $L_z=10, 15, 20,$ and $25$ is plotted in Fig.~\ref{ucf_EF_3d}(c-f) for each anisotropy parameter. The UCF converges only for $M_z/M_y=0.5$ in Fig.~\ref{ucf_EF_3d}(c), and is approaching convergence for $M_z/M_y=1.0$ in Fig.~\ref{ucf_EF_3d}(d). For $M_z/M_y>1.0$, the conductance fluctuation approaches convergence around the band edges and hints at convergence for larger system sizes near the band center. Again, the converged UCF is expected to depend on $E_F$, especially for anisotropy parameters $M_z/M_y>1.0$.

The results in this section hold some implications relevant to the experimental setup on UCF measurement versus the Fermi energy. The factor $c_d$ in Eq.~\ref{eq:deltaGdepend} suffers from the joint effects of anisotropy from both spatial dimension and Fermi surface, as has been discussed theoretically at the end of Section ~\ref{anisotropicBand}. Our numerical results suggest two ways to improve the device design for experiments designed to measure the dependence of $c_d$ on Fermi energy. First, it is found that to get a more significant signal of changes in UCF versus Fermi energy for a given material, an appropriate device should measure transport along the direction where the Fermi surface is compressed. For example, from Fig.~\ref{ucf_EF}(g) and (h), as well as Fig.~\ref{DOS_appendix}, the anisotropy parameters of $M_z/M_y=5.0$ and $0.2$ are describing the same material with the flipped direction of transport and equivalent in their density of states versus reduced energy. For $M_z/M_y=5.0$ in a 2D square device, transport along the z direction in Fig.~\ref{ucf_EF} (e) shows a doubled UCF amplitude. In contrast, transport along the y direction will be effectively described as $M_z/M_y=0.2$ and show a negligible change in UCF versus Fermi energy, as shown in Fig.~\ref{ucf_EF} (a). Second, a wider transverse length relative to the longitudinal length will effectively magnify the anisotropy effect, thus giving a more pronounced UCF dependence on Fermi energy, as shown in Fig.~\ref{2d_squ}(e-h) and Fig.~\ref{ucf_EF}(j). For experiments designed to measure the dependence of $\sqrt{\frac{ks^2}{\beta}}$ part in Eq.~\ref{eq:deltaGdepend} on Fermi energy, the opposite of the above two suggestions will be helpful to improve the device design. The idea is to pin the value of $c_d$ to the quasi-1D value such that it is independent of the perturbative anisotropic effects due to changes in Fermi energy. This is achieved through transport along the direction with a stretched Fermi surface as shown in Fig.~\ref{dG_EF} (a) and Fig.~\ref{deltaG3D}(a), or by designing a device that is much longer in the transport direction than the transverse directions. Once $c_d$ is fixed to the quasi-1D value, it is more convenient to analyze the effects of the symmetry indices. In this work, we do not study the case of a quasi-1D central region because the Fermi energy dependence of $c_d$ is expected to be weak.

\section{UCF in the 3D topological semimetal $\mathrm{Cd_3As_2}$}\label{ExpAndMaterial}
In previous sections, we demonstrated a Fermi-energy-dependent UCF due to a non-ellipsoidal Fermi surface in the tight-binding model of the band conductors. This suggests that an $E_F$-dependent UCF can be observed in real materials with complex Fermi surfaces that vary with chemical potential. Next, we study the $E_F$-dependence of UCF in the Dirac semimetal $\mathrm{Cd_3As_2}$, which has a much more complex Fermi surface. Theory~\cite{Hu2017,Hu2017a} predicts that UCF in $\mathrm{Cd_3As_2}$ is influenced by band anisotropy and Fermi energy, qualitatively consistent with recent experimental observations~\cite{Xiao23}. However, as noted in Ref.~\cite{Xiao23}, a quantitative understanding of Fermi surface anisotropy effects on UCF still requires further study using realistic material parameters. Thus, unlike previous theory studies, in the following, we perform calculations based on realistic material parameters of $\mathrm{Cd_3As_2}$ extracted from the first-principle study~\cite{chen2021quantum}. We confirm that the UCF increases when Fermi energy is away from the Dirac point. It is also found that the UCF amplitude shows distinct values when the transport takes place along $L_{x/y}$ and $L_z$ directions. 

\subsection{Tight binding model of $\mathrm{Cd_3As_2}$}
According to Ref.~\cite{wang13three,cano2017chiral,chen2021quantum}, the effective Weyl Hamiltonian for $\mathrm{Cd_3As_2}$ is given as follows 
\begin{equation}\label{cd3as2_h}
	H^\prime_0 = \varepsilon_0(\vec{k})\otimes\sigma_0 + M(\vec{k})\otimes\sigma_z + A ( k_x\otimes\sigma_x - k_y\otimes\sigma_y )
\end{equation}
where 
\begin{align}
	\varepsilon_0(\vec{k}) &= C_0 + C_1k_z^2 + C_2(k_x^2+k_y^2)\\
	M(\vec{k}) &= M_0 + M_1k_z^2 + M_2(k_x^2 + k_y^2),
\end{align}
$\sigma_{\alpha}$ is the Pauli matrices, and $\sigma_0$ is the identity matrix. $A$ is the spin-orbital coupling strength between orbitals. Here, we consider only the Weyl Hamiltonian to reduce the numerical cost and simplify the analysis. The two Weyl Hamiltonians that constitute the full Dirac Hamiltonian are related by time-reversal symmetry and double the conductance and its fluctuation of the following results~\cite{Hu2017}.

To facilitate the numerical calculations, we discretize the Hamiltonian into a tight-binding form,
\begin{equation}
H^\prime_0 = V^{\prime}\sum_\mathbf{r}a^{\prime\dagger}_\mathbf{r} a^\prime_\mathbf{r} + \Big(\sum_{\mathbf{r},\alpha=x,y,z}T^{\prime}_\alpha a^{\prime\dagger}_\mathbf{r} a^\prime_{\mathbf{r}+\delta\boldsymbol{\alpha}} + H.c.\Big) \label{cd3as2_h_discretize}
\end{equation}
with
\begin{align}
	V^\prime &= (C_0+\frac{2C_1}{c^2}+\frac{2C_2}{a^2}+\frac{2C_2}{b^2})\cdot\sigma_0\\ \notag
	&+ (M_0+\frac{2M_1}{c^2}+\frac{2M_2}{a^2}+\frac{2M_2}{b^2})\cdot\sigma_z
\end{align}
\begin{align}
	T^\prime_x &= \frac{-C_2}{a^2}\cdot\sigma_0 + \frac{-M_2}{a^2}\cdot\sigma_z + \frac{A}{2ia}\cdot\sigma_x\\
	T^\prime_y &= \frac{-C_2}{b^2}\cdot\sigma_0 + \frac{-M_2}{b^2}\cdot\sigma_z + \frac{A}{2ib}\cdot\sigma_y\\
	T^\prime_z &= \frac{-C_1}{c^2}\cdot\sigma_0 + \frac{-M_1}{c^2}\cdot\sigma_z
\end{align}
where $a^\prime_\mathbf{r}(a^{\prime\dagger}_\mathbf{r})$ represents the two-component electron annihilation(creation) operators on the site $\mathbf{r}$. Based on the parameters of the real material given in~\cite{chen2021quantum}, we set the parameters as $C_0=-0.0145 ~\rm{eV}$, $C_1=10.59 ~\rm{eV}\AA^2$, $C_2=11.5 ~\rm{eV}\AA^2$, $M_0=0.0205 ~\rm{eV}$, $M_1=-18.77 ~\rm{eV}\AA^2$, $M_2=-13.5 ~\rm{eV}\AA^2$, the SOC strength $A=0.889 ~\rm{eV}\AA$, and the lattice constant along the three directions is $a=b=3 ~\rm{\AA}, c=5 ~\rm{\AA}$. These parameters apply to both the central region and the two leads. The Fermi energy in the lead is set equal to that in the central region to ensure state alignment. Open boundary conditions are applied in the conductance calculations. The onsite disorder is added and averaged over at least 1200 ensembles to calculate $\Delta G$. In the following, $\rm{eV}$ and $\rm{\AA}$ are taken as units of energy and length, respectively, and will be omitted for simplicity hereafter.

\subsection{UCF versus Fermi energy along y and z transport directions}\label{yzTransport}
The largest system size considered is $L_x=L_y=L_z=20$, where $L_{\alpha}$ counts the number of unit cells along the $\alpha$ direction. Unlike in previous sections, the lattice constants are not unity. The system is thus spatially anisotropic with lengths $\hat{L}_x=\hat{L}_y=60~\rm{\AA}$ and $\hat{L}_z=100~\rm{\AA}$, where $\hat{L}_{\alpha}$ measures the sample lengths in units of $\rm{\AA}$ along the $\alpha$ direction. As discussed at the end of Section ~\ref{3D_EF_dep}, this geometry will favor a significant Fermi energy dependence of UCF for transport along the $x$ and $y$ directions by decreasing the diffusion time $t_{x/y}$ in Eq.~\ref{modifiedeigenvalue_new} and will be demonstrated below.

\begin{figure*}
	\centering
	\includegraphics[width=1.0\linewidth]{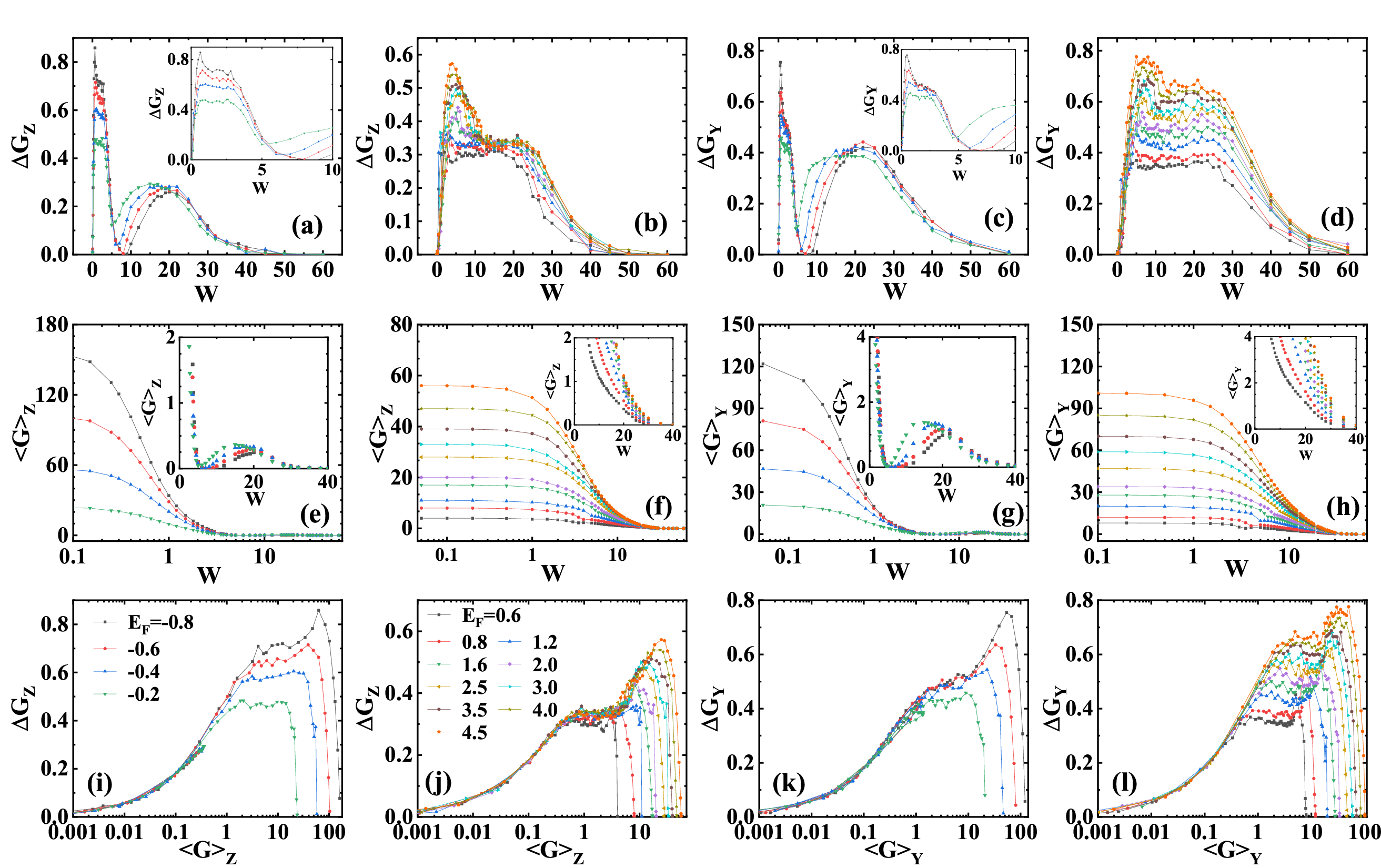}
	\caption{ (Color online) (a-d) The Conductance fluctuation $\Delta G$ and (e-h) the average conductance $\langle G\rangle$ (both in units of $\frac{e^2}{h}$) versus the disorder strength $W$ (in units of $eV$) in Dirac semimetal $\mathrm{Cd_3As_2}$. Fermi energy $E_F^{lead}$ (in units of $eV$) in the lead is the same as $E_F$ in the central region. (i-l) $\Delta G$ versus $\langle G\rangle$. For the left (right) two columns, transport is along the $z$ ($y$) direction. The system size is $L_x=L_y=L_z=20$. The subplots (a, e, i, c, g, k) share the same legend as in (i). The subplots (b, f, j, d, h, l) share the same legend as in (j).}\label{diff_direc_z}
\end{figure*}

First, we present results for the transport along the $z$ direction. The conductance fluctuations versus the disorder strength are shown in Fig.~\ref{diff_direc_z}(a) and (b) for the negative and positive Fermi energy, respectively. In Fig.~\ref{diff_direc_z}(a), below the Dirac point, two plateaus emerge from the fluctuation of conductance versus the disorder. $\Delta G$ first shows a peak of conductance fluctuation in disorder $W\approx 2.5$ and then a plateau at disorder $W \approx 15$ before vanishing in the localized regime. For Fermi energy $E_F=-0.2$ to $E_F=-0.8$ away from the Dirac point, the conductance fluctuations increase monotonically at $W \approx 2.5$ while remaining unchanged at $W \approx 15$. A sharp drop of conductance fluctuation towards zero appears between the two plateaus from $W \approx 5$ to $W \approx 10$. These sharp drops are accompanied by a decrease in conductances, as shown in the inset of Fig.~\ref{diff_direc_z}(e). So, it is necessary to identify the UCF plateau for the lower band below the Dirac point. 

\begin{figure*}
	\centering
	\includegraphics[width=1.0\linewidth]{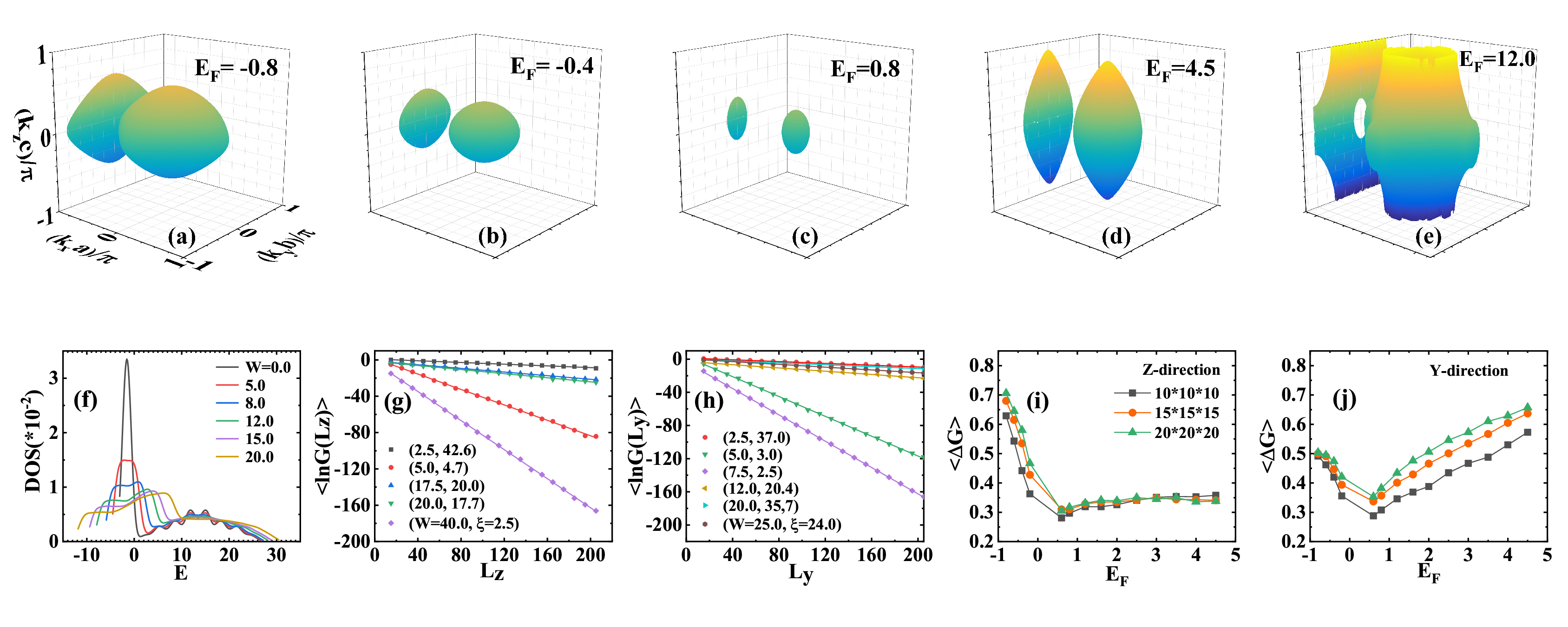}
	\caption{ (Color online) (a-e) The Fermi surface of $\mathrm{Cd_3As_2}$ calculated from the discretized Hamiltonian in Eq.~\ref{cd3as2_h_discretize} for Fermi energy $E_F=-0.8, -0.4, 0.8, 4.5$ and $12.0$. The axis labeling for (b-e) is the same as in (a). (f) Evolution of density of states versus the disorder strength. Localization length $\xi$ for several disorder strengths $W$ obtained by fitting to conductances for (g) $L_x=L_y=15$, $E_F=-0.2$ along the z direction and (h) $L_x=L_z=15$, $E_F=-0.8$ along the y direction. Plateau values of conductance fluctuations $\langle\Delta G\rangle$ versus Fermi energy for transport along (i) z-direction and (j) y-direction with $L_x=L_y=L_z=10, 15$ and $20$, respectively.}\label{cd3as2FermiSurface}
\end{figure*}

Considering that the Weyl Hamiltonian in Eq.~\ref{cd3as2_h} may experience disorder-induced phase transitions~\cite{pixley2015anderson,pixley2016rare,chen2015disorder,Liu2016Effect} before becoming an Anderson insulator, we calculate the evolutions of the density of states and the localization length to analyze this behavior further. In Fig.~\ref{cd3as2FermiSurface}(f), the lower bandwidth of the clean Hamiltonian is around $5$, and a disorder of strength $W=5$ effectively broadens the band below the Dirac point. Stronger disorders further mix the states in the lower band with those in the upper band. In Fig.~\ref{cd3as2FermiSurface}(g), the localization length for a smaller cross-section size of $L_x=L_y=15$ is calculated and shown for Fermi energy below the Dirac point at $E_F=-0.2$~\cite{zhang2009localization}. The system is localized for $W=5.0$ with a localization length $\xi=4.7$ much shorter than $L_z$. This indicates that the sudden drop in conductance and its fluctuation are due to disorder-induced localization. Several observations suggest that the first plateau corresponds to UCF. First, the inset in Fig.~\ref{diff_direc_z}(a) is reminiscent of the single-band 3D tight-binding model in Fig.~\ref{deltaG3D}, where the UCF plateau occurs in the diffusive regime around a disorder strength that is slightly smaller than its bandwidth of $4(M_x+M_y+M_z)$. Then localization starts for disorders larger than the bandwidth, and the $\Delta G - W$ curves tend to overlap and decay to zero. Second, for $W\approx 2.5$, the increase in conductance fluctuation with Fermi energy away from the Dirac point is aligned with the fact that the Fermi surface, meanwhile, is slightly stretched along the $k_{x/y}$ direction [see Fig.~\ref{cd3as2FermiSurface} (a) and (b)]. We also estimate the theoretical value of UCF by treating the Fermi surface in Fig.~\ref{cd3as2FermiSurface} (a) and (b) approximately as an ellipsoid and extracting the anisotropy parameters $\frac{M_z}{M_y}$ in analogy with Eq.~\ref{h0_continous}. Through Eq.~\ref{dgformula1} and Eq.~\ref{eq:deltaGdepend} with symmetry indices~\cite{Hu2017} $k=1, s=1, \beta=2$, it gives the theoretical UCF values around 0.48 and 0.44, respectively. The first plateau is Fermi energy sensitive and is reasonably above this estimate. Third, Fig.~\ref{diff_direc_z}(i) shows that the first plateaus occur for conductances larger than unity in the diffusive region. Meanwhile, Fig.~\ref{cd3as2FermiSurface} (g) shows the localization length at $W=2.5$, is $\xi=42.7$, which is much larger than the system size $L_z=15$. 

Interestingly, a further increase in disorder strength ($\xi=20$ for $W=17.5$, $E_F=-0.2$ with $L_x=L_y=15$) introduces reentry of the metallic regime with enhanced conductances, and the fluctuation of conductance shows a plateau, as shown in Fig.~\ref{diff_direc_z}(a). For the second plateau, the disorder is so strong relative to the bandwidth of $\approx 5 eV$ below the Dirac point that it distorts the band structure and leads to a Fermi-energy-insensitive UCF. In Fig.~\ref{diff_direc_z}(e), the average conductances corresponding to the second plateau are much smaller than unity. We thus take the first plateau as the UCF amplitude below the Dirac point and ignore the second plateau.

The different behaviors of conductance fluctuation for weak and strong disorders are also observed above the Dirac point. In Fig.~\ref{diff_direc_z}(b), $\Delta G$ also shows a peak of conductance fluctuation in weak disorder $W\approx 2.5$ and a plateau in moderate disorder $W \approx 15$. For Fermi energy $E_F=0.6$ to $E_F=4.5$ away from the Dirac point, the conductance fluctuations increase monotonically at $W < 10$ while remaining unchanged at $W \approx 15$. Unlike the conductance and its fluctuation below the Dirac point, $\Delta G$ does not decay to zero but shows a bent curve around $W=10$ between weak and strong disorders, and $\langle G\rangle$ is well above unity [see Fig.~\ref{diff_direc_z}(f)]. Fig.~\ref{diff_direc_z}(j) suggests that the plateau around $\langle G\rangle \approx 1$ is our desired amplitude of UCF while the fluctuation peak at weak disorder is non-universal. The fact that the UCF is pinned around a small amplitude is consistent with the quasi-1D shapes of the Fermi surface in Fig.~\ref{cd3as2FermiSurface} (c) and (d). We also extract the anisotropy parameters from these Fermi surfaces and find the theoretical estimate of UCF to be 0.31 and 0.28, respectively, for Fig.~\ref{cd3as2FermiSurface} (c) and (d). This is close to the plateau value at $W\approx 15$. 

Next, we show results for transport along the $y$ direction for system size $L_x=L_y=L_z=20$ in the right two columns of Fig.~\ref{diff_direc_z}. Below the Dirac point, in comparison with Fig.~\ref{diff_direc_z}(a) and (e), Fig.~\ref{diff_direc_z}(c) and (g) shows that the conductance and its fluctuations along the $y$ direction behave similarly at typical disorder strength of $W\approx 2.5, 10$ and $15$, respectively. We again perform a localization length calculation for a square cross-section of size $15\times15$ and find that $\xi=37, 3$ and $20.4$ at $E_F=-0.8$ for disorder strengths of $W=2.5, 5$ and $12$, respectively [see Fig.~\ref{cd3as2FermiSurface}(h)]. So, it again confirms the physics origin of the vanishing conductance fluctuations to be the Anderson localization, as in Fig.~\ref{diff_direc_z}(a). However, in Fig.~\ref{diff_direc_z}(k), the increase of UCF amplitude (between $1<\langle G\rangle < 10$) saturates versus $E_F$ slightly off the Dirac point and is less significant compared with that along the $z$ direction because Fermi surface is relatively stretched along the $k_{x/y}$ directions compared to the $k_z$ direction [see Fig.~\ref{cd3as2FermiSurface} (a) and (b)]. The theoretical estimates of UCFs are $0.35$ and $0.37$ along the $y$ transport directions, and the first plateaus are slightly above those estimations. It is worth noting that the second plateau in the re-entrant metallic regime is also developing with conductances larger than unity; however, we do not consider this regime relevant for the experiment because it still occurs at a disorder strength much larger than the lower bandwidth.

Above the Dirac point in Fig.~\ref{diff_direc_z}(d), $\Delta G$ increases when Fermi energy is away from the Dirac point and shows a plateau across a range of disorder from $W \approx 5$ to $W \approx 25$. This plateau is also split by $W\approx10$ for large $E_F$, but fortunately, that will not distract the analysis from concluding that UCF increases with the increase of $E_F$. This is also signified by the plateaus in the $\Delta G-\langle G\rangle$ plot in Fig.~\ref{diff_direc_z}(l). These results are again consistent with the joint effects of the Fermi surface shape in Fig.~\ref{cd3as2FermiSurface} (c) and (d) as well as the sample size $L_{\alpha}$. They collectively lead to a sensitive Fermi energy dependence of UCF. Using the ellipsoid approximation of the Fermi surface in Fig.~\ref{cd3as2FermiSurface} (c) and (d), the theoretical estimates of UCF are $0.44$ and $0.50$, respectively. The increase of UCF versus Fermi energy persists and saturates around $\Delta G=0.80$ until $E_F=12$ near the middle of the upper band, where the Fermi surface is severe anisotropic [see Fig.~\ref{cd3as2FermiSurface} (e)]. However, results for those $E_F$ are not shown in Fig.~\ref{diff_direc_z} because the energy is too high and beyond the applicability of the realistic modeling parameters close to the Dirac point.

We test the finite-size effect in Fig.~\ref{cd3as2FermiSurface}(i) and (j) for transport along $z$ and $y$ directions, respectively. Results for the central region with sizes of $L_x=L_y=L_z=10, 15$, and $20$ are compared. We take the average of conductance fluctuations for each size on the plateau. Specifically, in Fig.~\ref{diff_direc_z}, for the transport along the $z$ direction, the average is taken within $W\in[1.2, 2.8]$ below the Dirac point. Above the Dirac point, the region $W\in[9.5, 18]$ for $E_F=0.6\sim3.5$, and $W\in[11.0, 23.5]$ for $E_F=4.0\sim4.5$ is taken for average. For the transport along the $y$ direction, the average is taken within $W\in[1.5, 2.5]$ and $W\in[12, 26.5]$ below and above the Dirac point, respectively. The results have not yet converged but are approaching it. It is convincing that larger system sizes than $20$ will not change the above data trend or the conclusions on UCF's behavior. 

Suppose that our simulation is faithful to the experiment, there are several possible interpretations to connect our data and the experimental observation. We assume that the experiment is ergodically representative and the measured amplitude of conductance fluctuations is the UCF~\cite{liu2016conductance}. In our setup,  transport along the sample's $y$ direction shows an increased conductance fluctuation when the chemical potential moves away from the Dirac point. The UCF plateaus occur for disorder strength of $W\approx 2.5$ and $W\approx 15$ below and above the Dirac point, respectively. Experimentally, we expect a window of disorder strength within which the lower and upper bands are diffusive metals that manifest such UCFs. However, along the $z$ direction, the UCF increases significantly only for chemical potential below the Dirac point. The UCF rapidly saturates and is insensitive to the Fermi energy for chemical potentials above the Dirac point. The distinct behavior of UCF along different transport directions can be tested in future experiments.

In Ref.~\cite{Xiao23}, the experiment observes a UCF increase by $\sim 2.5$ for a Fermi energy that increases above the Dirac point. Their theory using toy model parameters of $\mathrm{Cd_3As_2}$ for Eq.~\ref{cd3as2_h} obtains an increase by a factor of $\sim 1.6$, and assumes that the transport is along the $z$ direction. To best mimic the experimental observations, our simulations suggest that transport is along the $y/x$ direction, giving a UCF increase by a factor of $\sim 2.0$ from $E_F=0.6 ~eV$ up to $E_F=12 ~eV$. The comparison between our results and the theory in Ref.~\cite{Xiao23} highlights the importance of parameter choices when analyzing the effect of anisotropy on UCF. Realistic material parameters and modeling are preferred even for qualitative comparisons because the Fermi energy sensitivity is closely related to the degree of anisotropies. However, we emphasize that previous theories do not contradict our simulations here, not only because the model parameters in Ref.~\cite{Xiao23} are very different from those in Eq.~\ref{cd3as2_h}, but also because the realistic parameters forbid a comparison on an equal footing. Our simulation uses the anisotropic lattice constants that lead to a spatially anisotropic sample. As a result, the sample is much longer in the $z$ direction, which is expected to enhance the Fermi energy sensitivity of the UCF for transport along the $x/y$ directions.

\section{Discussions and Conclusions}\label{conclusion}
Motivated by recent experimental observations~\cite{Xiao23} and theoretical progress~\cite{Hu2017}, this work comprehensively studies UCF in materials with anisotropic band dispersion. This introduces a significant Fermi-energy dependence of UCF, in contrast to the Altshuler-Lee-Stone theory~\cite{altshuler1985fluctuations,lee1985universal,lee1987universal}, whose quantitative result suggests a Fermi-energy independence.  

We revisit the Altshuler-Lee-Stone theory, which explicitly assumes an isotropic free electron gas model in its analytical derivation~\cite{lee1987universal}, and examine its generalization to electron gas with anisotropic parabolic dispersions, which reduces to an ellipsoidal Fermi surface~\cite{Hu2017} that has $2$-fold rotation symmetry. The generalized theory replaces the universal prefactor $c_d$ that determines the UCF through Eq.~\ref{eq:deltaGdepend} by $\tilde{c}_d$ in Eq.~\ref{dgformula1}. We find a consistency between the generalized UCF theory and the numerical calculations in tight-binding models when the Fermi energy is located near the band edges. However, when sweeping the Fermi energy across the entire band, the UCF amplitude significantly varies with Fermi energy when the transport occurs along the direction where the Fermi surface is compressed relative to the transverse directions. The sensitivity of UCF to Fermi energy is more pronounced for a stronger degree of anisotropy. We attribute this to the complex Fermi surface in the band structure, which is caused by the discretization in the tight-binding model but is also physically relevant in the energy bands of realistic materials. This complexity renormalizes the value of $\tilde{c}_d$ in Eq.~\ref{dgformula1} on a Fermi energy dependent basis. Our conclusion is supported by analyzing the symmetry indices and the finite-size effect. 

As an application, we calculate the conductance fluctuations in $\mathrm{Cd_3As_2}$ using first-principle parameters near the Dirac point. We consider a specific sample setup cleaved with the same number of unit cells along the $L_\alpha$($\alpha=x,y,z$) direction. It is found that the UCF increases when the chemical potential moves away from the Dirac point for transport along the $x$ or $y$ direction. In contrast, for transport along the $z$ directions, the UCF dependence on Fermi energy is much less sensitive above the Dirac point. The distinct behaviors of UCFs are testable in experiments. The scope of this work addresses the Fermi energy dependence of UCF in anisotropic materials. Our work does not consider the effects of the magnetic field~\cite{geim1992breakdown}, the long-range impurities~\cite{rossi2012universal}, decoherence~\cite{matsuo2013experimental,bergmann1994different,hoadley1999experimental,alagha2010universal,li2014indications}, finite temperature~\cite{gao1989temperature,Yang2012}, the changes of symmetry indices~\cite{hsu2018conductance,aslani2019conductance,wang2019non}, disorder-induced phase transitions~\cite{chen2015disorder,takagaki2012conductance}, non-ergodicity~\cite{liu2016conductance}, etc, which will be left for future research. In addition to $\mathrm{Cd_3As_2}$, anisotropic materials are prevalent~\cite{Rudenko2024} in nature and can be engineered by straining~\cite{pereira2009tight,nakatsuji2012uniaxial} or twisting ~\cite{carr2017twistronics}. With the well-developed techniques of tuning the chemical potential through gate voltage or doping, we expect our findings to be readily testable in future experiments. 

\section*{Acknowledgements}
The work is supported by the National Science Foundation of China (Grants Nos.12204432, 12304070, 12247106 and 11975126).

\section*{Data Availability}
The data supporting the findings of this article are available~\cite{Yang2025}, embargo periods may apply.

\section*{Appendix}

\appendix
\section{The theoretical calculation of $\tilde{c}_d$}\label{CalculateDeltaG}
\def\theequation{A\arabic{equation}}
\setcounter{equation}{0}

For completeness, we provide here the calculation formula for $\tilde{c}_d$ in the presence of band anisotropy:
\begin{eqnarray}
    \tilde{c}_d &=& 0.5 * \sqrt{2*{\frac{4}{\pi^2}}^2(F_1+F_2+F_3)}\label{dgformula1} \\
  F_1 &=& 2\sum^{\infty}_{m_x,m_y=0}\sum_{m_z=1,2,\ldots} \frac{1}{\tilde{\lambda}_m^2},\nonumber\label{dgformula2}\\
  F_2 &=& -8\sum_{m_x,m_y=0}^{\infty}\sum_{m_z=1,3\ldots}'\sum_{n_z=2,4\ldots}'
  \frac{f_{mn}^2}{\tilde{\lambda}_m\tilde{\lambda}_n}
  (\frac{1}{\tilde{\lambda}_m}+\frac{1}{\tilde{\lambda}_n}),\nonumber\label{dgformula3} \\
  F_3 &=& 24\sum_{m_x,m_y=0}^{\infty}\sum_{m_z,p_z=1,3\ldots}'\sum_{n_z,q_z=2,4\ldots}'
  \frac{f_{mn}f_{np}f_{pq}f_{qm}}{\tilde{\lambda}_m\tilde{\lambda}_n\tilde{\lambda}_p\tilde{\lambda}_q},\nonumber\label{dgformula4}
\end{eqnarray}
where $\tilde{\lambda}_m$ is from Eq.~\ref{modifiedeigenvalue_M} and 
\begin{equation}\label{diffsionEig}
  f_{mn}=4m_zn_z/\pi(m_z^2-n_z^2).\nonumber
\end{equation}
Equation ~\ref{dgformula1} is essentially the same as those in Eq.(2.9) of Ref.~\cite{lee1987universal}, except that the eigenvalues $\tilde{\lambda}_m$ take modified values due to band anisotropy. The infinite series summation consists of several layers of iteration. For each layer of iteration, the cutoff is triggered when the ratio of the new term to the partial sum in this layer is less than a threshold of $\eta=10^{-5}$. This gives reliable converged results up to at least the second digit.

\section{The Fermi surface and density of states}\label{fermisurfacefree}
\def\theequation{B\arabic{equation}}
\setcounter{equation}{0}

When the continuous model in Eq.~\ref{h0_continous} is discretized into Eq.~\ref{h0_discrete}, assuming periodic boundary condition, the Bloch state $\mathbf{k}$ on the Fermi surface satisfies the following relation
\begin{align}\label{A2_1}
	E_F &= M_0 + \frac{2M_x}{a^2}(1-\cos{k_xa}) \\ \notag
    &+ \frac{2M_y}{b^2}(1-\cos{k_yb}) + \frac{2M_z}{c^2}(1-\cos{k_zc}),
\end{align}
at Fermi energy $E_F$. For the lattice constants $a=b=c=1$, we have
\begin{align}\label{A2_2}
	\cos{k_z}=1 - \frac{E_F - M_0 - 2M_x(1-\cos{k_x}) - 2M_y(1-\cos{k_y})}{2M_z}
\end{align}
on the Fermi surface. The largest Fermi momentum in the $\alpha$ direction is $k_F^\alpha=\arccos(1-\frac{E_F-M_0}{2M_\alpha})$ for $\alpha=x, y, z$. 
In the calculations for Fig.~\ref{fermisurfaces}, we set the size of system to be $L_x=L_y=L_z=2000$, then search in the First Brillouin zone of $k_{\alpha}\in[-\frac{L_\alpha}{2}, \frac{L_\alpha}{2})*\frac{2\pi}{L_\alpha}$ for Bloch vectors $\vec{k}$ satisfying Eq.~\ref{A2_2}. The calculation of the Fermi surface in Fig.~\ref{cd3as2FermiSurface} for $\mathrm{Cd_3As_2}$ is similar and can be found in Ref.~\cite{Hu2017}. 

\begin{figure}
\centering
\includegraphics[width=1.0\linewidth]{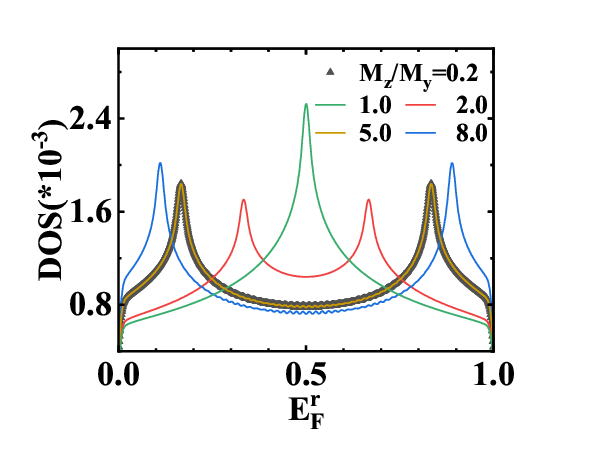}
\caption{\label{dos}(Color online) Density of states for the 2D square lattice for the anisotropy parameters $\frac{M_z}{M_y}=0.2, 1.0, 2.0, 5.0$ and $8.0$, respectively. The horizontal axis is the reduced Fermi energy $E^r_F\equiv E_F/[4(M_y+M_z)]$, $E_F$ divided by the bandwidth for each anisotropy parameter. The DOS for $\frac{M_z}{M_y}=0.2$ and $5.0$ overlap in such a plot. The size of the system is $L_y=L_z=80, L_x=1$.}\label{DOS_appendix}
\end{figure}

Following Ref.~\cite{Hu2021b}, the density of states (DOS) can be obtained through Gaussian broadening of the eigenvalues of Eq.~\ref{h0_discrete}. We take the broadening width to be 60 times the average level spacing. As a reference, we plot the DOS results for several anisotropy parameters in Fig.~\ref{DOS_appendix}. The DOS in Fig.~\ref{ucf_EF_3d}(a) is obtained similarly. The DOS versus disorder in Fig.~\ref{cd3as2FermiSurface}(f) is averaged over several ensembles.  

\section{Numerical results of UCF for the anisotropic free electron gas on a 3D lattice}\label{anisotropicBandNumerical3D}
\def\theequation{B\arabic{equation}}
\setcounter{equation}{0}

In Section. \ref{anisotropicBandNumerical}, the numerical results of UCF have been compared with theoretical predictions in 2D lattices. This section supplements the discussion with numerical results in 3D lattices. Specifically, we consider a cubic central region with $L_x=L_y=L_z=L$, and systematically compare the numerical universal conductance fluctuation (UCF) amplitudes with theoretical predictions for different anisotropy strengths $\frac{M_z}{M_y}$.

\begin{figure}
\centering
\includegraphics[width=1.0\linewidth]{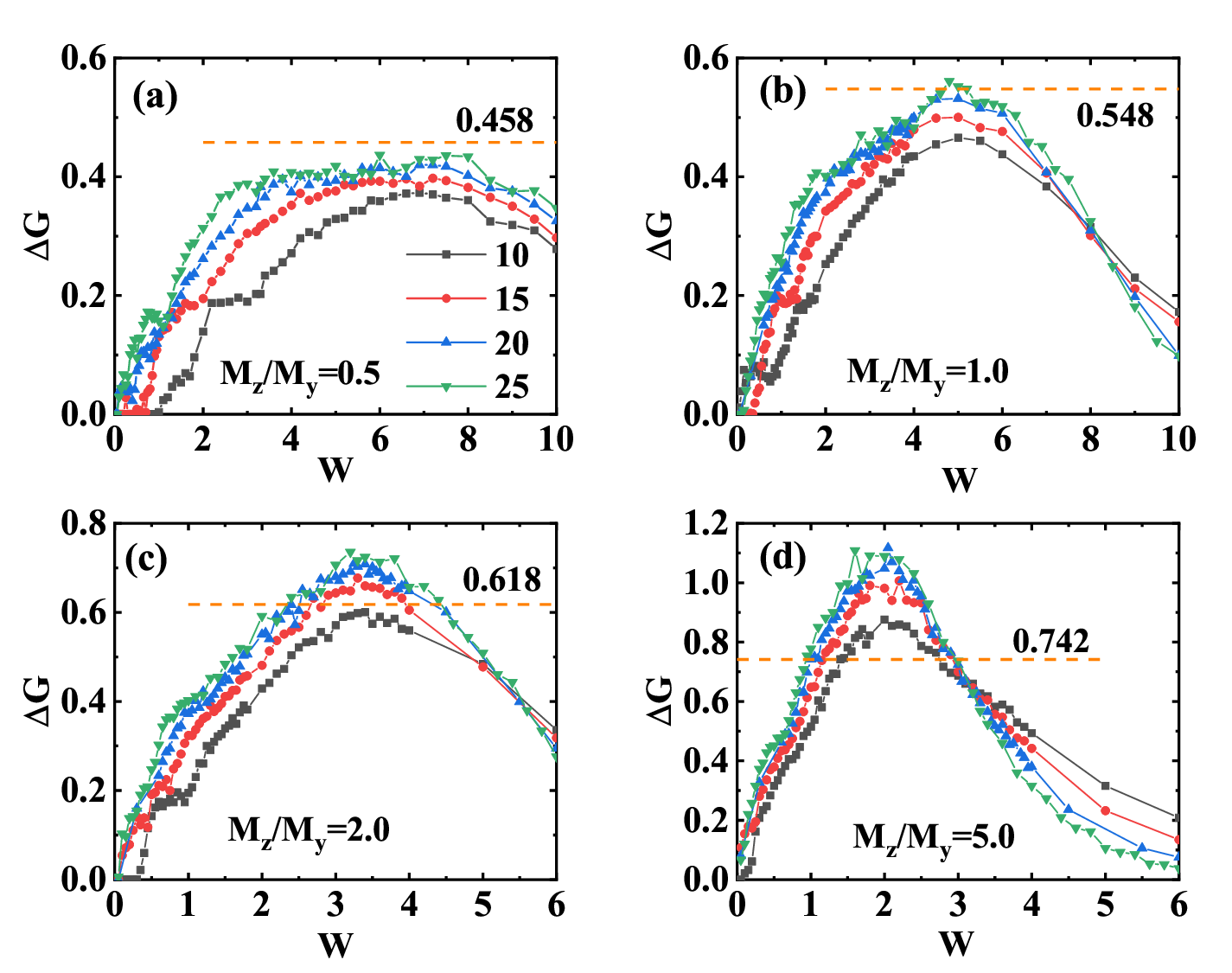}
\caption{(Color online) Numerical results of the conductance fluctuations $\Delta G$ versus the disorder strength W for different anisotropy parameters $\frac{M_z}{M_y}$ in the 3D cubic system with $L_x=L_y=L_z=L$ and with $M_z=0.5$, $M_x=M_y$. (a) $\frac{M_z}{M_y}=0.5$. (b) $\frac{M_z}{M_y}=1.0$. (c) $\frac{M_z}{M_y}=2.0$. (d) $\frac{M_z}{M_y}=5.0$. The Horizontal dashed lines in (a-d) correspond to the theoretical values of $0.458$, $0.548$, $0.618$, and $0.742$, respectively. For each $\frac{M_z}{M_y}$, the Fermi energy in the central region and lead are $E_F=\Delta$ at the band bottom and $E_F^{lead}=8\Delta$ at the band center with $\Delta = (M_x+M_y+M_z)/4$. Several system sizes $L=10, 15, 20$ and $25$ are calculated to show the convergence in the thermodynamic limit. The open boundary conditions are applied in the numerical calculations. The transport direction is along z, and the ensembles for the average are 1200.}\label{numerical_theory_3d}
\end{figure}

The numerical results confirm that for weak anisotropy of $\frac{M_z}{M_y} \le 1$ [see Fig.\ref{numerical_theory_3d} (a) and (b)], the UCF amplitudes converge well towards the theoretical predictions as the system size increases, consistent with the 2D case. For stronger anisotropy
($\frac{M_z}{M_y} > 1.0$), both theoretical and numerical UCF values increase significantly. In Fig.\ref{numerical_theory_3d} (c) and (d), the numerical UCF amplitudes become slightly larger than the theoretical values for the largest size $L=25$ considered here. This discrepancy is attributed to the sensitivity of UCF to the Fermi energy in anisotropic systems, and can be mitigated by positioning the Fermi level closer to the band bottom. However, due to computational cost constraints, we do not extend the calculations to larger system sizes. Nevertheless, the observed trends provide clear evidence supporting the theoretical dependence of UCF on anisotropy in 3D systems.  

\begin{thebibliography}{94}%
\makeatletter
\providecommand \@ifxundefined [1]{%
 \@ifx{#1\undefined}
}%
\providecommand \@ifnum [1]{%
 \ifnum #1\expandafter \@firstoftwo
 \else \expandafter \@secondoftwo
 \fi
}%
\providecommand \@ifx [1]{%
 \ifx #1\expandafter \@firstoftwo
 \else \expandafter \@secondoftwo
 \fi
}%
\providecommand \natexlab [1]{#1}%
\providecommand \enquote  [1]{``#1''}%
\providecommand \bibnamefont  [1]{#1}%
\providecommand \bibfnamefont [1]{#1}%
\providecommand \citenamefont [1]{#1}%
\providecommand \href@noop [0]{\@secondoftwo}%
\providecommand \href [0]{\begingroup \@sanitize@url \@href}%
\providecommand \@href[1]{\@@startlink{#1}\@@href}%
\providecommand \@@href[1]{\endgroup#1\@@endlink}%
\providecommand \@sanitize@url [0]{\catcode `\\12\catcode `\$12\catcode `\&12\catcode `\#12\catcode `\^12\catcode `\_12\catcode `\%12\relax}%
\providecommand \@@startlink[1]{}%
\providecommand \@@endlink[0]{}%
\providecommand \url  [0]{\begingroup\@sanitize@url \@url }%
\providecommand \@url [1]{\endgroup\@href {#1}{\urlprefix }}%
\providecommand \urlprefix  [0]{URL }%
\providecommand \Eprint [0]{\href }%
\providecommand \doibase [0]{https://doi.org/}%
\providecommand \selectlanguage [0]{\@gobble}%
\providecommand \bibinfo  [0]{\@secondoftwo}%
\providecommand \bibfield  [0]{\@secondoftwo}%
\providecommand \translation [1]{[#1]}%
\providecommand \BibitemOpen [0]{}%
\providecommand \bibitemStop [0]{}%
\providecommand \bibitemNoStop [0]{.\EOS\space}%
\providecommand \EOS [0]{\spacefactor3000\relax}%
\providecommand \BibitemShut  [1]{\csname bibitem#1\endcsname}%
\let\auto@bib@innerbib\@empty
\bibitem [{\citenamefont {Altshuler}(1985)}]{altshuler1985fluctuations}%
  \BibitemOpen
  \bibfield  {author} {\bibinfo {author} {\bibfnamefont {B.}~\bibnamefont {Altshuler}},\ }\bibfield  {title} {\bibinfo {title} {Fluctuations in the extrinsic conductivity of disordered conductors},\ }\href {http://jetpletters.ru/ps/1470/article_22425.shtml} {\bibfield  {journal} {\bibinfo  {journal} {JETP lett}\ }\textbf {\bibinfo {volume} {41}},\ \bibinfo {pages} {648} (\bibinfo {year} {1985})}\BibitemShut {NoStop}%
\bibitem [{\citenamefont {Lee}\ and\ \citenamefont {Stone}(1985)}]{lee1985universal}%
  \BibitemOpen
  \bibfield  {author} {\bibinfo {author} {\bibfnamefont {P.~A.}\ \bibnamefont {Lee}}\ and\ \bibinfo {author} {\bibfnamefont {A.~D.}\ \bibnamefont {Stone}},\ }\bibfield  {title} {\bibinfo {title} {Universal conductance fluctuations in metals},\ }\href {https://doi.org/10.1103/PhysRevLett.55.1622} {\bibfield  {journal} {\bibinfo  {journal} {Phys. Rev. Lett.}\ }\textbf {\bibinfo {volume} {55}},\ \bibinfo {pages} {1622} (\bibinfo {year} {1985})}\BibitemShut {NoStop}%
\bibitem [{\citenamefont {Stone}(1985)}]{stone1985magnetoresistance}%
  \BibitemOpen
  \bibfield  {author} {\bibinfo {author} {\bibfnamefont {A.~D.}\ \bibnamefont {Stone}},\ }\bibfield  {title} {\bibinfo {title} {Magnetoresistance fluctuations in mesoscopic wires and rings},\ }\href {https://doi.org/10.1103/PhysRevLett.54.2692} {\bibfield  {journal} {\bibinfo  {journal} {Phys. Rev. Lett.}\ }\textbf {\bibinfo {volume} {54}},\ \bibinfo {pages} {2692} (\bibinfo {year} {1985})}\BibitemShut {NoStop}%
\bibitem [{\citenamefont {Allen}(2006)}]{ALLEN2006}%
  \BibitemOpen
  \bibfield  {author} {\bibinfo {author} {\bibfnamefont {P.}~\bibnamefont {Allen}},\ }\bibfield  {title} {\bibinfo {title} {Chapter 6 electron transport},\ }in\ \href {https://doi.org/https://doi.org/10.1016/S1572-0934(06)02006-3} {\emph {\bibinfo {booktitle} {Conceptual Foundations of Materials}}},\ \bibinfo {series} {Contemporary Concepts of Condensed Matter Science}, Vol.~\bibinfo {volume} {2},\ \bibinfo {editor} {edited by\ \bibinfo {editor} {\bibfnamefont {S.~G.}\ \bibnamefont {Louie}}\ and\ \bibinfo {editor} {\bibfnamefont {M.~L.}\ \bibnamefont {Cohen}}}\ (\bibinfo  {publisher} {Elsevier},\ \bibinfo {year} {2006})\ pp.\ \bibinfo {pages} {165--218}\BibitemShut {NoStop}%
\bibitem [{\citenamefont {Akkermans}\ and\ \citenamefont {Montambaux}(2007)}]{akkermans2007mesoscopic}%
  \BibitemOpen
  \bibfield  {author} {\bibinfo {author} {\bibfnamefont {E.}~\bibnamefont {Akkermans}}\ and\ \bibinfo {author} {\bibfnamefont {G.}~\bibnamefont {Montambaux}},\ }\href {https://books.google.com.hk/books?id=cs7AVel15TAC} {\emph {\bibinfo {title} {Mesoscopic Physics of Electrons and Photons}}}\ (\bibinfo  {publisher} {Cambridge University Press},\ \bibinfo {year} {2007})\BibitemShut {NoStop}%
\bibitem [{\citenamefont {Islam}\ \emph {et~al.}(2023)\citenamefont {Islam}, \citenamefont {Shamim},\ and\ \citenamefont {Ghosh}}]{Saurav2023benchmarking}%
  \BibitemOpen
  \bibfield  {author} {\bibinfo {author} {\bibfnamefont {S.}~\bibnamefont {Islam}}, \bibinfo {author} {\bibfnamefont {S.}~\bibnamefont {Shamim}},\ and\ \bibinfo {author} {\bibfnamefont {A.}~\bibnamefont {Ghosh}},\ }\bibfield  {title} {\bibinfo {title} {Benchmarking noise and dephasing in emerging electrical materials for quantum technologies},\ }\href {https://doi.org/https://doi.org/10.1002/adma.202109671} {\bibfield  {journal} {\bibinfo  {journal} {Advanced Materials}\ }\textbf {\bibinfo {volume} {35}},\ \bibinfo {pages} {2109671} (\bibinfo {year} {2023})}\BibitemShut {NoStop}%
\bibitem [{\citenamefont {Lee}\ \emph {et~al.}(1987)\citenamefont {Lee}, \citenamefont {Stone},\ and\ \citenamefont {Fukuyama}}]{lee1987universal}%
  \BibitemOpen
  \bibfield  {author} {\bibinfo {author} {\bibfnamefont {P.~A.}\ \bibnamefont {Lee}}, \bibinfo {author} {\bibfnamefont {A.~D.}\ \bibnamefont {Stone}},\ and\ \bibinfo {author} {\bibfnamefont {H.}~\bibnamefont {Fukuyama}},\ }\bibfield  {title} {\bibinfo {title} {Universal conductance fluctuations in metals: Effects of finite temperature, interactions, and magnetic field},\ }\href {https://doi.org/10.1103/PhysRevB.35.1039} {\bibfield  {journal} {\bibinfo  {journal} {Phys. Rev. B}\ }\textbf {\bibinfo {volume} {35}},\ \bibinfo {pages} {1039} (\bibinfo {year} {1987})}\BibitemShut {NoStop}%
\bibitem [{\citenamefont {Altshuler}\ and\ \citenamefont {Shklovskii}(1986)}]{altshuler1986repulsion}%
  \BibitemOpen
  \bibfield  {author} {\bibinfo {author} {\bibfnamefont {B.~L.}\ \bibnamefont {Altshuler}}\ and\ \bibinfo {author} {\bibfnamefont {B.~I.}\ \bibnamefont {Shklovskii}},\ }\bibfield  {title} {\bibinfo {title} {Repulsion of energy levels and conductivity of small metal samples},\ }\href {http://jetp.ras.ru/cgi-bin/e/index/e/64/1/p127?a=list} {\bibfield  {journal} {\bibinfo  {journal} {Sov. Phys. JETP}\ }\textbf {\bibinfo {volume} {64}},\ \bibinfo {pages} {127} (\bibinfo {year} {1986})}\BibitemShut {NoStop}%
\bibitem [{\citenamefont {Beenakker}(1997)}]{beenakker1997random}%
  \BibitemOpen
  \bibfield  {author} {\bibinfo {author} {\bibfnamefont {C.~W.~J.}\ \bibnamefont {Beenakker}},\ }\bibfield  {title} {\bibinfo {title} {Random-matrix theory of quantum transport},\ }\href {https://doi.org/10.1103/RevModPhys.69.731} {\bibfield  {journal} {\bibinfo  {journal} {Rev. Mod. Phys.}\ }\textbf {\bibinfo {volume} {69}},\ \bibinfo {pages} {731} (\bibinfo {year} {1997})}\BibitemShut {NoStop}%
\bibitem [{\citenamefont {Debray}\ \emph {et~al.}(1989)\citenamefont {Debray}, \citenamefont {Pichard}, \citenamefont {Vicente},\ and\ \citenamefont {Tung}}]{debray1989reduction}%
  \BibitemOpen
  \bibfield  {author} {\bibinfo {author} {\bibfnamefont {P.}~\bibnamefont {Debray}}, \bibinfo {author} {\bibfnamefont {J.-L.}\ \bibnamefont {Pichard}}, \bibinfo {author} {\bibfnamefont {J.}~\bibnamefont {Vicente}},\ and\ \bibinfo {author} {\bibfnamefont {P.~N.}\ \bibnamefont {Tung}},\ }\bibfield  {title} {\bibinfo {title} {Reduction of mesoscopic conductance fluctuations due to zeeman splitting in a disordered conductor without spin-orbit scattering},\ }\href {https://doi.org/10.1103/PhysRevLett.63.2264} {\bibfield  {journal} {\bibinfo  {journal} {Phys. Rev. Lett.}\ }\textbf {\bibinfo {volume} {63}},\ \bibinfo {pages} {2264} (\bibinfo {year} {1989})}\BibitemShut {NoStop}%
\bibitem [{\citenamefont {Moon}\ \emph {et~al.}(1997)\citenamefont {Moon}, \citenamefont {Birge},\ and\ \citenamefont {Golding}}]{moon1997observation}%
  \BibitemOpen
  \bibfield  {author} {\bibinfo {author} {\bibfnamefont {J.~S.}\ \bibnamefont {Moon}}, \bibinfo {author} {\bibfnamefont {N.~O.}\ \bibnamefont {Birge}},\ and\ \bibinfo {author} {\bibfnamefont {B.}~\bibnamefont {Golding}},\ }\bibfield  {title} {\bibinfo {title} {Observation of universal conductance-fluctuation crossovers in mesoscopic li wires},\ }\href {https://doi.org/10.1103/PhysRevB.56.15124} {\bibfield  {journal} {\bibinfo  {journal} {Phys. Rev. B}\ }\textbf {\bibinfo {volume} {56}},\ \bibinfo {pages} {15124} (\bibinfo {year} {1997})}\BibitemShut {NoStop}%
\bibitem [{\citenamefont {Pal}\ \emph {et~al.}(2012)\citenamefont {Pal}, \citenamefont {Kochat},\ and\ \citenamefont {Ghosh}}]{pal2012direct}%
  \BibitemOpen
  \bibfield  {author} {\bibinfo {author} {\bibfnamefont {A.~N.}\ \bibnamefont {Pal}}, \bibinfo {author} {\bibfnamefont {V.}~\bibnamefont {Kochat}},\ and\ \bibinfo {author} {\bibfnamefont {A.}~\bibnamefont {Ghosh}},\ }\bibfield  {title} {\bibinfo {title} {Direct observation of valley hybridization and universal symmetry of graphene with mesoscopic conductance fluctuations},\ }\href {https://doi.org/10.1103/PhysRevLett.109.196601} {\bibfield  {journal} {\bibinfo  {journal} {Phys. Rev. Lett.}\ }\textbf {\bibinfo {volume} {109}},\ \bibinfo {pages} {196601} (\bibinfo {year} {2012})}\BibitemShut {NoStop}%
\bibitem [{\citenamefont {Wang}\ \emph {et~al.}(2016)\citenamefont {Wang}, \citenamefont {Wang}, \citenamefont {Li}, \citenamefont {Li}, \citenamefont {Yu},\ and\ \citenamefont {Liao}}]{wang2016universal}%
  \BibitemOpen
  \bibfield  {author} {\bibinfo {author} {\bibfnamefont {L.-X.}\ \bibnamefont {Wang}}, \bibinfo {author} {\bibfnamefont {S.}~\bibnamefont {Wang}}, \bibinfo {author} {\bibfnamefont {J.-G.}\ \bibnamefont {Li}}, \bibinfo {author} {\bibfnamefont {C.-Z.}\ \bibnamefont {Li}}, \bibinfo {author} {\bibfnamefont {D.}~\bibnamefont {Yu}},\ and\ \bibinfo {author} {\bibfnamefont {Z.-M.}\ \bibnamefont {Liao}},\ }\bibfield  {title} {\bibinfo {title} {Universal conductance fluctuations in dirac semimetal $\mathrm{C}{\mathrm{d}}_{3}\mathrm{A}{\mathrm{s}}_{2}$ nanowires},\ }\href {https://doi.org/10.1103/PhysRevB.94.161402} {\bibfield  {journal} {\bibinfo  {journal} {Phys. Rev. B}\ }\textbf {\bibinfo {volume} {94}},\ \bibinfo {pages} {161402} (\bibinfo {year} {2016})}\BibitemShut {NoStop}%
\bibitem [{\citenamefont {Zhang}\ \emph {et~al.}(2018)\citenamefont {Zhang}, \citenamefont {Pan}, \citenamefont {Li}, \citenamefont {Xie}, \citenamefont {Qin}, \citenamefont {Cao}, \citenamefont {Wang}, \citenamefont {Wang}, \citenamefont {Miao}, \citenamefont {Song},\ and\ \citenamefont {Wang}}]{zhang2018}%
  \BibitemOpen
  \bibfield  {author} {\bibinfo {author} {\bibfnamefont {S.}~\bibnamefont {Zhang}}, \bibinfo {author} {\bibfnamefont {X.-C.}\ \bibnamefont {Pan}}, \bibinfo {author} {\bibfnamefont {Z.}~\bibnamefont {Li}}, \bibinfo {author} {\bibfnamefont {F.}~\bibnamefont {Xie}}, \bibinfo {author} {\bibfnamefont {Y.}~\bibnamefont {Qin}}, \bibinfo {author} {\bibfnamefont {L.}~\bibnamefont {Cao}}, \bibinfo {author} {\bibfnamefont {X.}~\bibnamefont {Wang}}, \bibinfo {author} {\bibfnamefont {X.}~\bibnamefont {Wang}}, \bibinfo {author} {\bibfnamefont {F.}~\bibnamefont {Miao}}, \bibinfo {author} {\bibfnamefont {F.}~\bibnamefont {Song}},\ and\ \bibinfo {author} {\bibfnamefont {B.}~\bibnamefont {Wang}},\ }\bibfield  {title} {\bibinfo {title} {2 step of conductance fluctuations due to the broken time-reversal symmetry in bulk-insulating bisbtese2 devices},\ }\href {https://doi.org/10.1063/1.5031013} {\bibfield  {journal} {\bibinfo  {journal} {Applied Physics Letters}\ }\textbf {\bibinfo {volume} {112}},\ \bibinfo {pages} {243106}
  (\bibinfo {year} {2018})}\BibitemShut {NoStop}%
\bibitem [{\citenamefont {Hu}\ \emph {et~al.}(2017)\citenamefont {Hu}, \citenamefont {Liu}, \citenamefont {Jiang},\ and\ \citenamefont {Xie}}]{Hu2017}%
  \BibitemOpen
  \bibfield  {author} {\bibinfo {author} {\bibfnamefont {Y.}~\bibnamefont {Hu}}, \bibinfo {author} {\bibfnamefont {H.}~\bibnamefont {Liu}}, \bibinfo {author} {\bibfnamefont {H.}~\bibnamefont {Jiang}},\ and\ \bibinfo {author} {\bibfnamefont {X.~C.}\ \bibnamefont {Xie}},\ }\bibfield  {title} {\bibinfo {title} {Numerical study of universal conductance fluctuations in three-dimensional topological semimetals},\ }\href {https://doi.org/10.1103/PhysRevB.96.134201} {\bibfield  {journal} {\bibinfo  {journal} {Phys. Rev. B}\ }\textbf {\bibinfo {volume} {96}},\ \bibinfo {pages} {134201} (\bibinfo {year} {2017})}\BibitemShut {NoStop}%
\bibitem [{\citenamefont {R{\"u}hl{\"a}nder}\ \emph {et~al.}(2001)\citenamefont {R{\"u}hl{\"a}nder}, \citenamefont {Marko{\v{s}}},\ and\ \citenamefont {Soukoulis}}]{ruhlander2001symmetry}%
  \BibitemOpen
  \bibfield  {author} {\bibinfo {author} {\bibfnamefont {M.}~\bibnamefont {R{\"u}hl{\"a}nder}}, \bibinfo {author} {\bibfnamefont {P.}~\bibnamefont {Marko{\v{s}}}},\ and\ \bibinfo {author} {\bibfnamefont {C.~M.}\ \bibnamefont {Soukoulis}},\ }\bibfield  {title} {\bibinfo {title} {Symmetry, dimension, and the distribution of the conductance at the mobility edge},\ }\href {https://doi.org/10.1103/PhysRevB.64.212202} {\bibfield  {journal} {\bibinfo  {journal} {Phys. Rev. B}\ }\textbf {\bibinfo {volume} {64}},\ \bibinfo {pages} {212202} (\bibinfo {year} {2001})}\BibitemShut {NoStop}%
\bibitem [{\citenamefont {Takane}\ and\ \citenamefont {Ebisawa}(1992)}]{takane1992conductance}%
  \BibitemOpen
  \bibfield  {author} {\bibinfo {author} {\bibfnamefont {Y.}~\bibnamefont {Takane}}\ and\ \bibinfo {author} {\bibfnamefont {H.}~\bibnamefont {Ebisawa}},\ }\bibfield  {title} {\bibinfo {title} {Conductance and its fluctuations of mesoscopic wires in contact with a superconductor},\ }\href {https://doi.org/10.1143/JPSJ.61.2858} {\bibfield  {journal} {\bibinfo  {journal} {Journal of the Physical Society of Japan}\ }\textbf {\bibinfo {volume} {61}},\ \bibinfo {pages} {2858} (\bibinfo {year} {1992})}\BibitemShut {NoStop}%
\bibitem [{\citenamefont {Ryu}\ \emph {et~al.}(2007)\citenamefont {Ryu}, \citenamefont {Furusaki}, \citenamefont {Ludwig},\ and\ \citenamefont {Mudry}}]{ryu2007conductance}%
  \BibitemOpen
  \bibfield  {author} {\bibinfo {author} {\bibfnamefont {S.}~\bibnamefont {Ryu}}, \bibinfo {author} {\bibfnamefont {A.}~\bibnamefont {Furusaki}}, \bibinfo {author} {\bibfnamefont {A.}~\bibnamefont {Ludwig}},\ and\ \bibinfo {author} {\bibfnamefont {C.}~\bibnamefont {Mudry}},\ }\bibfield  {title} {\bibinfo {title} {Conductance fluctuations in disordered superconductors with broken time-reversal symmetry near two dimensions},\ }\href {https://doi.org/https://doi.org/10.1016/j.nuclphysb.2007.03.027} {\bibfield  {journal} {\bibinfo  {journal} {Nuclear Physics B}\ }\textbf {\bibinfo {volume} {780}},\ \bibinfo {pages} {105} (\bibinfo {year} {2007})}\BibitemShut {NoStop}%
\bibitem [{\citenamefont {Antonenko}\ \emph {et~al.}(2020)\citenamefont {Antonenko}, \citenamefont {Khalaf}, \citenamefont {Ostrovsky},\ and\ \citenamefont {Skvortsov}}]{antonenko2020mesoscopic}%
  \BibitemOpen
  \bibfield  {author} {\bibinfo {author} {\bibfnamefont {D.~S.}\ \bibnamefont {Antonenko}}, \bibinfo {author} {\bibfnamefont {E.}~\bibnamefont {Khalaf}}, \bibinfo {author} {\bibfnamefont {P.~M.}\ \bibnamefont {Ostrovsky}},\ and\ \bibinfo {author} {\bibfnamefont {M.~A.}\ \bibnamefont {Skvortsov}},\ }\bibfield  {title} {\bibinfo {title} {Mesoscopic conductance fluctuations and noise in disordered majorana wires},\ }\href {https://doi.org/10.1103/PhysRevB.102.195152} {\bibfield  {journal} {\bibinfo  {journal} {Phys. Rev. B}\ }\textbf {\bibinfo {volume} {102}},\ \bibinfo {pages} {195152} (\bibinfo {year} {2020})}\BibitemShut {NoStop}%
\bibitem [{\citenamefont {Zyuzin}\ and\ \citenamefont {Spivak}(1990)}]{zyuzin1990mesoscopic}%
  \BibitemOpen
  \bibfield  {author} {\bibinfo {author} {\bibfnamefont {A.~Y.}\ \bibnamefont {Zyuzin}}\ and\ \bibinfo {author} {\bibfnamefont {B.}~\bibnamefont {Spivak}},\ }\bibfield  {title} {\bibinfo {title} {Mesoscopic fluctuations of the resistance of point contact},\ }\href {http://jetp.ras.ru/cgi-bin/dn/e_071_03_0563.pdf} {\bibfield  {journal} {\bibinfo  {journal} {Sov. Phys. JETP}\ }\textbf {\bibinfo {volume} {71}},\ \bibinfo {pages} {563} (\bibinfo {year} {1990})}\BibitemShut {NoStop}%
\bibitem [{\citenamefont {Chandrasekhar}\ \emph {et~al.}(1991)\citenamefont {Chandrasekhar}, \citenamefont {Santhanam},\ and\ \citenamefont {Prober}}]{chandrasekhar1991weak}%
  \BibitemOpen
  \bibfield  {author} {\bibinfo {author} {\bibfnamefont {V.}~\bibnamefont {Chandrasekhar}}, \bibinfo {author} {\bibfnamefont {P.}~\bibnamefont {Santhanam}},\ and\ \bibinfo {author} {\bibfnamefont {D.~E.}\ \bibnamefont {Prober}},\ }\bibfield  {title} {\bibinfo {title} {Weak localization and conductance fluctuations in complex mesoscopic geometries},\ }\href {https://doi.org/10.1103/PhysRevB.44.11203} {\bibfield  {journal} {\bibinfo  {journal} {Phys. Rev. B}\ }\textbf {\bibinfo {volume} {44}},\ \bibinfo {pages} {11203} (\bibinfo {year} {1991})}\BibitemShut {NoStop}%
\bibitem [{\citenamefont {Trav{\v{e}}nec}(2004)}]{travvenec2004universal}%
  \BibitemOpen
  \bibfield  {author} {\bibinfo {author} {\bibfnamefont {I.}~\bibnamefont {Trav{\v{e}}nec}},\ }\bibfield  {title} {\bibinfo {title} {Universal conductance fluctuations in noninteger dimensions},\ }\href {https://doi.org/10.1103/PhysRevB.69.033104} {\bibfield  {journal} {\bibinfo  {journal} {Phys. Rev. B}\ }\textbf {\bibinfo {volume} {69}},\ \bibinfo {pages} {033104} (\bibinfo {year} {2004})}\BibitemShut {NoStop}%
\bibitem [{\citenamefont {Amin}\ \emph {et~al.}(2022)\citenamefont {Amin}, \citenamefont {Nagarajan}, \citenamefont {Pandit},\ and\ \citenamefont {Bid}}]{amin2022multifractal}%
  \BibitemOpen
  \bibfield  {author} {\bibinfo {author} {\bibfnamefont {K.~R.}\ \bibnamefont {Amin}}, \bibinfo {author} {\bibfnamefont {R.}~\bibnamefont {Nagarajan}}, \bibinfo {author} {\bibfnamefont {R.}~\bibnamefont {Pandit}},\ and\ \bibinfo {author} {\bibfnamefont {A.}~\bibnamefont {Bid}},\ }\bibfield  {title} {\bibinfo {title} {Multifractal conductance fluctuations in high-mobility graphene in the integer quantum hall regime},\ }\href {https://doi.org/10.1103/PhysRevLett.129.186802} {\bibfield  {journal} {\bibinfo  {journal} {Phys. Rev. Lett.}\ }\textbf {\bibinfo {volume} {129}},\ \bibinfo {pages} {186802} (\bibinfo {year} {2022})}\BibitemShut {NoStop}%
\bibitem [{\citenamefont {Asano}\ and\ \citenamefont {Bauer}(1996)}]{asano1996conductance}%
  \BibitemOpen
  \bibfield  {author} {\bibinfo {author} {\bibfnamefont {Y.}~\bibnamefont {Asano}}\ and\ \bibinfo {author} {\bibfnamefont {G.~E.~W.}\ \bibnamefont {Bauer}},\ }\bibfield  {title} {\bibinfo {title} {Conductance fluctuations near the ballistic-transport regime},\ }\href {https://doi.org/10.1103/PhysRevB.54.11602} {\bibfield  {journal} {\bibinfo  {journal} {Phys. Rev. B}\ }\textbf {\bibinfo {volume} {54}},\ \bibinfo {pages} {11602} (\bibinfo {year} {1996})}\BibitemShut {NoStop}%
\bibitem [{\citenamefont {Adam}\ \emph {et~al.}(2006)\citenamefont {Adam}, \citenamefont {Kindermann}, \citenamefont {Rahav},\ and\ \citenamefont {Brouwer}}]{adam2006mesoscopic}%
  \BibitemOpen
  \bibfield  {author} {\bibinfo {author} {\bibfnamefont {S.}~\bibnamefont {Adam}}, \bibinfo {author} {\bibfnamefont {M.}~\bibnamefont {Kindermann}}, \bibinfo {author} {\bibfnamefont {S.}~\bibnamefont {Rahav}},\ and\ \bibinfo {author} {\bibfnamefont {P.~W.}\ \bibnamefont {Brouwer}},\ }\bibfield  {title} {\bibinfo {title} {Mesoscopic anisotropic magnetoconductance fluctuations in ferromagnets},\ }\href {https://doi.org/10.1103/PhysRevB.73.212408} {\bibfield  {journal} {\bibinfo  {journal} {Phys. Rev. B}\ }\textbf {\bibinfo {volume} {73}},\ \bibinfo {pages} {212408} (\bibinfo {year} {2006})}\BibitemShut {NoStop}%
\bibitem [{\citenamefont {Bardarson}\ \emph {et~al.}(2007)\citenamefont {Bardarson}, \citenamefont {Adagideli},\ and\ \citenamefont {Jacquod}}]{bardarson2007mesoscopic}%
  \BibitemOpen
  \bibfield  {author} {\bibinfo {author} {\bibfnamefont {J.~H.}\ \bibnamefont {Bardarson}}, \bibinfo {author} {\bibfnamefont {{\.{I}}.}~\bibnamefont {Adagideli}},\ and\ \bibinfo {author} {\bibfnamefont {P.}~\bibnamefont {Jacquod}},\ }\bibfield  {title} {\bibinfo {title} {Mesoscopic spin hall effect},\ }\href {https://doi.org/10.1103/PhysRevLett.98.196601} {\bibfield  {journal} {\bibinfo  {journal} {Phys. Rev. Lett.}\ }\textbf {\bibinfo {volume} {98}},\ \bibinfo {pages} {196601} (\bibinfo {year} {2007})}\BibitemShut {NoStop}%
\bibitem [{\citenamefont {Scheid}\ \emph {et~al.}(2009)\citenamefont {Scheid}, \citenamefont {Adagideli}, \citenamefont {Nitta},\ and\ \citenamefont {Richter}}]{scheid2009anisotropic}%
  \BibitemOpen
  \bibfield  {author} {\bibinfo {author} {\bibfnamefont {M.}~\bibnamefont {Scheid}}, \bibinfo {author} {\bibfnamefont {I.}~\bibnamefont {Adagideli}}, \bibinfo {author} {\bibfnamefont {J.}~\bibnamefont {Nitta}},\ and\ \bibinfo {author} {\bibfnamefont {K.}~\bibnamefont {Richter}},\ }\bibfield  {title} {\bibinfo {title} {Anisotropic universal conductance fluctuations in disordered quantum wires with rashba and dresselhaus spin-orbit interaction and an applied in-plane magnetic field},\ }\href {https://doi.org/10.1088/0268-1242/24/6/064005} {\bibfield  {journal} {\bibinfo  {journal} {Semiconductor Science and Technology}\ }\textbf {\bibinfo {volume} {24}},\ \bibinfo {pages} {064005} (\bibinfo {year} {2009})}\BibitemShut {NoStop}%
\bibitem [{\citenamefont {Kaneko}\ \emph {et~al.}(2010)\citenamefont {Kaneko}, \citenamefont {Koshino},\ and\ \citenamefont {Ando}}]{kaneko2010symmetry}%
  \BibitemOpen
  \bibfield  {author} {\bibinfo {author} {\bibfnamefont {T.}~\bibnamefont {Kaneko}}, \bibinfo {author} {\bibfnamefont {M.}~\bibnamefont {Koshino}},\ and\ \bibinfo {author} {\bibfnamefont {T.}~\bibnamefont {Ando}},\ }\bibfield  {title} {\bibinfo {title} {Symmetry crossover in quantum wires with spin-orbit interaction},\ }\href {https://doi.org/10.1103/PhysRevB.81.155310} {\bibfield  {journal} {\bibinfo  {journal} {Phys. Rev. B}\ }\textbf {\bibinfo {volume} {81}},\ \bibinfo {pages} {155310} (\bibinfo {year} {2010})}\BibitemShut {NoStop}%
\bibitem [{\citenamefont {Ren}\ \emph {et~al.}(2006)\citenamefont {Ren}, \citenamefont {Qiao}, \citenamefont {Wang}, \citenamefont {Sun},\ and\ \citenamefont {Guo}}]{ren2006universal}%
  \BibitemOpen
  \bibfield  {author} {\bibinfo {author} {\bibfnamefont {W.}~\bibnamefont {Ren}}, \bibinfo {author} {\bibfnamefont {Z.}~\bibnamefont {Qiao}}, \bibinfo {author} {\bibfnamefont {J.}~\bibnamefont {Wang}}, \bibinfo {author} {\bibfnamefont {Q.}~\bibnamefont {Sun}},\ and\ \bibinfo {author} {\bibfnamefont {H.}~\bibnamefont {Guo}},\ }\bibfield  {title} {\bibinfo {title} {Universal spin-hall conductance fluctuations in two dimensions},\ }\href {https://doi.org/10.1103/PhysRevLett.97.066603} {\bibfield  {journal} {\bibinfo  {journal} {Phys. Rev. Lett.}\ }\textbf {\bibinfo {volume} {97}},\ \bibinfo {pages} {066603} (\bibinfo {year} {2006})}\BibitemShut {NoStop}%
\bibitem [{\citenamefont {Takagaki}(2012)}]{takagaki2012conductance}%
  \BibitemOpen
  \bibfield  {author} {\bibinfo {author} {\bibfnamefont {Y.}~\bibnamefont {Takagaki}},\ }\bibfield  {title} {\bibinfo {title} {Conductance fluctuations induced by bulk state in quasi-one-dimensional strips of topological insulator},\ }\href {https://doi.org/10.1103/PhysRevB.85.155308} {\bibfield  {journal} {\bibinfo  {journal} {Phys. Rev. B}\ }\textbf {\bibinfo {volume} {85}},\ \bibinfo {pages} {155308} (\bibinfo {year} {2012})}\BibitemShut {NoStop}%
\bibitem [{\citenamefont {Zhang}\ \emph {et~al.}(2014)\citenamefont {Zhang}, \citenamefont {Zhuang}, \citenamefont {Xing}, \citenamefont {Li}, \citenamefont {Wang},\ and\ \citenamefont {Guo}}]{zhang2014universal}%
  \BibitemOpen
  \bibfield  {author} {\bibinfo {author} {\bibfnamefont {L.}~\bibnamefont {Zhang}}, \bibinfo {author} {\bibfnamefont {J.}~\bibnamefont {Zhuang}}, \bibinfo {author} {\bibfnamefont {Y.}~\bibnamefont {Xing}}, \bibinfo {author} {\bibfnamefont {J.}~\bibnamefont {Li}}, \bibinfo {author} {\bibfnamefont {J.}~\bibnamefont {Wang}},\ and\ \bibinfo {author} {\bibfnamefont {H.}~\bibnamefont {Guo}},\ }\bibfield  {title} {\bibinfo {title} {Universal transport properties of three-dimensional topological insulator nanowires},\ }\href {https://doi.org/10.1103/PhysRevB.89.245107} {\bibfield  {journal} {\bibinfo  {journal} {Phys. Rev. B}\ }\textbf {\bibinfo {volume} {89}},\ \bibinfo {pages} {245107} (\bibinfo {year} {2014})}\BibitemShut {NoStop}%
\bibitem [{\citenamefont {Choe}\ and\ \citenamefont {Chang}(2015)}]{choe2015universal}%
  \BibitemOpen
  \bibfield  {author} {\bibinfo {author} {\bibfnamefont {D.-H.}\ \bibnamefont {Choe}}\ and\ \bibinfo {author} {\bibfnamefont {K.-J.}\ \bibnamefont {Chang}},\ }\bibfield  {title} {\bibinfo {title} {Universal conductance fluctuation in two-dimensional topological insulators},\ }\href {https://doi.org/10.1038/srep10997} {\bibfield  {journal} {\bibinfo  {journal} {Scientific reports}\ }\textbf {\bibinfo {volume} {5}},\ \bibinfo {pages} {10997} (\bibinfo {year} {2015})}\BibitemShut {NoStop}%
\bibitem [{\citenamefont {Vasconcelos}\ \emph {et~al.}(2016)\citenamefont {Vasconcelos}, \citenamefont {Ramos},\ and\ \citenamefont {Barbosa}}]{vasconcelos2016universal}%
  \BibitemOpen
  \bibfield  {author} {\bibinfo {author} {\bibfnamefont {T.~C.}\ \bibnamefont {Vasconcelos}}, \bibinfo {author} {\bibfnamefont {J.~G. G.~S.}\ \bibnamefont {Ramos}},\ and\ \bibinfo {author} {\bibfnamefont {A.~L.~R.}\ \bibnamefont {Barbosa}},\ }\bibfield  {title} {\bibinfo {title} {Universal spin hall conductance fluctuations in chaotic dirac quantum dots},\ }\href {https://doi.org/10.1103/PhysRevB.93.115120} {\bibfield  {journal} {\bibinfo  {journal} {Phys. Rev. B}\ }\textbf {\bibinfo {volume} {93}},\ \bibinfo {pages} {115120} (\bibinfo {year} {2016})}\BibitemShut {NoStop}%
\bibitem [{\citenamefont {Kechedzhi}\ \emph {et~al.}(2008)\citenamefont {Kechedzhi}, \citenamefont {Kashuba},\ and\ \citenamefont {Fal'ko}}]{kechedzhi2008quantum}%
  \BibitemOpen
  \bibfield  {author} {\bibinfo {author} {\bibfnamefont {K.}~\bibnamefont {Kechedzhi}}, \bibinfo {author} {\bibfnamefont {O.}~\bibnamefont {Kashuba}},\ and\ \bibinfo {author} {\bibfnamefont {V.~I.}\ \bibnamefont {Fal'ko}},\ }\bibfield  {title} {\bibinfo {title} {Quantum kinetic equation and universal conductance fluctuations in graphene},\ }\href {https://doi.org/10.1103/PhysRevB.77.193403} {\bibfield  {journal} {\bibinfo  {journal} {Phys. Rev. B}\ }\textbf {\bibinfo {volume} {77}},\ \bibinfo {pages} {193403} (\bibinfo {year} {2008})}\BibitemShut {NoStop}%
\bibitem [{\citenamefont {Kharitonov}\ and\ \citenamefont {Efetov}(2008)}]{kharitonov2008universal}%
  \BibitemOpen
  \bibfield  {author} {\bibinfo {author} {\bibfnamefont {M.~Y.}\ \bibnamefont {Kharitonov}}\ and\ \bibinfo {author} {\bibfnamefont {K.~B.}\ \bibnamefont {Efetov}},\ }\bibfield  {title} {\bibinfo {title} {Universal conductance fluctuations in graphene},\ }\href {https://doi.org/10.1103/PhysRevB.78.033404} {\bibfield  {journal} {\bibinfo  {journal} {Phys. Rev. B}\ }\textbf {\bibinfo {volume} {78}},\ \bibinfo {pages} {033404} (\bibinfo {year} {2008})}\BibitemShut {NoStop}%
\bibitem [{\citenamefont {Qiao}\ \emph {et~al.}(2008)\citenamefont {Qiao}, \citenamefont {Wang}, \citenamefont {Wei},\ and\ \citenamefont {Guo}}]{qiao2008universal}%
  \BibitemOpen
  \bibfield  {author} {\bibinfo {author} {\bibfnamefont {Z.}~\bibnamefont {Qiao}}, \bibinfo {author} {\bibfnamefont {J.}~\bibnamefont {Wang}}, \bibinfo {author} {\bibfnamefont {Y.}~\bibnamefont {Wei}},\ and\ \bibinfo {author} {\bibfnamefont {H.}~\bibnamefont {Guo}},\ }\bibfield  {title} {\bibinfo {title} {Universal quantized spin-hall conductance fluctuation in graphene},\ }\href {https://doi.org/10.1103/PhysRevLett.101.016804} {\bibfield  {journal} {\bibinfo  {journal} {Phys. Rev. Lett.}\ }\textbf {\bibinfo {volume} {101}},\ \bibinfo {pages} {016804} (\bibinfo {year} {2008})}\BibitemShut {NoStop}%
\bibitem [{\citenamefont {Rycerz}\ \emph {et~al.}(2007)\citenamefont {Rycerz}, \citenamefont {Tworzydło},\ and\ \citenamefont {Beenakker}}]{rycerz2007anomalously}%
  \BibitemOpen
  \bibfield  {author} {\bibinfo {author} {\bibfnamefont {A.}~\bibnamefont {Rycerz}}, \bibinfo {author} {\bibfnamefont {J.}~\bibnamefont {Tworzydło}},\ and\ \bibinfo {author} {\bibfnamefont {C.~W.~J.}\ \bibnamefont {Beenakker}},\ }\bibfield  {title} {\bibinfo {title} {Anomalously large conductance fluctuations in weakly disordered graphene},\ }\href {https://doi.org/10.1209/0295-5075/79/57003} {\bibfield  {journal} {\bibinfo  {journal} {Europhysics Letters}\ }\textbf {\bibinfo {volume} {79}},\ \bibinfo {pages} {57003} (\bibinfo {year} {2007})}\BibitemShut {NoStop}%
\bibitem [{\citenamefont {Rossi}\ \emph {et~al.}(2012)\citenamefont {Rossi}, \citenamefont {Bardarson}, \citenamefont {Fuhrer},\ and\ \citenamefont {Das~Sarma}}]{rossi2012universal}%
  \BibitemOpen
  \bibfield  {author} {\bibinfo {author} {\bibfnamefont {E.}~\bibnamefont {Rossi}}, \bibinfo {author} {\bibfnamefont {J.~H.}\ \bibnamefont {Bardarson}}, \bibinfo {author} {\bibfnamefont {M.~S.}\ \bibnamefont {Fuhrer}},\ and\ \bibinfo {author} {\bibfnamefont {S.}~\bibnamefont {Das~Sarma}},\ }\bibfield  {title} {\bibinfo {title} {Universal conductance fluctuations in dirac materials in the presence of long-range disorder},\ }\href {https://doi.org/10.1103/PhysRevLett.109.096801} {\bibfield  {journal} {\bibinfo  {journal} {Phys. Rev. Lett.}\ }\textbf {\bibinfo {volume} {109}},\ \bibinfo {pages} {096801} (\bibinfo {year} {2012})}\BibitemShut {NoStop}%
\bibitem [{\citenamefont {Hsu}\ \emph {et~al.}(2018)\citenamefont {Hsu}, \citenamefont {Kleftogiannis}, \citenamefont {Guo},\ and\ \citenamefont {Gopar}}]{hsu2018conductance}%
  \BibitemOpen
  \bibfield  {author} {\bibinfo {author} {\bibfnamefont {H.-C.}\ \bibnamefont {Hsu}}, \bibinfo {author} {\bibfnamefont {I.}~\bibnamefont {Kleftogiannis}}, \bibinfo {author} {\bibfnamefont {G.-Y.}\ \bibnamefont {Guo}},\ and\ \bibinfo {author} {\bibfnamefont {V.~A.}\ \bibnamefont {Gopar}},\ }\bibfield  {title} {\bibinfo {title} {Conductance fluctuations in disordered 2d topological insulator wires: From quantum spin-hall to ordinary phases},\ }\href {https://doi.org/10.7566/JPSJ.87.034701} {\bibfield  {journal} {\bibinfo  {journal} {Journal of the Physical Society of Japan}\ }\textbf {\bibinfo {volume} {87}},\ \bibinfo {pages} {034701} (\bibinfo {year} {2018})}\BibitemShut {NoStop}%
\bibitem [{\citenamefont {Aslani}\ \emph {et~al.}(2019)\citenamefont {Aslani}, \citenamefont {Sisakht}, \citenamefont {Fazileh}, \citenamefont {Ghorbanfekr-Kalashami},\ and\ \citenamefont {Peeters}}]{aslani2019conductance}%
  \BibitemOpen
  \bibfield  {author} {\bibinfo {author} {\bibfnamefont {Z.}~\bibnamefont {Aslani}}, \bibinfo {author} {\bibfnamefont {E.~T.}\ \bibnamefont {Sisakht}}, \bibinfo {author} {\bibfnamefont {F.}~\bibnamefont {Fazileh}}, \bibinfo {author} {\bibfnamefont {H.}~\bibnamefont {Ghorbanfekr-Kalashami}},\ and\ \bibinfo {author} {\bibfnamefont {F.~M.}\ \bibnamefont {Peeters}},\ }\bibfield  {title} {\bibinfo {title} {Conductance fluctuations of monolayer ${\mathrm{gesnh}}_{2}$ in the topological phase using a low-energy effective tight-binding hamiltonian},\ }\href {https://doi.org/10.1103/PhysRevB.99.115421} {\bibfield  {journal} {\bibinfo  {journal} {Phys. Rev. B}\ }\textbf {\bibinfo {volume} {99}},\ \bibinfo {pages} {115421} (\bibinfo {year} {2019})}\BibitemShut {NoStop}%
\bibitem [{\citenamefont {Li}\ and\ \citenamefont {Shi}(2009)}]{li2009dorokhov}%
  \BibitemOpen
  \bibfield  {author} {\bibinfo {author} {\bibfnamefont {D.}~\bibnamefont {Li}}\ and\ \bibinfo {author} {\bibfnamefont {J.}~\bibnamefont {Shi}},\ }\bibfield  {title} {\bibinfo {title} {Dorokhov-mello-pereyra-kumar equation for the edge transport of a quantum spin hall insulator},\ }\href {https://doi.org/10.1103/PhysRevB.79.241303} {\bibfield  {journal} {\bibinfo  {journal} {Phys. Rev. B}\ }\textbf {\bibinfo {volume} {79}},\ \bibinfo {pages} {241303} (\bibinfo {year} {2009})}\BibitemShut {NoStop}%
\bibitem [{\citenamefont {S\'a}\ \emph {et~al.}(2020)\citenamefont {S\'a}, \citenamefont {Barbosa},\ and\ \citenamefont {Ramos}}]{Louis2020}%
  \BibitemOpen
  \bibfield  {author} {\bibinfo {author} {\bibfnamefont {L.~G. C.~S.}\ \bibnamefont {S\'a}}, \bibinfo {author} {\bibfnamefont {A.~L.~R.}\ \bibnamefont {Barbosa}},\ and\ \bibinfo {author} {\bibfnamefont {J.~G. G.~S.}\ \bibnamefont {Ramos}},\ }\bibfield  {title} {\bibinfo {title} {Conductance peak density in disordered graphene topological insulators},\ }\href {https://doi.org/10.1103/PhysRevB.102.115105} {\bibfield  {journal} {\bibinfo  {journal} {Phys. Rev. B}\ }\textbf {\bibinfo {volume} {102}},\ \bibinfo {pages} {115105} (\bibinfo {year} {2020})}\BibitemShut {NoStop}%
\bibitem [{\citenamefont {Shafiei}\ \emph {et~al.}(2024)\citenamefont {Shafiei}, \citenamefont {Fazileh}, \citenamefont {Peeters},\ and\ \citenamefont {Milo{\v{s}}evi{\'c}}}]{shafiei2024tailoring}%
  \BibitemOpen
  \bibfield  {author} {\bibinfo {author} {\bibfnamefont {M.}~\bibnamefont {Shafiei}}, \bibinfo {author} {\bibfnamefont {F.}~\bibnamefont {Fazileh}}, \bibinfo {author} {\bibfnamefont {F.~M.}\ \bibnamefont {Peeters}},\ and\ \bibinfo {author} {\bibfnamefont {M.~V.}\ \bibnamefont {Milo{\v{s}}evi{\'c}}},\ }\bibfield  {title} {\bibinfo {title} {Tailoring weak and metallic phases in a strong topological insulator by strain and disorder: Conductance fluctuations signatures},\ }\href {https://journals.aps.org/prb/abstract/10.1103/PhysRevB.109.045129} {\bibfield  {journal} {\bibinfo  {journal} {Physical Review B}\ }\textbf {\bibinfo {volume} {109}},\ \bibinfo {pages} {045129} (\bibinfo {year} {2024})}\BibitemShut {NoStop}%
\bibitem [{\citenamefont {Chen}\ \emph {et~al.}(2010)\citenamefont {Chen}, \citenamefont {Bae}, \citenamefont {Chialvo}, \citenamefont {Dirks}, \citenamefont {Bezryadin},\ and\ \citenamefont {Mason}}]{chen2010magnetoresistance}%
  \BibitemOpen
  \bibfield  {author} {\bibinfo {author} {\bibfnamefont {Y.-F.}\ \bibnamefont {Chen}}, \bibinfo {author} {\bibfnamefont {M.-H.}\ \bibnamefont {Bae}}, \bibinfo {author} {\bibfnamefont {C.}~\bibnamefont {Chialvo}}, \bibinfo {author} {\bibfnamefont {T.}~\bibnamefont {Dirks}}, \bibinfo {author} {\bibfnamefont {A.}~\bibnamefont {Bezryadin}},\ and\ \bibinfo {author} {\bibfnamefont {N.}~\bibnamefont {Mason}},\ }\bibfield  {title} {\bibinfo {title} {Magnetoresistance in single-layer graphene: weak localization and universal conductance fluctuation studies},\ }\href {https://doi.org/10.1088/0953-8984/22/20/205301} {\bibfield  {journal} {\bibinfo  {journal} {Journal of Physics: Condensed Matter}\ }\textbf {\bibinfo {volume} {22}},\ \bibinfo {pages} {205301} (\bibinfo {year} {2010})}\BibitemShut {NoStop}%
\bibitem [{\citenamefont {Kandala}\ \emph {et~al.}(2013)\citenamefont {Kandala}, \citenamefont {Richardella}, \citenamefont {Zhang}, \citenamefont {Flanagan},\ and\ \citenamefont {Samarth}}]{kandala2013surface}%
  \BibitemOpen
  \bibfield  {author} {\bibinfo {author} {\bibfnamefont {A.}~\bibnamefont {Kandala}}, \bibinfo {author} {\bibfnamefont {A.}~\bibnamefont {Richardella}}, \bibinfo {author} {\bibfnamefont {D.}~\bibnamefont {Zhang}}, \bibinfo {author} {\bibfnamefont {T.~C.}\ \bibnamefont {Flanagan}},\ and\ \bibinfo {author} {\bibfnamefont {N.}~\bibnamefont {Samarth}},\ }\bibfield  {title} {\bibinfo {title} {Surface-sensitive two-dimensional magneto-fingerprint in mesoscopic bi2se3 channels},\ }\href {https://doi.org/10.1021/nl4012358} {\bibfield  {journal} {\bibinfo  {journal} {Nano Letters}\ }\textbf {\bibinfo {volume} {13}},\ \bibinfo {pages} {2471} (\bibinfo {year} {2013})}\BibitemShut {NoStop}%
\bibitem [{\citenamefont {Li}\ \emph {et~al.}(2014)\citenamefont {Li}, \citenamefont {Meng}, \citenamefont {Pan}, \citenamefont {Chen}, \citenamefont {Hong}, \citenamefont {Li}, \citenamefont {Wang}, \citenamefont {Song},\ and\ \citenamefont {Wang}}]{li2014indications}%
  \BibitemOpen
  \bibfield  {author} {\bibinfo {author} {\bibfnamefont {Z.}~\bibnamefont {Li}}, \bibinfo {author} {\bibfnamefont {Y.}~\bibnamefont {Meng}}, \bibinfo {author} {\bibfnamefont {J.}~\bibnamefont {Pan}}, \bibinfo {author} {\bibfnamefont {T.}~\bibnamefont {Chen}}, \bibinfo {author} {\bibfnamefont {X.}~\bibnamefont {Hong}}, \bibinfo {author} {\bibfnamefont {S.}~\bibnamefont {Li}}, \bibinfo {author} {\bibfnamefont {X.}~\bibnamefont {Wang}}, \bibinfo {author} {\bibfnamefont {F.}~\bibnamefont {Song}},\ and\ \bibinfo {author} {\bibfnamefont {B.}~\bibnamefont {Wang}},\ }\bibfield  {title} {\bibinfo {title} {Indications of topological transport by universal conductance fluctuations in bi2te2se microflakes},\ }\href {https://doi.org/10.7567/APEX.7.065202} {\bibfield  {journal} {\bibinfo  {journal} {Applied Physics Express}\ }\textbf {\bibinfo {volume} {7}},\ \bibinfo {pages} {065202} (\bibinfo {year} {2014})}\BibitemShut {NoStop}%
\bibitem [{\citenamefont {Trivedi}\ \emph {et~al.}(2016)\citenamefont {Trivedi}, \citenamefont {Sonde}, \citenamefont {Movva},\ and\ \citenamefont {Banerjee}}]{trivedi2016weak}%
  \BibitemOpen
  \bibfield  {author} {\bibinfo {author} {\bibfnamefont {T.}~\bibnamefont {Trivedi}}, \bibinfo {author} {\bibfnamefont {S.}~\bibnamefont {Sonde}}, \bibinfo {author} {\bibfnamefont {H.~C.~P.}\ \bibnamefont {Movva}},\ and\ \bibinfo {author} {\bibfnamefont {S.~K.}\ \bibnamefont {Banerjee}},\ }\bibfield  {title} {\bibinfo {title} {Weak antilocalization and universal conductance fluctuations in bismuth telluro-sulfide topological insulators},\ }\href {https://doi.org/10.1063/1.4941265} {\bibfield  {journal} {\bibinfo  {journal} {Journal of Applied Physics}\ }\textbf {\bibinfo {volume} {119}},\ \bibinfo {pages} {055706} (\bibinfo {year} {2016})}\BibitemShut {NoStop}%
\bibitem [{\citenamefont {Islam}\ \emph {et~al.}(2018)\citenamefont {Islam}, \citenamefont {Bhattacharyya}, \citenamefont {Nhalil}, \citenamefont {Elizabeth},\ and\ \citenamefont {Ghosh}}]{Islam2018}%
  \BibitemOpen
  \bibfield  {author} {\bibinfo {author} {\bibfnamefont {S.}~\bibnamefont {Islam}}, \bibinfo {author} {\bibfnamefont {S.}~\bibnamefont {Bhattacharyya}}, \bibinfo {author} {\bibfnamefont {H.}~\bibnamefont {Nhalil}}, \bibinfo {author} {\bibfnamefont {S.}~\bibnamefont {Elizabeth}},\ and\ \bibinfo {author} {\bibfnamefont {A.}~\bibnamefont {Ghosh}},\ }\bibfield  {title} {\bibinfo {title} {Universal conductance fluctuations and direct observation of crossover of symmetry classes in topological insulators},\ }\href {https://doi.org/10.1103/PhysRevB.97.241412} {\bibfield  {journal} {\bibinfo  {journal} {Phys. Rev. B}\ }\textbf {\bibinfo {volume} {97}},\ \bibinfo {pages} {241412} (\bibinfo {year} {2018})}\BibitemShut {NoStop}%
\bibitem [{\citenamefont {Andersen}\ \emph {et~al.}(2023)\citenamefont {Andersen}, \citenamefont {Mikheev}, \citenamefont {Rosen}, \citenamefont {Tai}, \citenamefont {Zhang}, \citenamefont {Wang}, \citenamefont {Kastner},\ and\ \citenamefont {Goldhaber-Gordon}}]{andersen2023universal}%
  \BibitemOpen
  \bibfield  {author} {\bibinfo {author} {\bibfnamefont {M.~P.}\ \bibnamefont {Andersen}}, \bibinfo {author} {\bibfnamefont {E.}~\bibnamefont {Mikheev}}, \bibinfo {author} {\bibfnamefont {I.~T.}\ \bibnamefont {Rosen}}, \bibinfo {author} {\bibfnamefont {L.}~\bibnamefont {Tai}}, \bibinfo {author} {\bibfnamefont {P.}~\bibnamefont {Zhang}}, \bibinfo {author} {\bibfnamefont {K.~L.}\ \bibnamefont {Wang}}, \bibinfo {author} {\bibfnamefont {M.~A.}\ \bibnamefont {Kastner}},\ and\ \bibinfo {author} {\bibfnamefont {D.}~\bibnamefont {Goldhaber-Gordon}},\ }\bibfield  {title} {\bibinfo {title} {Universal conductance fluctuations in a mnbi2te4 thin film},\ }\href {https://doi.org/10.1021/acs.nanolett.3c02932} {\bibfield  {journal} {\bibinfo  {journal} {Nano Letters}\ }\textbf {\bibinfo {volume} {23}},\ \bibinfo {pages} {10802} (\bibinfo {year} {2023})}\BibitemShut {NoStop}%
\bibitem [{\citenamefont {Matsuo}\ \emph {et~al.}(2013)\citenamefont {Matsuo}, \citenamefont {Chida}, \citenamefont {Chiba}, \citenamefont {Ono}, \citenamefont {Slevin}, \citenamefont {Kobayashi}, \citenamefont {Ohtsuki}, \citenamefont {Chang}, \citenamefont {He}, \citenamefont {Ma},\ and\ \citenamefont {Xue}}]{matsuo2013experimental}%
  \BibitemOpen
  \bibfield  {author} {\bibinfo {author} {\bibfnamefont {S.}~\bibnamefont {Matsuo}}, \bibinfo {author} {\bibfnamefont {K.}~\bibnamefont {Chida}}, \bibinfo {author} {\bibfnamefont {D.}~\bibnamefont {Chiba}}, \bibinfo {author} {\bibfnamefont {T.}~\bibnamefont {Ono}}, \bibinfo {author} {\bibfnamefont {K.}~\bibnamefont {Slevin}}, \bibinfo {author} {\bibfnamefont {K.}~\bibnamefont {Kobayashi}}, \bibinfo {author} {\bibfnamefont {T.}~\bibnamefont {Ohtsuki}}, \bibinfo {author} {\bibfnamefont {C.-Z.}\ \bibnamefont {Chang}}, \bibinfo {author} {\bibfnamefont {K.}~\bibnamefont {He}}, \bibinfo {author} {\bibfnamefont {X.-C.}\ \bibnamefont {Ma}},\ and\ \bibinfo {author} {\bibfnamefont {Q.-K.}\ \bibnamefont {Xue}},\ }\bibfield  {title} {\bibinfo {title} {Experimental proof of universal conductance fluctuation in quasi-one-dimensional epitaxial bi${}_{2}$se${}_{3}$ wires},\ }\href {https://doi.org/10.1103/PhysRevB.88.155438} {\bibfield  {journal} {\bibinfo  {journal} {Phys. Rev. B}\ }\textbf {\bibinfo {volume} {88}},\
  \bibinfo {pages} {155438} (\bibinfo {year} {2013})}\BibitemShut {NoStop}%
\bibitem [{\citenamefont {Xiao}\ \emph {et~al.}(2023)\citenamefont {Xiao}, \citenamefont {Islam}, \citenamefont {Yanez}, \citenamefont {Ou}, \citenamefont {Liu}, \citenamefont {Xie}, \citenamefont {Chamorro}, \citenamefont {McQueen},\ and\ \citenamefont {Samarth}}]{Xiao23}%
  \BibitemOpen
  \bibfield  {author} {\bibinfo {author} {\bibfnamefont {R.}~\bibnamefont {Xiao}}, \bibinfo {author} {\bibfnamefont {S.}~\bibnamefont {Islam}}, \bibinfo {author} {\bibfnamefont {W.}~\bibnamefont {Yanez}}, \bibinfo {author} {\bibfnamefont {Y.}~\bibnamefont {Ou}}, \bibinfo {author} {\bibfnamefont {H.}~\bibnamefont {Liu}}, \bibinfo {author} {\bibfnamefont {X.}~\bibnamefont {Xie}}, \bibinfo {author} {\bibfnamefont {J.}~\bibnamefont {Chamorro}}, \bibinfo {author} {\bibfnamefont {T.~M.}\ \bibnamefont {McQueen}},\ and\ \bibinfo {author} {\bibfnamefont {N.}~\bibnamefont {Samarth}},\ }\bibfield  {title} {\bibinfo {title} {Influence of magnetic and electric fields on universal conductance fluctuations in thin films of the dirac semimetal cd3as2},\ }\href {https://doi.org/10.1021/acs.nanolett.3c01174} {\bibfield  {journal} {\bibinfo  {journal} {Nano Letters}\ }\textbf {\bibinfo {volume} {23}},\ \bibinfo {pages} {5634} (\bibinfo {year} {2023})}\BibitemShut {NoStop}%
\bibitem [{\citenamefont {Staley}\ \emph {et~al.}(2008)\citenamefont {Staley}, \citenamefont {Puls},\ and\ \citenamefont {Liu}}]{staley2008suppression}%
  \BibitemOpen
  \bibfield  {author} {\bibinfo {author} {\bibfnamefont {N.~E.}\ \bibnamefont {Staley}}, \bibinfo {author} {\bibfnamefont {C.~P.}\ \bibnamefont {Puls}},\ and\ \bibinfo {author} {\bibfnamefont {Y.}~\bibnamefont {Liu}},\ }\bibfield  {title} {\bibinfo {title} {Suppression of conductance fluctuation in weakly disordered mesoscopic graphene samples near the charge neutral point},\ }\href {https://doi.org/10.1103/PhysRevB.77.155429} {\bibfield  {journal} {\bibinfo  {journal} {Phys. Rev. B}\ }\textbf {\bibinfo {volume} {77}},\ \bibinfo {pages} {155429} (\bibinfo {year} {2008})}\BibitemShut {NoStop}%
\bibitem [{\citenamefont {Hu}(2017)}]{Hu2017a}%
  \BibitemOpen
  \bibfield  {author} {\bibinfo {author} {\bibfnamefont {Y.}~\bibnamefont {Hu}},\ }\emph {\bibinfo {title} {Numerical Study of Universal Conductance Fluctuations in 3D Topological Semimetals}},\ \href {https://drm.lib.pku.edu.cn/pdfindex1.jsp?fid=b504ca11d299d46a0f2b70390284cd94} {\bibinfo {type} {Master's thesis}},\ \bibinfo  {school} {Peking University} (\bibinfo {year} {2017})\BibitemShut {NoStop}%
\bibitem [{\citenamefont {Wang}\ \emph {et~al.}(2019)\citenamefont {Wang}, \citenamefont {Yan},\ and\ \citenamefont {Wang}}]{wang2019non}%
  \BibitemOpen
  \bibfield  {author} {\bibinfo {author} {\bibfnamefont {C.}~\bibnamefont {Wang}}, \bibinfo {author} {\bibfnamefont {P.}~\bibnamefont {Yan}},\ and\ \bibinfo {author} {\bibfnamefont {X.~R.}\ \bibnamefont {Wang}},\ }\bibfield  {title} {\bibinfo {title} {Non-wigner-dyson level statistics and fractal wave function of disordered weyl semimetals},\ }\href {https://doi.org/10.1103/PhysRevB.99.205140} {\bibfield  {journal} {\bibinfo  {journal} {Phys. Rev. B}\ }\textbf {\bibinfo {volume} {99}},\ \bibinfo {pages} {205140} (\bibinfo {year} {2019})}\BibitemShut {NoStop}%
\bibitem [{\citenamefont {Chen}\ \emph {et~al.}(2015)\citenamefont {Chen}, \citenamefont {Song}, \citenamefont {Jiang}, \citenamefont {Sun}, \citenamefont {Wang},\ and\ \citenamefont {Xie}}]{chen2015disorder}%
  \BibitemOpen
  \bibfield  {author} {\bibinfo {author} {\bibfnamefont {C.-Z.}\ \bibnamefont {Chen}}, \bibinfo {author} {\bibfnamefont {J.}~\bibnamefont {Song}}, \bibinfo {author} {\bibfnamefont {H.}~\bibnamefont {Jiang}}, \bibinfo {author} {\bibfnamefont {Q.-f.}\ \bibnamefont {Sun}}, \bibinfo {author} {\bibfnamefont {Z.}~\bibnamefont {Wang}},\ and\ \bibinfo {author} {\bibfnamefont {X.~C.}\ \bibnamefont {Xie}},\ }\bibfield  {title} {\bibinfo {title} {Disorder and metal-insulator transitions in weyl semimetals},\ }\href {https://doi.org/10.1103/PhysRevLett.115.246603} {\bibfield  {journal} {\bibinfo  {journal} {Phys. Rev. Lett.}\ }\textbf {\bibinfo {volume} {115}},\ \bibinfo {pages} {246603} (\bibinfo {year} {2015})}\BibitemShut {NoStop}%
\bibitem [{\citenamefont {Liu}\ \emph {et~al.}(2016{\natexlab{a}})\citenamefont {Liu}, \citenamefont {Akis}, \citenamefont {Ferry}, \citenamefont {Bohra}, \citenamefont {Somphonsane}, \citenamefont {Ramamoorthy},\ and\ \citenamefont {Bird}}]{liu2016conductance}%
  \BibitemOpen
  \bibfield  {author} {\bibinfo {author} {\bibfnamefont {B.}~\bibnamefont {Liu}}, \bibinfo {author} {\bibfnamefont {R.}~\bibnamefont {Akis}}, \bibinfo {author} {\bibfnamefont {D.~K.}\ \bibnamefont {Ferry}}, \bibinfo {author} {\bibfnamefont {G.}~\bibnamefont {Bohra}}, \bibinfo {author} {\bibfnamefont {R.}~\bibnamefont {Somphonsane}}, \bibinfo {author} {\bibfnamefont {H.}~\bibnamefont {Ramamoorthy}},\ and\ \bibinfo {author} {\bibfnamefont {J.~P.}\ \bibnamefont {Bird}},\ }\bibfield  {title} {\bibinfo {title} {Conductance fluctuations in graphene in the presence of long-range disorder},\ }\href {https://doi.org/10.1088/0953-8984/28/13/135302} {\bibfield  {journal} {\bibinfo  {journal} {Journal of Physics: Condensed Matter}\ }\textbf {\bibinfo {volume} {28}},\ \bibinfo {pages} {135302} (\bibinfo {year} {2016}{\natexlab{a}})}\BibitemShut {NoStop}%
\bibitem [{\citenamefont {Alagha}\ \emph {et~al.}(2010)\citenamefont {Alagha}, \citenamefont {Hernández}, \citenamefont {Bl{\"o}mers}, \citenamefont {Stoica}, \citenamefont {Calarco},\ and\ \citenamefont {Schäpers}}]{alagha2010universal}%
  \BibitemOpen
  \bibfield  {author} {\bibinfo {author} {\bibfnamefont {S.}~\bibnamefont {Alagha}}, \bibinfo {author} {\bibfnamefont {S.~E.}\ \bibnamefont {Hernández}}, \bibinfo {author} {\bibfnamefont {C.}~\bibnamefont {Bl{\"o}mers}}, \bibinfo {author} {\bibfnamefont {T.}~\bibnamefont {Stoica}}, \bibinfo {author} {\bibfnamefont {R.}~\bibnamefont {Calarco}},\ and\ \bibinfo {author} {\bibfnamefont {T.}~\bibnamefont {Schäpers}},\ }\bibfield  {title} {\bibinfo {title} {Universal conductance fluctuations and localization effects in inn nanowires connected in parallel},\ }\href {https://doi.org/10.1063/1.3516216} {\bibfield  {journal} {\bibinfo  {journal} {Journal of Applied Physics}\ }\textbf {\bibinfo {volume} {108}},\ \bibinfo {pages} {113704} (\bibinfo {year} {2010})}\BibitemShut {NoStop}%
\bibitem [{\citenamefont {Cao}\ \emph {et~al.}(2018{\natexlab{a}})\citenamefont {Cao}, \citenamefont {Fatemi}, \citenamefont {Fang}, \citenamefont {Watanabe}, \citenamefont {Taniguchi}, \citenamefont {Kaxiras},\ and\ \citenamefont {Jarillo-Herrero}}]{cao2018unconventional}%
  \BibitemOpen
  \bibfield  {author} {\bibinfo {author} {\bibfnamefont {Y.}~\bibnamefont {Cao}}, \bibinfo {author} {\bibfnamefont {V.}~\bibnamefont {Fatemi}}, \bibinfo {author} {\bibfnamefont {S.}~\bibnamefont {Fang}}, \bibinfo {author} {\bibfnamefont {K.}~\bibnamefont {Watanabe}}, \bibinfo {author} {\bibfnamefont {T.}~\bibnamefont {Taniguchi}}, \bibinfo {author} {\bibfnamefont {E.}~\bibnamefont {Kaxiras}},\ and\ \bibinfo {author} {\bibfnamefont {P.}~\bibnamefont {Jarillo-Herrero}},\ }\bibfield  {title} {\bibinfo {title} {Unconventional superconductivity in magic-angle graphene superlattices},\ }\href {https://www.nature.com/articles/nature26160} {\bibfield  {journal} {\bibinfo  {journal} {Nature}\ }\textbf {\bibinfo {volume} {556}},\ \bibinfo {pages} {43} (\bibinfo {year} {2018}{\natexlab{a}})}\BibitemShut {NoStop}%
\bibitem [{\citenamefont {Cao}\ \emph {et~al.}(2018{\natexlab{b}})\citenamefont {Cao}, \citenamefont {Fatemi}, \citenamefont {Demir}, \citenamefont {Fang}, \citenamefont {Tomarken}, \citenamefont {Luo}, \citenamefont {Sanchez-Yamagishi}, \citenamefont {Watanabe}, \citenamefont {Taniguchi}, \citenamefont {Kaxiras} \emph {et~al.}}]{cao2018correlated}%
  \BibitemOpen
  \bibfield  {author} {\bibinfo {author} {\bibfnamefont {Y.}~\bibnamefont {Cao}}, \bibinfo {author} {\bibfnamefont {V.}~\bibnamefont {Fatemi}}, \bibinfo {author} {\bibfnamefont {A.}~\bibnamefont {Demir}}, \bibinfo {author} {\bibfnamefont {S.}~\bibnamefont {Fang}}, \bibinfo {author} {\bibfnamefont {S.~L.}\ \bibnamefont {Tomarken}}, \bibinfo {author} {\bibfnamefont {J.~Y.}\ \bibnamefont {Luo}}, \bibinfo {author} {\bibfnamefont {J.~D.}\ \bibnamefont {Sanchez-Yamagishi}}, \bibinfo {author} {\bibfnamefont {K.}~\bibnamefont {Watanabe}}, \bibinfo {author} {\bibfnamefont {T.}~\bibnamefont {Taniguchi}}, \bibinfo {author} {\bibfnamefont {E.}~\bibnamefont {Kaxiras}}, \emph {et~al.},\ }\bibfield  {title} {\bibinfo {title} {Correlated insulator behaviour at half-filling in magic-angle graphene superlattices},\ }\href {https://www.nature.com/articles/nature26154} {\bibfield  {journal} {\bibinfo  {journal} {Nature}\ }\textbf {\bibinfo {volume} {556}},\ \bibinfo {pages} {80} (\bibinfo {year}
  {2018}{\natexlab{b}})}\BibitemShut {NoStop}%
\bibitem [{\citenamefont {Edelberg}\ \emph {et~al.}(2020)\citenamefont {Edelberg}, \citenamefont {Kumar}, \citenamefont {Shenoy}, \citenamefont {Ochoa},\ and\ \citenamefont {Pasupathy}}]{edelberg2020tunable}%
  \BibitemOpen
  \bibfield  {author} {\bibinfo {author} {\bibfnamefont {D.}~\bibnamefont {Edelberg}}, \bibinfo {author} {\bibfnamefont {H.}~\bibnamefont {Kumar}}, \bibinfo {author} {\bibfnamefont {V.}~\bibnamefont {Shenoy}}, \bibinfo {author} {\bibfnamefont {H.}~\bibnamefont {Ochoa}},\ and\ \bibinfo {author} {\bibfnamefont {A.~N.}\ \bibnamefont {Pasupathy}},\ }\bibfield  {title} {\bibinfo {title} {Tunable strain soliton networks confine electrons in van der waals materials},\ }\href {https://www.nature.com/articles/s41567-020-0953-2} {\bibfield  {journal} {\bibinfo  {journal} {Nature Physics}\ }\textbf {\bibinfo {volume} {16}},\ \bibinfo {pages} {1097} (\bibinfo {year} {2020})}\BibitemShut {NoStop}%
\bibitem [{\citenamefont {Bai}\ \emph {et~al.}(2020)\citenamefont {Bai}, \citenamefont {Zhou}, \citenamefont {Wang}, \citenamefont {Wu}, \citenamefont {McGilly}, \citenamefont {Halbertal}, \citenamefont {Lo}, \citenamefont {Liu}, \citenamefont {Ardelean}, \citenamefont {Rivera} \emph {et~al.}}]{bai2020excitons}%
  \BibitemOpen
  \bibfield  {author} {\bibinfo {author} {\bibfnamefont {Y.}~\bibnamefont {Bai}}, \bibinfo {author} {\bibfnamefont {L.}~\bibnamefont {Zhou}}, \bibinfo {author} {\bibfnamefont {J.}~\bibnamefont {Wang}}, \bibinfo {author} {\bibfnamefont {W.}~\bibnamefont {Wu}}, \bibinfo {author} {\bibfnamefont {L.~J.}\ \bibnamefont {McGilly}}, \bibinfo {author} {\bibfnamefont {D.}~\bibnamefont {Halbertal}}, \bibinfo {author} {\bibfnamefont {C.~F.~B.}\ \bibnamefont {Lo}}, \bibinfo {author} {\bibfnamefont {F.}~\bibnamefont {Liu}}, \bibinfo {author} {\bibfnamefont {J.}~\bibnamefont {Ardelean}}, \bibinfo {author} {\bibfnamefont {P.}~\bibnamefont {Rivera}}, \emph {et~al.},\ }\bibfield  {title} {\bibinfo {title} {Excitons in strain-induced one-dimensional moir{\'e} potentials at transition metal dichalcogenide heterojunctions},\ }\href {https://www.nature.com/articles/s41563-020-0730-8} {\bibfield  {journal} {\bibinfo  {journal} {Nature Materials}\ }\textbf {\bibinfo {volume} {19}},\ \bibinfo {pages} {1068} (\bibinfo {year}
  {2020})}\BibitemShut {NoStop}%
\bibitem [{\citenamefont {Kennes}\ \emph {et~al.}(2021)\citenamefont {Kennes}, \citenamefont {Claassen}, \citenamefont {Xian}, \citenamefont {Georges}, \citenamefont {Millis}, \citenamefont {Hone}, \citenamefont {Dean}, \citenamefont {Basov}, \citenamefont {Pasupathy},\ and\ \citenamefont {Rubio}}]{kennes2021moire}%
  \BibitemOpen
  \bibfield  {author} {\bibinfo {author} {\bibfnamefont {D.~M.}\ \bibnamefont {Kennes}}, \bibinfo {author} {\bibfnamefont {M.}~\bibnamefont {Claassen}}, \bibinfo {author} {\bibfnamefont {L.}~\bibnamefont {Xian}}, \bibinfo {author} {\bibfnamefont {A.}~\bibnamefont {Georges}}, \bibinfo {author} {\bibfnamefont {A.~J.}\ \bibnamefont {Millis}}, \bibinfo {author} {\bibfnamefont {J.}~\bibnamefont {Hone}}, \bibinfo {author} {\bibfnamefont {C.~R.}\ \bibnamefont {Dean}}, \bibinfo {author} {\bibfnamefont {D.}~\bibnamefont {Basov}}, \bibinfo {author} {\bibfnamefont {A.~N.}\ \bibnamefont {Pasupathy}},\ and\ \bibinfo {author} {\bibfnamefont {A.}~\bibnamefont {Rubio}},\ }\bibfield  {title} {\bibinfo {title} {Moir{\'e} heterostructures as a condensed-matter quantum simulator},\ }\href {https://www.nature.com/articles/s41567-020-01154-3} {\bibfield  {journal} {\bibinfo  {journal} {Nature Physics}\ }\textbf {\bibinfo {volume} {17}},\ \bibinfo {pages} {155} (\bibinfo {year} {2021})}\BibitemShut {NoStop}%
\bibitem [{\citenamefont {Hou}\ \emph {et~al.}(2024)\citenamefont {Hou}, \citenamefont {Hu},\ and\ \citenamefont {Yang}}]{Hou24}%
  \BibitemOpen
  \bibfield  {author} {\bibinfo {author} {\bibfnamefont {Z.}~\bibnamefont {Hou}}, \bibinfo {author} {\bibfnamefont {Y.-Y.}\ \bibnamefont {Hu}},\ and\ \bibinfo {author} {\bibfnamefont {G.-W.}\ \bibnamefont {Yang}},\ }\bibfield  {title} {\bibinfo {title} {Moir\'e pattern assisted commensuration resonance in disordered twisted bilayer graphene},\ }\href {https://doi.org/10.1103/PhysRevB.109.085412} {\bibfield  {journal} {\bibinfo  {journal} {Phys. Rev. B}\ }\textbf {\bibinfo {volume} {109}},\ \bibinfo {pages} {085412} (\bibinfo {year} {2024})}\BibitemShut {NoStop}%
\bibitem [{\citenamefont {Meir}\ and\ \citenamefont {Wingreen}(1992)}]{Meir92Landauer}%
  \BibitemOpen
  \bibfield  {author} {\bibinfo {author} {\bibfnamefont {Y.}~\bibnamefont {Meir}}\ and\ \bibinfo {author} {\bibfnamefont {N.~S.}\ \bibnamefont {Wingreen}},\ }\bibfield  {title} {\bibinfo {title} {Landauer formula for the current through an interacting electron region},\ }\href {https://doi.org/10.1103/PhysRevLett.68.2512} {\bibfield  {journal} {\bibinfo  {journal} {Phys. Rev. Lett.}\ }\textbf {\bibinfo {volume} {68}},\ \bibinfo {pages} {2512} (\bibinfo {year} {1992})}\BibitemShut {NoStop}%
\bibitem [{\citenamefont {Datta}(1997)}]{Dat}%
  \BibitemOpen
  \bibfield  {author} {\bibinfo {author} {\bibfnamefont {S.}~\bibnamefont {Datta}},\ }\href@noop {} {\emph {\bibinfo {title} {Electronic Transport in Mesoscopic Systems}}}\ (\bibinfo  {publisher} {Cambridge University Press, Cambridge},\ \bibinfo {year} {1997})\BibitemShut {NoStop}%
\bibitem [{\citenamefont {Landauer}(1970)}]{Landauer}%
  \BibitemOpen
  \bibfield  {author} {\bibinfo {author} {\bibfnamefont {R.}~\bibnamefont {Landauer}},\ }\bibfield  {title} {\bibinfo {title} {Electrical resistance of disordered one-dimensional lattices},\ }\href {https://doi.org/10.1080/14786437008238472} {\bibfield  {journal} {\bibinfo  {journal} {The Philosophical Magazine: A Journal of Theoretical Experimental and Applied Physics}\ }\textbf {\bibinfo {volume} {21}},\ \bibinfo {pages} {863} (\bibinfo {year} {1970})}\BibitemShut {NoStop}%
\bibitem [{\citenamefont {Fisher}\ and\ \citenamefont {Lee}(1981)}]{Fisher}%
  \BibitemOpen
  \bibfield  {author} {\bibinfo {author} {\bibfnamefont {D.~S.}\ \bibnamefont {Fisher}}\ and\ \bibinfo {author} {\bibfnamefont {P.~A.}\ \bibnamefont {Lee}},\ }\bibfield  {title} {\bibinfo {title} {Relation between conductivity and transmission matrix},\ }\href {https://doi.org/10.1103/PhysRevB.23.6851} {\bibfield  {journal} {\bibinfo  {journal} {Phys. Rev. B}\ }\textbf {\bibinfo {volume} {23}},\ \bibinfo {pages} {6851} (\bibinfo {year} {1981})}\BibitemShut {NoStop}%
\bibitem [{\citenamefont {B\"uttiker}(1988)}]{Butti}%
  \BibitemOpen
  \bibfield  {author} {\bibinfo {author} {\bibfnamefont {M.}~\bibnamefont {B\"uttiker}},\ }\bibfield  {title} {\bibinfo {title} {Absence of backscattering in the quantum hall effect in multiprobe conductors},\ }\href {https://doi.org/10.1103/PhysRevB.38.9375} {\bibfield  {journal} {\bibinfo  {journal} {Phys. Rev. B}\ }\textbf {\bibinfo {volume} {38}},\ \bibinfo {pages} {9375} (\bibinfo {year} {1988})}\BibitemShut {NoStop}%
\bibitem [{\citenamefont {Zhang}\ and\ \citenamefont {Liu}(2019)}]{zhang2019electronic}%
  \BibitemOpen
  \bibfield  {author} {\bibinfo {author} {\bibfnamefont {X.~W.}\ \bibnamefont {Zhang}}\ and\ \bibinfo {author} {\bibfnamefont {Y.~L.}\ \bibnamefont {Liu}},\ }\bibfield  {title} {\bibinfo {title} {Electronic transport and spatial current patterns of 2d electronic system: A recursive green’s function method study},\ }\href {https://doi.org/10.1063/1.5130534} {\bibfield  {journal} {\bibinfo  {journal} {AIP Advances}\ }\textbf {\bibinfo {volume} {9}},\ \bibinfo {pages} {115209} (\bibinfo {year} {2019})}\BibitemShut {NoStop}%
\bibitem [{\citenamefont {Mackinnon}(1985)}]{Mac}%
  \BibitemOpen
  \bibfield  {author} {\bibinfo {author} {\bibfnamefont {A.}~\bibnamefont {Mackinnon}},\ }\bibfield  {title} {\bibinfo {title} {The calculation of transport properties and density of states of disordered solids},\ }\href {https://doi.org/10.1007/BF01328846} {\bibfield  {journal} {\bibinfo  {journal} {Zeitschrift für Physik B Condensed Matter}\ }\textbf {\bibinfo {volume} {59}},\ \bibinfo {pages} {385} (\bibinfo {year} {1985})}\BibitemShut {NoStop}%
\bibitem [{\citenamefont {Metalidis}\ and\ \citenamefont {Bruno}(2005)}]{Met}%
  \BibitemOpen
  \bibfield  {author} {\bibinfo {author} {\bibfnamefont {G.}~\bibnamefont {Metalidis}}\ and\ \bibinfo {author} {\bibfnamefont {P.}~\bibnamefont {Bruno}},\ }\bibfield  {title} {\bibinfo {title} {Green's function technique for studying electron flow in two-dimensional mesoscopic samples},\ }\href {https://doi.org/10.1103/PhysRevB.72.235304} {\bibfield  {journal} {\bibinfo  {journal} {Phys. Rev. B}\ }\textbf {\bibinfo {volume} {72}},\ \bibinfo {pages} {235304} (\bibinfo {year} {2005})}\BibitemShut {NoStop}%
\bibitem [{\citenamefont {Abrikosov}\ \emph {et~al.}(2012)\citenamefont {Abrikosov}, \citenamefont {Gorkov}, \citenamefont {Dzyaloshinski},\ and\ \citenamefont {Silverman}}]{abrikosov2012methods}%
  \BibitemOpen
  \bibfield  {author} {\bibinfo {author} {\bibfnamefont {A.}~\bibnamefont {Abrikosov}}, \bibinfo {author} {\bibfnamefont {L.}~\bibnamefont {Gorkov}}, \bibinfo {author} {\bibfnamefont {I.}~\bibnamefont {Dzyaloshinski}},\ and\ \bibinfo {author} {\bibfnamefont {R.}~\bibnamefont {Silverman}},\ }\href {https://books.google.com/books?id=JYTCAgAAQBAJ} {\emph {\bibinfo {title} {Methods of Quantum Field Theory in Statistical Physics}}},\ Dover Books on Physics\ (\bibinfo  {publisher} {Dover Publications},\ \bibinfo {year} {2012})\BibitemShut {NoStop}%
\bibitem [{\citenamefont {Rudenko}\ and\ \citenamefont {Katsnelson}(2024)}]{Rudenko2024}%
  \BibitemOpen
  \bibfield  {author} {\bibinfo {author} {\bibfnamefont {A.~N.}\ \bibnamefont {Rudenko}}\ and\ \bibinfo {author} {\bibfnamefont {M.~I.}\ \bibnamefont {Katsnelson}},\ }\bibfield  {title} {\bibinfo {title} {Anisotropic effects in two-dimensional materials},\ }\href {https://doi.org/10.1088/2053-1583/ad64e1} {\bibfield  {journal} {\bibinfo  {journal} {2D Materials}\ }\textbf {\bibinfo {volume} {11}},\ \bibinfo {pages} {042002} (\bibinfo {year} {2024})}\BibitemShut {NoStop}%
\bibitem [{\citenamefont {Yuan}\ \emph {et~al.}(2015)\citenamefont {Yuan}, \citenamefont {Rudenko},\ and\ \citenamefont {Katsnelson}}]{Yuan15}%
  \BibitemOpen
  \bibfield  {author} {\bibinfo {author} {\bibfnamefont {S.}~\bibnamefont {Yuan}}, \bibinfo {author} {\bibfnamefont {A.~N.}\ \bibnamefont {Rudenko}},\ and\ \bibinfo {author} {\bibfnamefont {M.~I.}\ \bibnamefont {Katsnelson}},\ }\bibfield  {title} {\bibinfo {title} {Transport and optical properties of single- and bilayer black phosphorus with defects},\ }\href {https://doi.org/10.1103/PhysRevB.91.115436} {\bibfield  {journal} {\bibinfo  {journal} {Phys. Rev. B}\ }\textbf {\bibinfo {volume} {91}},\ \bibinfo {pages} {115436} (\bibinfo {year} {2015})}\BibitemShut {NoStop}%
\bibitem [{\citenamefont {Ashcroft}\ and\ \citenamefont {Mermin}(1976)}]{ashcroft1976solid}%
  \BibitemOpen
  \bibfield  {author} {\bibinfo {author} {\bibfnamefont {N.}~\bibnamefont {Ashcroft}}\ and\ \bibinfo {author} {\bibfnamefont {N.}~\bibnamefont {Mermin}},\ }\href {https://books.google.com/books?id=oXIfAQAAMAAJ} {\emph {\bibinfo {title} {Solid State Physics}}},\ HRW international editions\ (\bibinfo  {publisher} {Holt, Rinehart and Winston},\ \bibinfo {year} {1976})\BibitemShut {NoStop}%
\bibitem [{\citenamefont {Giordano}(1988)}]{giordano1988conductance}%
  \BibitemOpen
  \bibfield  {author} {\bibinfo {author} {\bibfnamefont {N.}~\bibnamefont {Giordano}},\ }\bibfield  {title} {\bibinfo {title} {Conductance fluctuations in disordered systems: Dependence on the degree of disorder},\ }\href {https://doi.org/10.1103/PhysRevB.38.4746} {\bibfield  {journal} {\bibinfo  {journal} {Phys. Rev. B}\ }\textbf {\bibinfo {volume} {38}},\ \bibinfo {pages} {4746} (\bibinfo {year} {1988})}\BibitemShut {NoStop}%
\bibitem [{\citenamefont {Hu}\ \emph {et~al.}(2021)\citenamefont {Hu}, \citenamefont {Ge}, \citenamefont {Zhang},\ and\ \citenamefont {Jain}}]{Hu2021b}%
  \BibitemOpen
  \bibfield  {author} {\bibinfo {author} {\bibfnamefont {Y.}~\bibnamefont {Hu}}, \bibinfo {author} {\bibfnamefont {Y.}~\bibnamefont {Ge}}, \bibinfo {author} {\bibfnamefont {J.-X.}\ \bibnamefont {Zhang}},\ and\ \bibinfo {author} {\bibfnamefont {J.~K.}\ \bibnamefont {Jain}},\ }\bibfield  {title} {\bibinfo {title} {Crystalline solutions of the kohn-sham equations in the fractional quantum hall regime},\ }\href {https://doi.org/10.1103/PhysRevB.104.035122} {\bibfield  {journal} {\bibinfo  {journal} {Phys. Rev. B}\ }\textbf {\bibinfo {volume} {104}},\ \bibinfo {pages} {035122} (\bibinfo {year} {2021})}\BibitemShut {NoStop}%
\bibitem [{\citenamefont {Qiao}\ \emph {et~al.}(2010)\citenamefont {Qiao}, \citenamefont {Xing},\ and\ \citenamefont {Wang}}]{qiao2010universal}%
  \BibitemOpen
  \bibfield  {author} {\bibinfo {author} {\bibfnamefont {Z.}~\bibnamefont {Qiao}}, \bibinfo {author} {\bibfnamefont {Y.}~\bibnamefont {Xing}},\ and\ \bibinfo {author} {\bibfnamefont {J.}~\bibnamefont {Wang}},\ }\bibfield  {title} {\bibinfo {title} {Universal conductance fluctuation of mesoscopic systems in the metal-insulator crossover regime},\ }\href {https://doi.org/10.1103/PhysRevB.81.085114} {\bibfield  {journal} {\bibinfo  {journal} {Phys. Rev. B}\ }\textbf {\bibinfo {volume} {81}},\ \bibinfo {pages} {085114} (\bibinfo {year} {2010})}\BibitemShut {NoStop}%
\bibitem [{\citenamefont {Chen}\ \emph {et~al.}(2021)\citenamefont {Chen}, \citenamefont {Wang}, \citenamefont {Liu}, \citenamefont {Lu},\ and\ \citenamefont {Xie}}]{chen2021quantum}%
  \BibitemOpen
  \bibfield  {author} {\bibinfo {author} {\bibfnamefont {R.}~\bibnamefont {Chen}}, \bibinfo {author} {\bibfnamefont {C.~M.}\ \bibnamefont {Wang}}, \bibinfo {author} {\bibfnamefont {T.}~\bibnamefont {Liu}}, \bibinfo {author} {\bibfnamefont {H.-Z.}\ \bibnamefont {Lu}},\ and\ \bibinfo {author} {\bibfnamefont {X.~C.}\ \bibnamefont {Xie}},\ }\bibfield  {title} {\bibinfo {title} {Quantum hall effect originated from helical edge states in ${\mathrm{cd}}_{3}{\mathrm{as}}_{2}$},\ }\href {https://doi.org/10.1103/PhysRevResearch.3.033227} {\bibfield  {journal} {\bibinfo  {journal} {Phys. Rev. Res.}\ }\textbf {\bibinfo {volume} {3}},\ \bibinfo {pages} {033227} (\bibinfo {year} {2021})}\BibitemShut {NoStop}%
\bibitem [{\citenamefont {Wang}\ \emph {et~al.}(2013)\citenamefont {Wang}, \citenamefont {Weng}, \citenamefont {Wu}, \citenamefont {Dai},\ and\ \citenamefont {Fang}}]{wang13three}%
  \BibitemOpen
  \bibfield  {author} {\bibinfo {author} {\bibfnamefont {Z.}~\bibnamefont {Wang}}, \bibinfo {author} {\bibfnamefont {H.}~\bibnamefont {Weng}}, \bibinfo {author} {\bibfnamefont {Q.}~\bibnamefont {Wu}}, \bibinfo {author} {\bibfnamefont {X.}~\bibnamefont {Dai}},\ and\ \bibinfo {author} {\bibfnamefont {Z.}~\bibnamefont {Fang}},\ }\bibfield  {title} {\bibinfo {title} {Three-dimensional dirac semimetal and quantum transport in cd${}_{3}$as${}_{2}$},\ }\href {https://doi.org/10.1103/PhysRevB.88.125427} {\bibfield  {journal} {\bibinfo  {journal} {Phys. Rev. B}\ }\textbf {\bibinfo {volume} {88}},\ \bibinfo {pages} {125427} (\bibinfo {year} {2013})}\BibitemShut {NoStop}%
\bibitem [{\citenamefont {Cano}\ \emph {et~al.}(2017)\citenamefont {Cano}, \citenamefont {Bradlyn}, \citenamefont {Wang}, \citenamefont {Hirschberger}, \citenamefont {Ong},\ and\ \citenamefont {Bernevig}}]{cano2017chiral}%
  \BibitemOpen
  \bibfield  {author} {\bibinfo {author} {\bibfnamefont {J.}~\bibnamefont {Cano}}, \bibinfo {author} {\bibfnamefont {B.}~\bibnamefont {Bradlyn}}, \bibinfo {author} {\bibfnamefont {Z.}~\bibnamefont {Wang}}, \bibinfo {author} {\bibfnamefont {M.}~\bibnamefont {Hirschberger}}, \bibinfo {author} {\bibfnamefont {N.~P.}\ \bibnamefont {Ong}},\ and\ \bibinfo {author} {\bibfnamefont {B.~A.}\ \bibnamefont {Bernevig}},\ }\bibfield  {title} {\bibinfo {title} {Chiral anomaly factory: Creating weyl fermions with a magnetic field},\ }\href {https://doi.org/10.1103/PhysRevB.95.161306} {\bibfield  {journal} {\bibinfo  {journal} {Phys. Rev. B}\ }\textbf {\bibinfo {volume} {95}},\ \bibinfo {pages} {161306} (\bibinfo {year} {2017})}\BibitemShut {NoStop}%
\bibitem [{\citenamefont {Pixley}\ \emph {et~al.}(2015)\citenamefont {Pixley}, \citenamefont {Goswami},\ and\ \citenamefont {Das~Sarma}}]{pixley2015anderson}%
  \BibitemOpen
  \bibfield  {author} {\bibinfo {author} {\bibfnamefont {J.~H.}\ \bibnamefont {Pixley}}, \bibinfo {author} {\bibfnamefont {P.}~\bibnamefont {Goswami}},\ and\ \bibinfo {author} {\bibfnamefont {S.}~\bibnamefont {Das~Sarma}},\ }\bibfield  {title} {\bibinfo {title} {Anderson localization and the quantum phase diagram of three dimensional disordered dirac semimetals},\ }\href {https://doi.org/10.1103/PhysRevLett.115.076601} {\bibfield  {journal} {\bibinfo  {journal} {Phys. Rev. Lett.}\ }\textbf {\bibinfo {volume} {115}},\ \bibinfo {pages} {076601} (\bibinfo {year} {2015})}\BibitemShut {NoStop}%
\bibitem [{\citenamefont {Pixley}\ \emph {et~al.}(2016)\citenamefont {Pixley}, \citenamefont {Huse},\ and\ \citenamefont {Das~Sarma}}]{pixley2016rare}%
  \BibitemOpen
  \bibfield  {author} {\bibinfo {author} {\bibfnamefont {J.~H.}\ \bibnamefont {Pixley}}, \bibinfo {author} {\bibfnamefont {D.~A.}\ \bibnamefont {Huse}},\ and\ \bibinfo {author} {\bibfnamefont {S.}~\bibnamefont {Das~Sarma}},\ }\bibfield  {title} {\bibinfo {title} {Rare-region-induced avoided quantum criticality in disordered three-dimensional dirac and weyl semimetals},\ }\href {https://doi.org/10.1103/PhysRevX.6.021042} {\bibfield  {journal} {\bibinfo  {journal} {Phys. Rev. X}\ }\textbf {\bibinfo {volume} {6}},\ \bibinfo {pages} {021042} (\bibinfo {year} {2016})}\BibitemShut {NoStop}%
\bibitem [{\citenamefont {Liu}\ \emph {et~al.}(2016{\natexlab{b}})\citenamefont {Liu}, \citenamefont {Ohtsuki},\ and\ \citenamefont {Shindou}}]{Liu2016Effect}%
  \BibitemOpen
  \bibfield  {author} {\bibinfo {author} {\bibfnamefont {S.}~\bibnamefont {Liu}}, \bibinfo {author} {\bibfnamefont {T.}~\bibnamefont {Ohtsuki}},\ and\ \bibinfo {author} {\bibfnamefont {R.}~\bibnamefont {Shindou}},\ }\bibfield  {title} {\bibinfo {title} {Effect of disorder in a three-dimensional layered chern insulator},\ }\href {https://doi.org/10.1103/PhysRevLett.116.066401} {\bibfield  {journal} {\bibinfo  {journal} {Phys. Rev. Lett.}\ }\textbf {\bibinfo {volume} {116}},\ \bibinfo {pages} {066401} (\bibinfo {year} {2016}{\natexlab{b}})}\BibitemShut {NoStop}%
\bibitem [{\citenamefont {Zhang}\ \emph {et~al.}(2009)\citenamefont {Zhang}, \citenamefont {Hu}, \citenamefont {Bernevig}, \citenamefont {Wang}, \citenamefont {Xie},\ and\ \citenamefont {Liu}}]{zhang2009localization}%
  \BibitemOpen
  \bibfield  {author} {\bibinfo {author} {\bibfnamefont {Y.-Y.}\ \bibnamefont {Zhang}}, \bibinfo {author} {\bibfnamefont {J.}~\bibnamefont {Hu}}, \bibinfo {author} {\bibfnamefont {B.~A.}\ \bibnamefont {Bernevig}}, \bibinfo {author} {\bibfnamefont {X.~R.}\ \bibnamefont {Wang}}, \bibinfo {author} {\bibfnamefont {X.~C.}\ \bibnamefont {Xie}},\ and\ \bibinfo {author} {\bibfnamefont {W.~M.}\ \bibnamefont {Liu}},\ }\bibfield  {title} {\bibinfo {title} {Localization and the kosterlitz-thouless transition in disordered graphene},\ }\href {https://doi.org/10.1103/PhysRevLett.102.106401} {\bibfield  {journal} {\bibinfo  {journal} {Phys. Rev. Lett.}\ }\textbf {\bibinfo {volume} {102}},\ \bibinfo {pages} {106401} (\bibinfo {year} {2009})}\BibitemShut {NoStop}%
\bibitem [{\citenamefont {Geim}\ \emph {et~al.}(1992)\citenamefont {Geim}, \citenamefont {Main}, \citenamefont {Beton}, \citenamefont {Eaves}, \citenamefont {Beaumont},\ and\ \citenamefont {Wilkinson}}]{geim1992breakdown}%
  \BibitemOpen
  \bibfield  {author} {\bibinfo {author} {\bibfnamefont {A.~K.}\ \bibnamefont {Geim}}, \bibinfo {author} {\bibfnamefont {P.~C.}\ \bibnamefont {Main}}, \bibinfo {author} {\bibfnamefont {P.~H.}\ \bibnamefont {Beton}}, \bibinfo {author} {\bibfnamefont {L.}~\bibnamefont {Eaves}}, \bibinfo {author} {\bibfnamefont {S.~P.}\ \bibnamefont {Beaumont}},\ and\ \bibinfo {author} {\bibfnamefont {C.~D.~W.}\ \bibnamefont {Wilkinson}},\ }\bibfield  {title} {\bibinfo {title} {Breakdown of universal scaling of conductance fluctuations in high magnetic fields},\ }\href {https://doi.org/10.1103/PhysRevLett.69.1248} {\bibfield  {journal} {\bibinfo  {journal} {Phys. Rev. Lett.}\ }\textbf {\bibinfo {volume} {69}},\ \bibinfo {pages} {1248} (\bibinfo {year} {1992})}\BibitemShut {NoStop}%
\bibitem [{\citenamefont {Bergmann}(1994)}]{bergmann1994different}%
  \BibitemOpen
  \bibfield  {author} {\bibinfo {author} {\bibfnamefont {G.}~\bibnamefont {Bergmann}},\ }\bibfield  {title} {\bibinfo {title} {Different coherence lengths in universal conductance fluctuations: A numerical calculation for the quasi-two-dimensional case},\ }\href {https://doi.org/10.1103/PhysRevB.49.8377} {\bibfield  {journal} {\bibinfo  {journal} {Phys. Rev. B}\ }\textbf {\bibinfo {volume} {49}},\ \bibinfo {pages} {8377} (\bibinfo {year} {1994})}\BibitemShut {NoStop}%
\bibitem [{\citenamefont {Hoadley}\ \emph {et~al.}(1999)\citenamefont {Hoadley}, \citenamefont {McConville},\ and\ \citenamefont {Birge}}]{hoadley1999experimental}%
  \BibitemOpen
  \bibfield  {author} {\bibinfo {author} {\bibfnamefont {D.}~\bibnamefont {Hoadley}}, \bibinfo {author} {\bibfnamefont {P.}~\bibnamefont {McConville}},\ and\ \bibinfo {author} {\bibfnamefont {N.~O.}\ \bibnamefont {Birge}},\ }\bibfield  {title} {\bibinfo {title} {Experimental comparison of the phase-breaking lengths in weak localization and universal conductance fluctuations},\ }\href {https://doi.org/10.1103/PhysRevB.60.5617} {\bibfield  {journal} {\bibinfo  {journal} {Phys. Rev. B}\ }\textbf {\bibinfo {volume} {60}},\ \bibinfo {pages} {5617} (\bibinfo {year} {1999})}\BibitemShut {NoStop}%
\bibitem [{\citenamefont {Gao}\ \emph {et~al.}(1989)\citenamefont {Gao}, \citenamefont {Caro}, \citenamefont {Verbruggen}, \citenamefont {Radelaar},\ and\ \citenamefont {Middelhoek}}]{gao1989temperature}%
  \BibitemOpen
  \bibfield  {author} {\bibinfo {author} {\bibfnamefont {J.~R.}\ \bibnamefont {Gao}}, \bibinfo {author} {\bibfnamefont {J.}~\bibnamefont {Caro}}, \bibinfo {author} {\bibfnamefont {A.~H.}\ \bibnamefont {Verbruggen}}, \bibinfo {author} {\bibfnamefont {S.}~\bibnamefont {Radelaar}},\ and\ \bibinfo {author} {\bibfnamefont {J.}~\bibnamefont {Middelhoek}},\ }\bibfield  {title} {\bibinfo {title} {Temperature dependence of universal conductance fluctuations in narrow mesoscopic si inversion layers},\ }\href {https://doi.org/10.1103/PhysRevB.40.11676} {\bibfield  {journal} {\bibinfo  {journal} {Phys. Rev. B}\ }\textbf {\bibinfo {volume} {40}},\ \bibinfo {pages} {11676} (\bibinfo {year} {1989})}\BibitemShut {NoStop}%
\bibitem [{\citenamefont {Yang}\ \emph {et~al.}(2012)\citenamefont {Yang}, \citenamefont {Wang}, \citenamefont {Hsu},\ and\ \citenamefont {Lin}}]{Yang2012}%
  \BibitemOpen
  \bibfield  {author} {\bibinfo {author} {\bibfnamefont {P.-Y.}\ \bibnamefont {Yang}}, \bibinfo {author} {\bibfnamefont {L.~Y.}\ \bibnamefont {Wang}}, \bibinfo {author} {\bibfnamefont {Y.-W.}\ \bibnamefont {Hsu}},\ and\ \bibinfo {author} {\bibfnamefont {J.-J.}\ \bibnamefont {Lin}},\ }\bibfield  {title} {\bibinfo {title} {Universal conductance fluctuations in indium tin oxide nanowires},\ }\href {https://doi.org/10.1103/PhysRevB.85.085423} {\bibfield  {journal} {\bibinfo  {journal} {Phys. Rev. B}\ }\textbf {\bibinfo {volume} {85}},\ \bibinfo {pages} {085423} (\bibinfo {year} {2012})}\BibitemShut {NoStop}%
\bibitem [{\citenamefont {Pereira}\ \emph {et~al.}(2009)\citenamefont {Pereira}, \citenamefont {Castro~Neto},\ and\ \citenamefont {Peres}}]{pereira2009tight}%
  \BibitemOpen
  \bibfield  {author} {\bibinfo {author} {\bibfnamefont {V.~M.}\ \bibnamefont {Pereira}}, \bibinfo {author} {\bibfnamefont {A.~H.}\ \bibnamefont {Castro~Neto}},\ and\ \bibinfo {author} {\bibfnamefont {N.~M.~R.}\ \bibnamefont {Peres}},\ }\bibfield  {title} {\bibinfo {title} {Tight-binding approach to uniaxial strain in graphene},\ }\href {https://doi.org/10.1103/PhysRevB.80.045401} {\bibfield  {journal} {\bibinfo  {journal} {Phys. Rev. B}\ }\textbf {\bibinfo {volume} {80}},\ \bibinfo {pages} {045401} (\bibinfo {year} {2009})}\BibitemShut {NoStop}%
\bibitem [{\citenamefont {Nakatsuji}\ \emph {et~al.}(2012)\citenamefont {Nakatsuji}, \citenamefont {Yoshimura}, \citenamefont {Komori}, \citenamefont {Morita},\ and\ \citenamefont {Tanaka}}]{nakatsuji2012uniaxial}%
  \BibitemOpen
  \bibfield  {author} {\bibinfo {author} {\bibfnamefont {K.}~\bibnamefont {Nakatsuji}}, \bibinfo {author} {\bibfnamefont {T.}~\bibnamefont {Yoshimura}}, \bibinfo {author} {\bibfnamefont {F.}~\bibnamefont {Komori}}, \bibinfo {author} {\bibfnamefont {K.}~\bibnamefont {Morita}},\ and\ \bibinfo {author} {\bibfnamefont {S.}~\bibnamefont {Tanaka}},\ }\bibfield  {title} {\bibinfo {title} {Uniaxial deformation of graphene dirac cone on a vicinal sic substrate},\ }\href {https://doi.org/10.1103/PhysRevB.85.195416} {\bibfield  {journal} {\bibinfo  {journal} {Phys. Rev. B}\ }\textbf {\bibinfo {volume} {85}},\ \bibinfo {pages} {195416} (\bibinfo {year} {2012})}\BibitemShut {NoStop}%
\bibitem [{\citenamefont {Carr}\ \emph {et~al.}(2017)\citenamefont {Carr}, \citenamefont {Massatt}, \citenamefont {Fang}, \citenamefont {Cazeaux}, \citenamefont {Luskin},\ and\ \citenamefont {Kaxiras}}]{carr2017twistronics}%
  \BibitemOpen
  \bibfield  {author} {\bibinfo {author} {\bibfnamefont {S.}~\bibnamefont {Carr}}, \bibinfo {author} {\bibfnamefont {D.}~\bibnamefont {Massatt}}, \bibinfo {author} {\bibfnamefont {S.}~\bibnamefont {Fang}}, \bibinfo {author} {\bibfnamefont {P.}~\bibnamefont {Cazeaux}}, \bibinfo {author} {\bibfnamefont {M.}~\bibnamefont {Luskin}},\ and\ \bibinfo {author} {\bibfnamefont {E.}~\bibnamefont {Kaxiras}},\ }\bibfield  {title} {\bibinfo {title} {Twistronics: Manipulating the electronic properties of two-dimensional layered structures through their twist angle},\ }\href {https://doi.org/10.1103/PhysRevB.95.075420} {\bibfield  {journal} {\bibinfo  {journal} {Phys. Rev. B}\ }\textbf {\bibinfo {volume} {95}},\ \bibinfo {pages} {075420} (\bibinfo {year} {2017})}\BibitemShut {NoStop}%
\bibitem [{\citenamefont {Yang}(2025)}]{Yang2025}%
  \BibitemOpen
  \bibfield  {author} {\bibinfo {author} {\bibfnamefont {Q.}~\bibnamefont {Yang}},\ }\bibfield  {title} {\bibinfo {title} {Data for “fermi-energy-sensitive universal conductance fluctuations in anisotropic materials”}\ }\href {https://doi.org/https://doi.org/10.6084/m9.figshare.29117462.v1} {https://doi.org/10.6084/m9.figshare.29117462.v1} (\bibinfo {year} {2025})\BibitemShut {NoStop}%
\end{thebibliography}

%

\end{document}